\documentclass[reprint,
 superscriptaddress,
 showpacs,
 nofootinbib,
 amsmath,amssymb, aps,
 prd,
floatfix,
]{revtex4-1}

\usepackage{graphicx}
\usepackage[caption=false]{subfig}
\usepackage{dcolumn}
\usepackage{multirow}
\usepackage{tabularx}
\usepackage{bm}
\usepackage{hyperref}
\usepackage{amsmath}
\usepackage{amssymb}
\usepackage{comment}
\usepackage{color}
\usepackage{enumerate}
\usepackage{booktabs}
\usepackage{siunitx}

\newcommand{\GeV}{\text{\ GeV}}

\newcommand{\keV}{\text{\ keV}}

\newcommand{\cm}{\ \text{cm}}
\newcommand{\second}{\ \text{s}}

\newcommand{\Var}{\operatorname{Var}}
\newcommand{\Mean}[1]{\left\langle #1 \right\rangle}
\newcommand{\Hyp}{\mathcal{H}}

\newcolumntype{L}[1]{>{\hsize=#1\hsize\raggedright\arraybackslash}X}%
\newcolumntype{R}[1]{>{\hsize=#1\hsize\raggedleft\arraybackslash}X}%
\newcolumntype{C}[1]{>{\hsize=#1\hsize\centering\arraybackslash}X}%

\makeatletter
\newcommand{\thickhline}{%
    \noalign {\ifnum 0=`}\fi \hrule height 2pt
    \futurelet \reserved@a \@xhline
}
\newcolumntype{"}{@{\hskip\tabcolsep\vrule width 2pt\hskip\tabcolsep}}
\makeatother

\newcommand{\Eq}[1]{Eq.~(\ref{#1})}
\newcommand{\Eqs}[2]{Eqs.~(\ref{#1})~\&~(\ref{#2})}
\newcommand{\Fig}[1]{Fig.~\ref{#1}}

\newcommand{\Sec}[1]{Sec.~\ref{#1}}

\begin{document}

\title{
Towards a Bullet-proof test for indirect signals of dark matter\\
}

\author{Peter W. Graham}\email{pwgraham@stanford.edu}\affiliation{Stanford Institute for Theoretical Physics, Stanford University, Stanford, CA 94305, USA}
\author{Surjeet Rajendran}\email{surjeet@berkeley.edu}\affiliation{Berkeley Center for Theoretical Physics, University of California, Berkeley, CA 94720, USA}
\author{Ken Van Tilburg}\email{kenvt@stanford.edu}\affiliation{Stanford Institute for Theoretical Physics, Stanford University, Stanford, CA 94305, USA}
\author{Timothy D. Wiser}\email{tdwiser@stanford.edu}\affiliation{Stanford Institute for Theoretical Physics, Stanford University, Stanford, CA 94305, USA}

\date{\today}

\begin{abstract}
Merging galaxy clusters such as the Bullet Cluster provide a powerful testing ground for indirect detection of dark matter. The spatial distribution of the dark matter is both directly measurable through gravitational lensing and substantially different from the distribution of potential astrophysical backgrounds. We propose to use this spatial information to identify the origin of indirect detection signals, and we show that even statistical excesses of a few sigma can be robustly tested for consistency---or inconsistency---with a dark matter source. For example, our methods, combined with already-existing observations of the Coma Cluster, would allow the $3.55 \keV$ line to be tested for compatibility with a dark matter origin. We also discuss the optimal spatial reweighting of photons for indirect detection searches. The current discovery rate of merging galaxy clusters and associated lensing maps strongly motivates deep exposures in these dark matter targets for both current and upcoming indirect detection experiments in the X-ray and gamma-ray bands.

\end{abstract}
\maketitle

\tableofcontents

\section{Introduction}\label{sec:intro}

Over the course of the last century, astrophysical and cosmological observations have provided conclusive evidence of a non-luminous form of matter---dark matter (DM)---which comprises most of the matter content in our Universe.  
Signatures of the gravitational influence of the DM energy density have been seen in the rotation curves of galaxies, orbital velocities of galaxies in galaxy clusters, strong and weak gravitational lensing, the cosmic microwave background, and large scale structure~\cite{Bertone:2004pz}.  
Nevertheless, the mass and possible non-gravitational couplings of the DM particle(s) remain unknown. 
There exist relatively weak model-independent upper and lower bounds on the DM mass, as well as a variety of constraints on its non-gravitational interactions from astrophysical observations, collider searches, direct detection experiments, and indirect detection experiments (see Refs.~\cite{Gelmini:2015zpa,Essig:2013goa,Funk:2013gxa,Askew:2014kqa,Cushman:2013zza} for reviews).

The latter category of experiments searches for photons\footnote{High-energy neutrinos~\cite{Barger:2001ur}, or electron-positron pairs~\cite{Cirelli:2008pk} are also interesting, but the methods in our paper are not relevant for these detection channels because of their lack of angular resolution.} produced via decays or (semi-)annihilations of DM particles in astrophysical objects such as the galactic center (GC), dwarf spheroidal galaxies, or clusters of galaxies. 
For example, a DM particle decaying to two photons would produce a monochromatic excess of photons with energy equal to half the DM mass---a line in the photon energy spectrum. 
A variety of experiments have been performed or are currently underway, particularly in the X-ray and gamma-ray energy bands.

Most of the above searches have so far only yielded constraints, though several tantalizing anomalies have occasionally appeared in the data. 
Evidence for an intense line at $511 \keV$ from the GC has accumulated over the last 30 years (see Ref.~\cite{2011RvMP...83.1001P} for a review).
Several groups have identified a broad excess of $1$--$10\GeV$ photons originating in the GC in the Fermi-LAT data~\cite{Goodenough:2009gk,Hooper:2010mq,Hooper:2011ti,Daylan:2014rsa,Abazajian:2012pn,Macias:2013vya,Abazajian:2014fta}.  
Most recently, a line-shaped excess of $3.55\keV$ photons has been identified in a spectrum of stacked galaxy clusters~\cite{0004-637X-789-1-13} in both XMM-Newton and Chandra data, and independently in the Andromeda galaxy and the Perseus cluster~\cite{Boyarsky:2014jta} and the galactic center~\cite{Boyarsky:2014ska} with XMM-Newton data.

A monochromatic photon line is usually considered to be a ``smoking gun'' signature of dark matter, but the energy spectrum of the signal may not be sufficient to confirm a dark matter origin of an excess.
The positrons needed to fuel the bright $511\keV$ line in the GC may be provided by radioactive decays of unstable nuclei produced in supernovae or massive stars, though a significant contribution from DM sources is not excluded~\cite{2011RvMP...83.1001P}.
While consistent with DM annihilation, the broad excess at $1$--$10\GeV$ in the galactic center could be due to other astrophysical sources such as unresolved millisecond pulsars~\cite{Abazajian:2012pn}.
Lastly, the $3.55\keV$ line is consistent with, e.g., a $7.1\keV$ sterile neutrino decay, but could also arise from known emission lines in the same energy range~\cite{Jeltema:2014qfa}.

When does a positive signal constitute a discovery? Additional input beyond the energy spectrum alone may be necessary to distinguish a DM signal from other astrophysical processes.
The spatial distribution of the signal in the sky is the most obvious handle for this purpose. 
One can mask known point sources from the analysis (e.g.~Sgr A* in the GC), or test whether an excess is consistent with DM annihilation or decay given a certain model for the DM mass density $\rho(r)$ in the Milky Way (MW). 
However, significant uncertainties arise in any quantitative analysis because the radial DM density profile in the inner regions of the MW has not been experimentally measured and can only be estimated from $N$-body simulations~\cite{Springel:2008cc}. 
In addition, the spatial distribution of possible background sources such as pulsars and supernova remnants in the MW remains poorly constrained.
Hence, the spatial distribution of an excess (as compared to the distribution of DM and background sources) in the MW cannot be considered a definitive test of a DM signal. The same is true for sources such as globular clusters and nearby dwarf spheroidal galaxies (dSphs), whose DM content has only been measured indirectly (e.g. by the motion of their stars): the spatial profile of DM in these objects is typically an input for the calculation of their total mass. In galaxy clusters, however, the mass distribution of DM can be measured with weak gravitational lensing. Moreover, the major source of backgrounds in clusters---the intracluster medium (ICM)---emits X-ray radiation and therefore has a measurable spatial distribution.

{\bf We advocate performing indirect detection studies in \emph{merging} galaxy clusters, where the dark matter is physically separated from the baryons---and thus most backgrounds.} The spatial separation between the DM and the bulk of the baryons can be exploited to construct powerful, quantitative tests of a potential DM signal. Although the GC and dSphs are expected to be much brighter sources of DM, and therefore more promising for the initial detection of a signal, merging clusters offer the unique possibility of comparing the spatial distribution of a signal with the known, distinctive distribution of DM.

\begin{figure}[t]%
\centering
\subfloat[]{\includegraphics[width=0.45\textwidth]{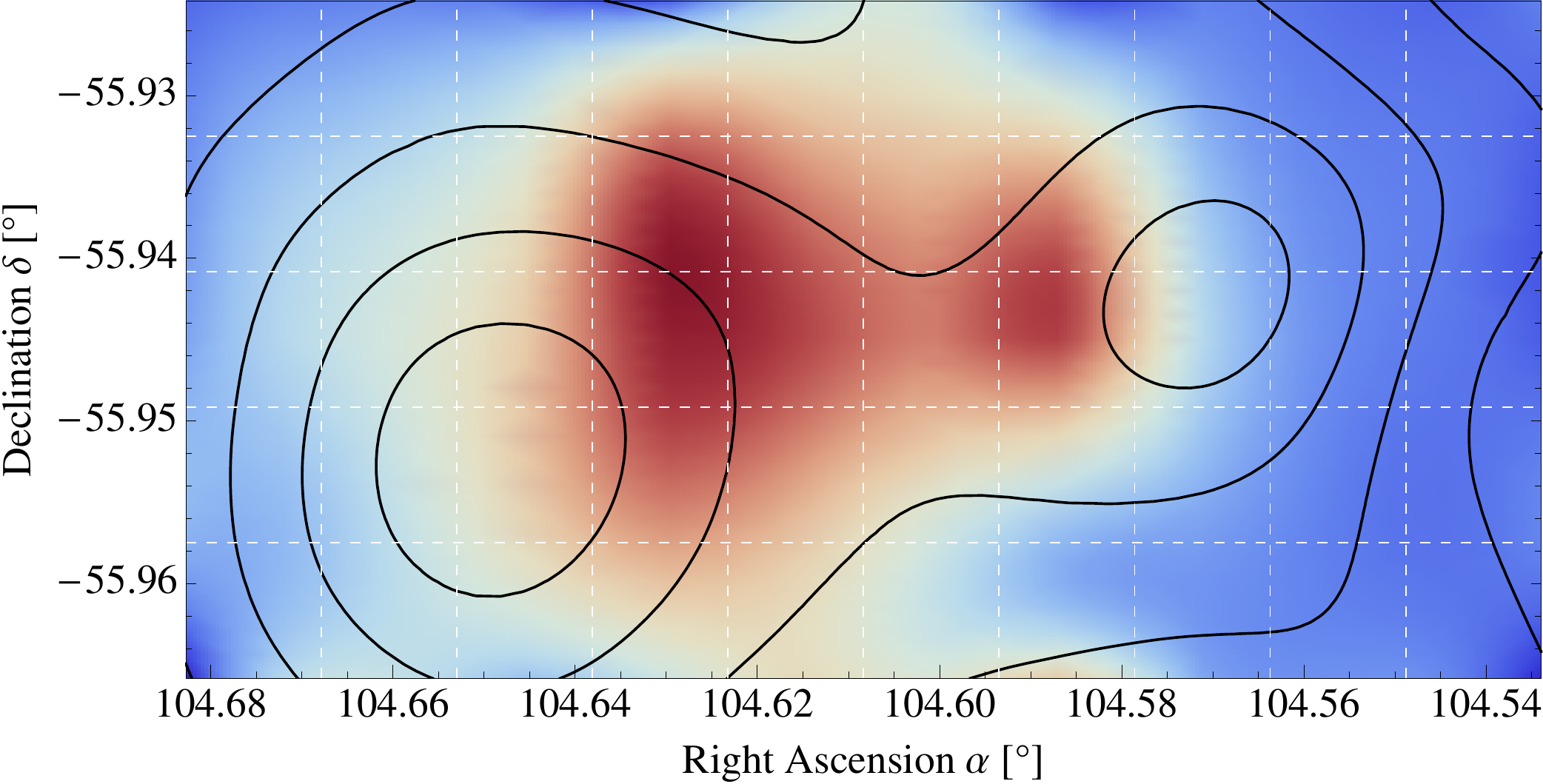} \label{fig:bulletmap}}\\
\subfloat[]{\includegraphics[width=0.45\textwidth]{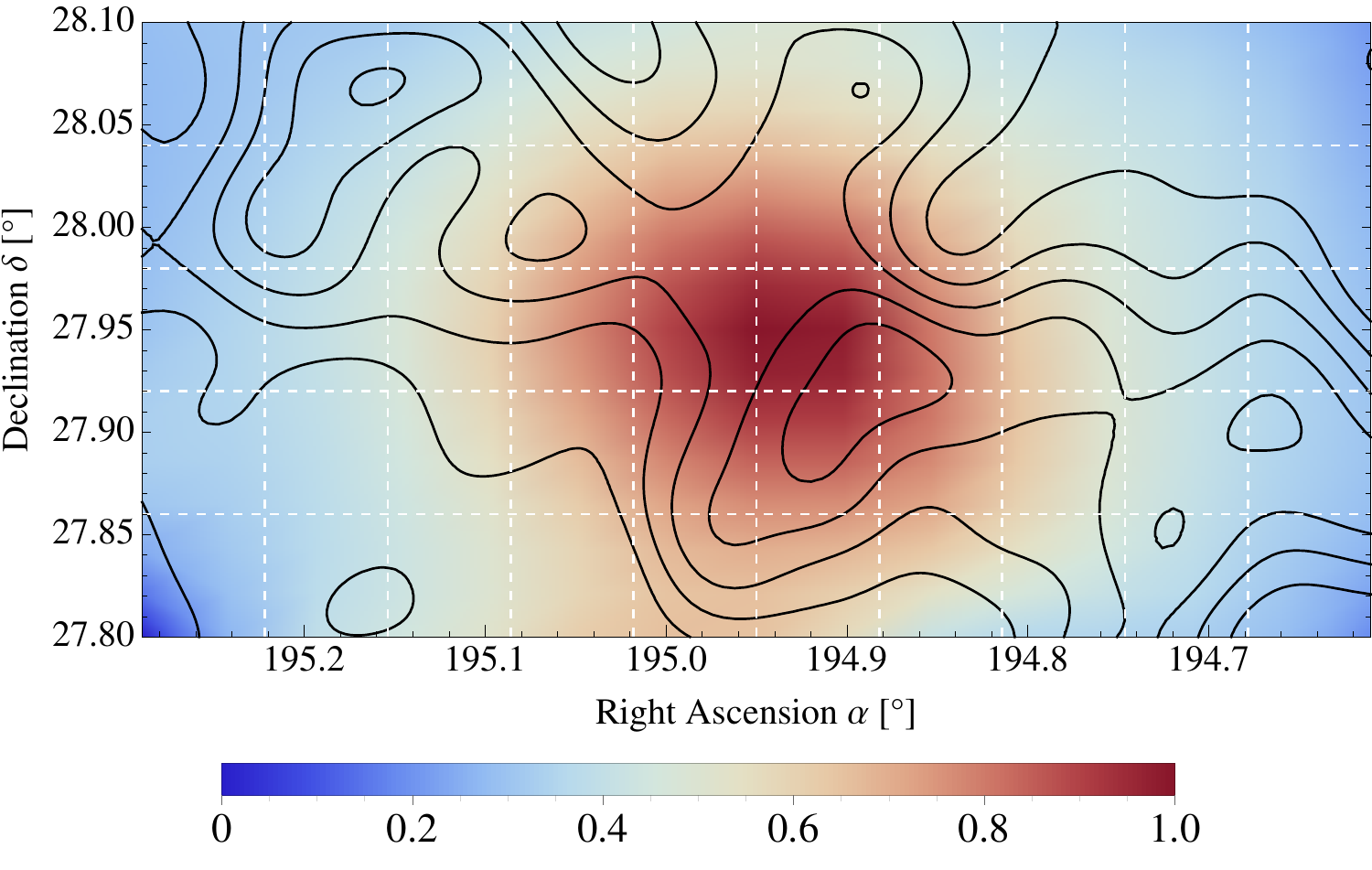} \label{fig:comamap}}
\caption{(Color online) Weak gravitational lensing contours and X-ray luminosity maps of (a) the Bullet Cluster~\cite{Clowe:2006eq,data:bullet_lens,data:xmm_0112980201} and (b) the Coma Cluster~\cite{Okabe:2013oza,data:rosat_RH800242A04}, displaying the partial separation of the dark matter and the intracluster medium. The lensing maps of (a)~\&~(b) were smoothed with Gaussian kernels of $\sigma = 20''$~\&~$1.7'$, respectively; contours signify integer multiples of signal-to-noise ratios (with $S/N = 1$ at the outermost contours). The X-ray density maps are proportional to the intensity of X-ray photons in the (a) 0.2--15~keV and (b) 0.2--2~keV energy band. Gridlines represent the binning ($30''$ \& $4'$, respectively) to be used in quantitative results in Sec.~\ref{sec:results}.}\label{fig:maps}
\end{figure}

The first, spectacular evidence of a merger event causing separation of dark and baryonic matter was observed in the galaxy cluster 1E 0657-558, the Bullet Cluster~\cite{Clowe:2006eq}.
Since then, lensing map reconstructions have identified many similar clusters. {Major merger events typically lead to increased radio emission on the outskirts of galaxies~\cite{2003ApJ...583..695G,2003ApJ...584..190F}. Giant radio arcs are thus an easy ``tag'' of merging clusters, leading to the rapidly increasing availability of weak lensing maps for this class of clusters.} Most notably, the nearby A1656---the Coma Cluster---was discovered to have undergone a merger leading to a separation of the DM and ICM~\cite{1996ApJ...473L..71H,Okabe:2013oza,2009A&A...498L..33G}. In Table~\ref{tab:clusterinfo}, we list relevant properties of the Bullet and Coma Clusters as well as several other merging clusters. We will use the Bullet and Coma Clusters as the two benchmark clusters of our paper; their lensing contours and X-ray luminosity maps are shown in \Fig{fig:maps}.

\newcolumntype{d}[1]{D{.}{.}{#1} }
\begin{center}
\begin{table*}[t]
\renewcommand{\arraystretch}{1.1}
\setlength{\tabcolsep}{14pt}
\begin{minipage}[t]{0.99\textwidth}
  \begin{tabular*}{\textwidth}{l | d{1.2} | d{1.4} | d{2.1} | d{1.1} | d{1.1} }
 &  \multicolumn{1}{c|}{$d_L$\footnotemark[1]} & \multicolumn{1}{c|}{$z$} &  \multicolumn{1}{c|}{$\Delta \theta_{\text{DM}}$}  &  \multicolumn{1}{c|}{$M_{200}$}  &  \multicolumn{1}{c}{$r_{200}$} \\ 
 &  \multicolumn{1}{c|}{[Gpc]} &  &  \multicolumn{1}{c|}{[arcmin]} &  \multicolumn{1}{c|}{$[h^{-1} 10^{15} M_\odot]$} &   \multicolumn{1}{c}{$[h^{-1} \text{\,Mpc}]$} \\
\hline\hline
A1656 ``Coma"~\cite{Kubo:2007wt,Okabe:2013oza}
			& 0.10 & 0.0231      	& 30     	& 1.9 			& 2.0 	\\
1E 0657-558 ``Bullet"~\cite{2004ApJ...604..596C}    	
			& 1.5	 & 0.296   		& 2.7		& 2.2\footnotemark[2]		& 1.6\footnotemark[2]\\ \hline
A3376~\cite{Durret:2013mna,Machado:2013jq}					 	
			& 0.20  & 0.046 		& \multicolumn{1}{D{,}{\phantom{.}}{2.1}|}{10,\mkern-\medmuskip\footnotemark[3]}			& 0.3			& 1.1		\\
			A520 ``Train Wreck"~\cite{Jee:2014hja,2012ApJ...758..128C}	
			& 0.97   & 0.199  		& 3			& 0.7			& 1.3	\\
A2163~\cite{Okabe:2011dt}					 	
			& 0.99   & 0.201  		& 1.6		& 2.0		 	& 1.8	\\
A1758~\cite{Ragozzine:2011qt}					 	
			& 1.4  & 0.279 		& 1		 	& 1.5\footnotemark[2]			& 1.6\footnotemark[2]	\\ 
A2744 ``Pandora"~\cite{2012AASP....2...56B,Merten:2011wj}	 	
			& 1.6   & 0.308  		& 2			& 1.5			& 1.7	\\
DLSCL J0916.2+2951 ``Musket Ball"~\cite{2012ApJ...747L..42D}
			& 3.1   & 0.533     	& 2.8		& 2.2		& 0.7 	\\
MACS J0025.4-1222~\cite{Bradac:2008eu}			
			& 3.4   & 0.586 		& 1.3		& \multicolumn{1}{c|}{-}			& \multicolumn{1}{c}{-}		\\ 
  \end{tabular*}
  \footnotetext[1]{assuming $H_0 = 70 \text{\,km}\text{\,s}^{-1}\text{\,Mpc}^{-1}$ $(h = 0.7)$, $\Omega_{\Lambda} = 0.7$, and $\Omega_{m} = 0.3$.}
  \footnotetext[2]{main subcluster only}
  \footnotetext[3]{no lensing data available; estimate from the distribution of brightest galaxies}
  \end{minipage}
  \caption{List of merging galaxy clusters, with their luminosity distance $d_L$, redshift $z$, angular dark matter separation $\Delta \theta_\textrm{DM}$ (as estimated from the distance between the two most prominent maxima in the weak lensing map), mass $M_{200}$, and radius $r_{200}$.  The Bullet and Coma Clusters are the two benchmark clusters in this paper. $M_{200}$ and $r_{200}$ should be thought of as rough estimates for the mass and radius of the cluster, as the determination of these parameters assumes spherical symmetry that merging clusters do not have. We refer the reader to the references for details of the uncertainties and assumptions that enter into these parameters.
  }\label{tab:clusterinfo}
\end{table*}
\end{center}

We will show how the weak lensing map and the X-ray luminosity map, representative of the DM density and ICM density spatial distributions, respectively, can be used to improve indirect detection studies. While the basic idea is simple---cluster mergers separate DM and ICM so that they can be distinguished---there is still a considerable amount of overlap between the distributions, and some statistical techniques are necessary to fully utilize the spatial information. To that end, we develop three major procedures in this paper:
\begin{itemize}
\item {\bf Method A:} If an excess of photons is observed in a merging cluster, the spatial distribution of the signal can be used to discriminate between DM decay and ICM emission (or any other pair of known spatial distributions) as the origin of the excess. We develop a quantitative procedure for this discrimination and show that it is quite powerful in realistic scenarios (e.g.~the $3.55\keV$ line).

\item {\bf Method B:} We develop a procedure to characterize the spatial distributions of potential DM excesses by fitting to a combination of multiple spatial templates. This procedure is useful when the excess could be from a combination of sources, and we  specifically use it to allow for uncertainty in the spatial distribution of annihilating DM. This procedure generalizes Method A and is equally powerful.

\item {\bf Method C:} Even if no definitive excess is observed, one can reweight the observed photons based on their spatial position to either enhance potential spectral features or to improve the exclusion limit (on, e.g., the DM annihilation cross-section or lifetime). Whereas previous studies~\cite{Boyarsky:2006kc,Weniger:2012tx} have masked out regions with low expected signal-to-background, reweighting allows all of the photons to contribute in an optimal way, improving the expected discovery reach for DM in merging clusters.

\end{itemize}
These procedures only depend on the spatial distributions of background and signal and the angular resolutions of the observing instrument and lensing map, and they apply equally well to both broad excesses and monochromatic lines.

While lensing maps do have uncertainties (a dramatic example is the ``dark peaks'' issue in A520 \cite{Jee:2014hja,2012ApJ...758..128C}) we have attempted to make our methods robust against such uncertainties by using the entire lensing map rather than one or more individual peaks. We analyze the robustness of our procedure against lensing map uncertainties in \Sec{sec:statproc_test} and conclude that they are unimportant for the small signals typical of potential DM excesses.

Prior studies have used spatial information as a diagnostic for potential DM signals, but always in the GC or non-merging clusters, and based on expectations from simulations rather than measured distributions~\cite{2010MNRAS.407.1188B,2010arXiv1001.4055K}. Recently, a spatial likelihood method similar to one we describe in \Sec{sec:statproc_test} was used in an attempt to characterize the origin of the $3.55\text{ keV}$ excess in the GC and in the Perseus cluster~\cite{Carlson:2014lla}. We advocate instead performing these spatial tests on merging clusters (especially the nearby Coma Cluster), in which the spatial distribution of DM is both measured (by weak lensing) and quite different from the distribution of potential backgrounds. 
Similarly, Ref.~\cite{Maurin:2012tv} proposed a method to distinguish DM decays or annihilations into gamma rays from the cosmic ray-induced background in non-merging clusters. Their method is based on comparing the slope of the signal distribution with various models of DM and ICM spatial profiles, whereas our methods are based on the measured surface mass density and use all of the available spatial information in an optimal way.

Spatial information has also been used in an attempt to strengthen limits and boost discovery potential: the weak lensing map and DM--ICM separation of the Bullet Cluster were used to set limits on sterile neutrinos~\cite{Boyarsky:2006kc}. However, their procedure actually resulted in a less stringent limit overall. Our reweighting procedure, described in \Sec{sec:statproc_weight} and applied to the Bullet Cluster in \Sec{sec:results_weight}, takes all of the DM into account and can strengthen the Bullet Cluster limit on the lifetime of a sterile neutrino by a factor of $\sim$1.2. An analogous reweighting procedure, based on both spatial information and overall signal-to-background ratio in a stack of dSph galaxies was performed in Ref.~\cite{Geringer-Sameth:2014qqa} to search for DM annihilating to gamma rays. By contrast, our reweighting procedure (Method C), which is based on a measured weak lensing map, cannot directly improve searches for annihilating DM. However, once observed, an annihilating DM signal could be tested by our template fit (Method B).

We describe our statistical methods~A,~B, and~C~in Sections~\ref{sec:statproc_test},~\ref{sec:statproc_fit}, and~\ref{sec:statproc_weight}, respectively. In Section~\ref{sec:application} we describe the range of applicability of our methods to indirect detection experiments in various energy ranges.
In Sections~\ref{sec:results_test},~\ref{sec:results_fit}, and~\ref{sec:results_weight}, we demonstrate the power of our methods in a few realistic scenarios. 
Finally, we discuss the implications of the techniques proposed here for future indirect detection studies in Sec.~\ref{sec:discussion}.
Appendix \ref{sec:app_conslim} details an extension to our statistical reweighting procedure for optimizing conservative (as opposed to background-subtracted) limits.

\section{Statistical procedure} 
\label{sec:statproc}

\subsection{Spatial tests} \label{sec:statproc_test}
In this section, we study to what extent we can use the spatial information of a signal from a galaxy cluster to test for consistency with different spatial distributions. We will first consider the simple case of distinguishing between two definite spatial distributions (leaving a generalization to more than two to \Sec{sec:statproc_fit}). This technique can be used to differentiate between two possible signal origins with known spatial distributions, such as DM decay (true signal) vs.~ICM emission (background). 

Concretely, let us assume the signal sits on top of a large background with normalized, fractional spatial distribution $b_i$, where $i$ labels spatial bins; e.g., a spatially uniform distribution with $k$ number of bins would have $b_i = 1/k$ for all $i$.
For simplicity, we take the photon counts in each spatial bin $i$ to be independent random variables, which is a reasonable assumption as long as the bin size is greater than or equal to the angular resolution of the observing instrument.

Suppose an excess of photons is observed in a particular energy band, with a significance of $s$ number of standard deviations; assuming large statistics, this amounts to an observation of $N + s\sqrt{N}$ photons where only $N$ were expected.
If the excess of photons is due to a component with a fractional spatial distribution $f_i$ (with $\sum_i f_i = 1$), then the data---given by the number of counts $x_i$ in each bin---should be consistent with being a Poisson random variable:
\begin{align}
\mathcal{H}_0 : \; x_i \sim \operatorname{Pois}\left( N b_i + s \sqrt{N} f_i \right),
\label{eq:H0}
\end{align}
a scenario which we call the null hypothesis $\mathcal{H}_0$.  The alternative hypothesis $\mathcal{H}_1$ is that the excess follows some other spatial distribution $g_i$, in which case the data should be distributed according to
\begin{align}
\mathcal{H}_1 : \; x_i \sim \operatorname{Pois}\left(  N b_i + s \sqrt{N} g_i \right),
\label{eq:H1}
\end{align}
where $\sum_i g_i = 1$.

We devise a simple statistical test to distinguish between these two hypotheses.  
As neither $\mathcal{H}_0$ nor $\mathcal{H}_1$ have unknown parameters, the Neyman-Pearson lemma states that the most powerful statistical test to distinguish between them is the likelihood ratio test, which rejects $\mathcal{H}_0$ in favor of $\mathcal{H}_1$ when
\begin{align}
\Lambda(\vec{x}) = F\left(\frac{\text{Prob}(\vec{x}|\mathcal{H}_1 )}{\text{Prob}(\vec{x}|\mathcal{H}_0)}\right) \ge \Lambda^*,
\label{eq:Lambda1}
\end{align}
where $F$ is any monotonically increasing function, and $\Lambda^*$ is determined by the test size $\alpha$ via $\text{Prob} \left( \Lambda(\vec{x}) \ge \Lambda^* | \mathcal{H}_0 \right) = \alpha.$
In other words, the Neyman-Pearson lemma ensures that $\Lambda(\vec{x})$ (or any monotonic function of the likelihood ratio) is the variable for which the pdfs under the two hypotheses $\mathcal{H}_0$ and $\mathcal{H}_1$ have the minimal amount of overlap, and therefore provides the strongest statistical discrimination between the two scenarios.

Assuming sufficiently large statistics\footnote{In particular, we do not require the number of signal counts per bin, $s\sqrt{N}f_i$ or $s\sqrt{N}g_i$, to be large. We discuss the practicality of this assumption further in \Sec{sec:application}.} in each spatially resolved bin---$N b_i \gg 1$ for each $i$---we can treat the Poisson distributions in \Eqs{eq:H0}{eq:H1} as Gaussians, and the most powerful discriminant takes the simple form
\begin{align}
\Lambda(\vec{x}) = \frac{1}{2N^{3/2}}\sum_i \frac{(g_i - f_i)}{b_i^2} x_i^2,
\label{eq:Lambda2}
\end{align}
where we took $F$ in \Eq{eq:Lambda1} to be a rescaled logarithmic function to simplify future expressions. The probability distribution of $\Lambda(\vec{x})$ is a weighted non-central $\chi^2$ distribution.  
There is no simple closed-form expression for this distribution, but each term in \Eq{eq:Lambda2} rapidly approaches a Gaussian for large $\Mean{x_i}$,\footnote{The skewness of each individual term in \Eq{eq:Lambda2} is parametrically suppressed as $\mathcal{O}(N b_i)^{-1/2}$, and the excess kurtosis as $\mathcal{O}(N b_i)^{-1}$.}
and the sum over spatial bins $i$ makes the pdf of $\Lambda$ converge even faster to a normal distribution (by the central limit theorem).
We will treat it as a Gaussian in the rest of the paper.\footnote{We have checked via Monte Carlo that the formula in \Eq{eq:Jvalue} for the statistical significance $\tilde{s}$ is a good approximation up to very large values of $s$ for all cases of interest. In the example considered in Sec.~\ref{sec:results_test}, the fractional error on $\tilde s$ is $\lesssim 1\%$ for $s \le 8$.}

Let us now extract the behavior of the $\Lambda$ test statistic of \Eq{eq:Lambda2} in the large-$N b_i$ limit. 
Assuming for the moment that $b_i$, $f_i$, and $g_i$ are perfectly known, we have for the expectation values and variance of $\Lambda$:
\begin{widetext}
\begin{alignat}{7}
	\Mean{\Lambda|\Hyp_1} - \Mean{\Lambda|\Hyp_0}
	&\simeq&~ \frac{1}{2N^{3/2}} \sum_{i=1}^k \frac{g_i-f_i}{b_i^2} \left[(Nb_i + s\sqrt{N}g_i)^2 - (Nb_i+s\sqrt{N}f_i)^2\right]
	&\simeq&~ s \sum_{i=1}^k\frac{(g_i-f_i)^2}{b_i} 
	&\equiv&~ T^2 s,\label{eq:parametric1}  \\
	\Var(\Lambda|\Hyp_{0,1})
	&\simeq&~ \frac{1}{4N^3}\sum_{i=1}^k \frac{(g_i-f_i)^2}{b_i^4}\Var(x_i^2) \simeq \frac{1}{4N^3}\sum_{i=1}^k \frac{(g_i-f_i)^2}{b_i^4}4(Nb_i)^3
	&\simeq&~ \sum_{i=1}^k \frac{(g_i-f_i)^2}{b_i}
	&\equiv&~ T^2. \label{eq:parametric2}
\end{alignat}
\end{widetext}

Because $p(\Lambda)$ is well-approximated by a normal distribution, the expected significance of the likelihood ratio test can be expressed as the number of standard deviations:
\begin{align}
\tilde{s} \simeq \frac{ \Mean{ \Lambda | \Hyp_1 } - \Mean{\Lambda | \Hyp_0}}{\sqrt{\Var(\Lambda|\Hyp_0)}} \simeq T s; \quad T \equiv \sqrt{\sum_{i=1}^k \frac{(g_i - f_i)^2}{b_i}}.\label{eq:Jvalue}
\end{align}
In the case of $\mathcal{O}(1)$ separation---$|g_i - f_i| \sim \mathcal{O}(1/k)$---the overlap factor $T$ can be close to unity (or even exceed it).  
In addition to being the statistically most powerful discriminant, the test statistic $\Lambda$ has the desirable property of being insensitive to the number of background photons $N$, and the details of spatial binning, as long as the gross substructure of $f_i$ and $g_i$ is resolved.
We shall see in Section~\ref{sec:results_test} that $\Lambda$ is a powerful discriminant between $\mathcal{H}_0$ and $\mathcal{H}_1$ in realistic scenarios such as discriminating between DM and ICM origins of an X-ray excess.  

So far, we have only considered statistical contributions to the variance of $x_i$ (and thus $\Lambda$), ignoring possible systematic uncertainty in the spatial distributions $b_i$, $f_i$, and $g_i$.  
To leading order in the number of photons $N$, the statistical variance of the counts in each bin is the same for both $\mathcal{H}_0$ and $\mathcal{H}_1$: $\text{Var}_\text{stat}(x_i) \simeq N b_i$.  
The systematic contribution to the variance of $x_i$ from an imperfect knowledge of $b_i$ and $f_i$ given the null hypothesis $\mathcal{H}_0$ is: 
\begin{align}
\text{Var}_\text{sys}(x_i) \simeq N^2 \text{Var}(b_i) + s^2 N \text{Var}(f_i) \label{eq:varsys},
\end{align}
with a similar expression (with $f_i \to g_i$) for the alternative $\Hyp_1$.
In principle, this uncertainty should be incorporated in the formula of \Eq{eq:Jvalue}: a larger variance in the pdf of $\Lambda$ given $\mathcal{H}_0$ would degrade the expected power of the test (to exclude $\mathcal{H}_0$), while a larger variance of $\Lambda$ given $\mathcal{H}_1$ implies a larger spread in the expected significance $\tilde{s}$ but with the same expected value $\langle \tilde{s}\rangle$. However, the first term in \Eq{eq:varsys} is smaller than the statistical variance $N b_i$ given a satisfactory background model $b_i$ (which, for example, can be measured from the side energy bands). The second term in \Eq{eq:varsys} can dominate over the statistical variance, but only for very large signals $s \gtrsim \sqrt{b_i/\text{Var}(f_i)} \sim \text{SNR}(f_i)\sqrt{k}$, and similarly for $g_i$. For example, the lensing maps in \Fig{fig:maps} have an average per-bin signal-to-noise ratio $\langle \text{SNR}(\kappa_i) \rangle$ greater than 2, for $k = 50$ number of bins. The uncertainty in the lensing map would thus only become important for signals with significance greater than about $s \sim 15$ in this case.


\subsection{Spatial template fits}\label{sec:statproc_fit}

The hypothesis test procedure described above is optimal (by the Neyman-Pearson lemma) for discriminating between two hypotheses with no free parameters. In a more realistic scenario there may be more than two spatial distributions of interest, e.g.~the temperature, number density of galaxies, etc., in addition to the mass and ICM distributions considered above. More crucially for our discussion, the spatial profile of annihilating DM in galaxy clusters is not well known, but is likely described by a mixture of a sharply peaked \emph{core} component that follows the mass density distribution raised to some power, and a \emph{substructure} component with a more uniform distribution, as we will discuss further in Sections~\ref{sec:application} and~\ref{sec:results_fit}.
In these circumstances, one way to proceed is by fitting to a linear interpolation of the relevant spatial templates. As above, we will work under the assumption that the errors on all bins can be considered to be Gaussian; in that case the maximum-likelihood estimate of the interpolation parameters coincides with the minimum-$\chi^2$ estimate. Additionally, the covariance matrix of the best-fit parameters (and, from that, the confidence intervals) can be extracted from the second derivatives of $\chi^2$ at the minimum.

Although the Neyman-Pearson lemma does not apply to this procedure, we will show that in particular cases the fitting procedure reproduces the results of the hypothesis test exactly. As a result, we expect that the fitting procedure adequately captures all of the spatial information available in the signal.

First, we describe the general case with an arbitrary number of candidate spatial templates, generalizing the two distributions $f$ and $g$ from above.
Suppose that we consider fitting an $s$ sigma excess (atop $N$ background photons) to $n$ spatial templates $f^a$, $a=1\ldots n$. (We always take spatial templates to be normalized to unit sum, $\sum_i f^a_i=1$.) Then we have $n$ parameters $\theta_a$ representing the fractional contribution of map $f^a$ to the excess. In other words, our model function for the distribution of counts $x_i$ (where $i$ labels the spatial bin) is
\begin{align}
	\mu_i(\{\theta_a\}) \equiv \Mean{ x_i | \{\theta_a\} } = Nb_i + s\sqrt{N}(\theta_a f^a_i).
\end{align}
We will constrain $\sum_a \theta_a = 1$, so that only $n-1$ of the parameters are free.

The best fit parameters $\hat\theta_a$ are those that minimize the chi-squared:
\begin{align}
	\chi^2(\{\theta_a\}) = \sum_i\frac{(x_i-\mu_i(\{\theta_a\}))^2}{\sigma_i^2},
\end{align}
and the $(n-1)\times(n-1)$ covariance matrix $V_{ab}\equiv \operatorname{Cov}(\theta_a,\theta_b)$, with $a,b = 1,\dots,n-1$, is calculated through its inverse as:
\begin{align}
	(V_{ab})^{-1} &= \left.\frac{1}{2}\frac{\partial^2\chi^2(\{\theta_c\})}{\partial\theta_a\partial\theta_b}\right|_{\theta_c=\hat\theta_c} \\
	&\simeq s^2 \sum_i \frac{(f^a_i-f^n_i)(f^b_i-f^n_i)}{b_i},\label{eq:covariancematrix}
\end{align}
where the second line follows from taking $Nb_i$ large.

The $1\sigma$ confidence interval for $\theta_a$ is $\hat\theta_a\pm\sqrt{\Var(\theta_a)}$; $\Var(\theta_a)=V_{aa}$ for $a=1\ldots n-1$, and $\Var(\theta_n) =\sum_{a,b=1}^{n-1}V_{ab}$. Note that $\Var(\theta_a)$ depends on {\em all} of the $f^a$, $a=1\ldots n$, through the matrix inverse. The covariance matrix is still determined by geometric quantities only: $s^2V_{ab}$ depends only on the spatial distributions, not their relative amplitudes. However, the degree to which a particular hypothesis (characterized by $\theta^0_a$) is excluded does depend on the best-fit amplitudes according to the formula
\begin{align}
	\tilde s = \sqrt{(\theta^0_a-\hat\theta_a)V^{-1}_{ab}(\theta^0_b-\hat\theta_b)},\label{eq:gen_exclusion}
\end{align}
where $\tilde s$ is the number of standard deviations to which the hypothesis $\theta^0_a$ is excluded. Taken as a function of $\theta^0_a$, \Eq{eq:gen_exclusion} defines the $\tilde s$-sigma error ellipsoid of the fit parameters. Since $V^{-1}_{ab}\propto s^2$, $\tilde s \propto s$ exactly as in \Eq{eq:Jvalue}.

Now we consider the strength of the fitting procedure applied to two spatial distributions and show that we reproduce the result of the optimal Neyman-Pearson hypothesis test.
For the case of two spatial distributions, the fit procedure provides a slight generalization to the hypothesis testing procedure by allowing for a single free parameter $\theta$ describing the fraction of the excess originating from the spatial distribution $f^1\equiv g$, while $1-\theta$ is the fraction of the excess originating from $f^2\equiv f$. In this case, \Eq{eq:covariancematrix} reduces from a matrix to a single number:
\begin{align}
	(\Var(\theta))^{-1} \simeq s^2 \sum_i \frac{(g_i-f_i)^2}{b_i}.
\end{align}
The expected best-fit parameter is $\hat\theta=0$ for the gas-only scenario (cf.~\Eq{eq:H0}) and $\hat\theta=1$ for the DM-only scenario (cf.~\Eq{eq:H1}). In either case, the expected significance of the result (exclusion of the alternative hypothesis) is simply
\begin{align}
	\tilde s \simeq \frac{1}{\sqrt{\Var(\theta)}} \simeq \sqrt{\sum_i \frac{(g_i-f_i)^2}{b_i}} s,
\end{align}
and we recover the result of \Eq{eq:Jvalue}: $\tilde s = T s$.

In fact, the fitting procedure is optimal (i.e.~equivalent to the Neyman-Pearson test) in a generalized case where we compare one distribution (WLOG $f^n$) to an arbitrary mixture of the rest. With null hypothesis $\theta^0_n=1$ and expected best-fit point $\hat\theta_n=0$, \Eq{eq:gen_exclusion} becomes
\begin{align}
	\tilde s \simeq \sqrt{\sum_i \frac{(\hat\theta_a f^a_i - f^n_i)^2}{b_i}}s.
\end{align}
Since $\hat\theta_a f^a_i$ is a properly normalized spatial template, we recognize \Eq{eq:Jvalue} with the best-fit signal distribution in place of $g_i$; in other words, the fitting procedure reproduces the optimal result of the hypothesis test despite not knowing the correct signal distribution in advance. 


\subsection{Spatial reweighting} \label{sec:statproc_weight}

In the absence of an obvious excess in the spectrum of photons from a galaxy cluster, spatial information of the DM distribution and the background distribution can still be useful to enhance a \emph{potential} signal relative to the background by spatially reweighting the photons.  A reweighting procedure that boosts any signal correlated with the dark matter distribution would aid in the identification of potential features and strengthen DM exclusion limits if none are found. In Sec.~\ref{sec:results_weight}, we will demonstrate the approach outlined below on a realistic (though hypothetical) example of an indirect detection analysis.

Suppose a total number of photons $B=\sum_{i=1}^k B_i$ are observed from a cluster, with $B_i$ counts in each spatial bin $i$, and that a DM signal would amount to $S=\sum_{i=1}^k S_i$ photons, with $S_i \propto g_i$ (the fractional spatial distribution of the signal).  If the total photon count is dominated by the background in each bin, we expect $B_i \propto b_i$ (the fractional spatial distribution of the background). Our goal is to determine the optimal weights $w_i$ such that $\hat S=\sum_i w_i S_i$ and $\hat B = \sum_i w_i B_i$ yield the maximum statistical significance. If there is a component of the data which follows the DM signal distribution, it will be enhanced relative to the background with appropriately chosen weights.

To maximize the statistical significance of a signal, we choose weights $w_i$ to maximize
\begin{align} 
\Sigma(\lbrace w_i \rbrace) \equiv \frac{\langle \hat S \rangle}{\sqrt{\Var \hat B}} = \frac{\sum_i w_i S_i}{\sqrt{\sum_i w_i^2 B_i}},\label{eq:targetfn}
\end{align}
the expected number of standard deviations of the excess above the background.  \Eq{eq:targetfn} applies under the assumptions that the dominant source of error is statistical and that each $B_i$ is large enough to be an approximately Gaussian random variable. The target function $\Sigma(\lbrace w_i \rbrace)$ is invariant under rescaling the weights, so we choose the normalization condition $\sum^k_{i=1} w_i = k$ so that the average of the weights $w_i$ is unity. Maximizing $\Sigma(\lbrace w_i \rbrace)$ subject to this constraint gives the optimal weights:
\begin{align}
 w^*_{i} = c \frac{S_i}{B_i},
\label{eq:SoverBweights}\end{align}
with $c = k/(\sum_{i=1}^k S_i/B_i)$ chosen to satisfy the normalization condition. Crucially, the $w^*_i$ are insensitive to uniformly rescaling $S_i$ (as any such rescaling will be absorbed into $c$), so the total flux produced by DM need not be known a priori to determine the optimal weights.

The boost in significance provided by this reweighting, relative to a simple integration of all bins ($w_i=1$ for all $i$), is given by
\begin{align} 
R \equiv  \frac{\Sigma(\lbrace w^*_i \rbrace)}{\Sigma(\lbrace  1 \rbrace)} = \frac{\sqrt{\sum_i S_i^2/B_i}}{\sum_i S_i / \sqrt{\sum_i B_i}} = \sqrt{\sum_{i=1}^k \frac{g_i^2}{b_i}}.\end{align}
The enhancement factor $R$ is insensitive to rescalings of both $S_i$ and $B_i$; like $T$ (the strength of the hypothesis test procedure in Sec.~\ref{sec:statproc_test}), $R$ is a constant that depends only on the spatial distributions of signal and background. In Sec.~\ref{sec:results_weight}, we discuss the practical application of this technique, including typical values of $R$ for several scenarios.

Spatial reweighting can also strengthen the exclusion limits derived from data in the absence of a signal. For limits derived using a background-subtraction procedure, the optimal weights are also given by \Eq{eq:SoverBweights}. For conservative limits that do not perform background subtraction, a different weighting procedure is optimal; see Appendix \ref{sec:app_conslim}.

\section{Application to dark matter searches}
\label{sec:application}
Having developed the statistical underpinnings of our methods in \Sec{sec:statproc}, we now discuss their applicability to indirect dark matter searches in galaxy clusters. Three criteria need to be simultaneously satisfied for our methods to be relevant: (1) applicability in well-motivated photon energy ranges, (2) detectable and resolvable flux from galaxy clusters, and (3) sufficient knowledge of the spatial distributions of the dominant background and possible dark matter signals. In this section, we show that these criteria are indeed fulfilled.

The photons produced by DM decay or annihilations typically have energy on the same order as DM particle mass, which for fermions must be above a keV due to phase space~\cite{Boyarsky:2008ju} and structure formation~\cite{Viel:2013fqw} constraints. Therefore, the X-ray and gamma-ray regimes are the most interesting energy ranges to look for photons from fermionic DM. Bosonic DM may be much lighter than a keV~\cite{Marsh:2013ywa} but indirect detection in the optical, infrared, and radio bands is subject to large astrophysical backgrounds; the X-ray and gamma-ray backgrounds are somewhat cleaner and their spatial distribution is understood. Moreover, a DM particle (bosonic or fermionic) with a mass near the electroweak scale and a self-annihilation cross section of $\langle \sigma v  \rangle \sim 3 \cdot 10^{-26} \cm^3 \second^{-1}$ is thermally produced in the early universe with the correct DM relic abundance---the so-called WIMP miracle---motivating searches for annihilations into gamma rays. 

The second criterion for the applicability of our methods is that the DM-induced photon flux from galaxy clusters should be observably large, and resolvable by the observing instrument. While our Galactic Center at 8 kpc is the brightest object in the sky in terms of photons produced in any DM model, dwarf spheroidals and galaxy clusters also are interesting targets.  
Clusters are much farther away but make up for it with their large overall mass (see Table~\ref{tab:clusterinfo}), large DM fraction of overall mass~\cite{Voit:2004ah},  potentially large substructure boosts for annihilations (see below), and the fact that the astrophysical backgrounds emanating from the galaxies in the clusters are correspondingly lower due to their large distances from Earth. 

\begin{figure}[t]
\begin{centering}
\includegraphics[width=0.45\textwidth]{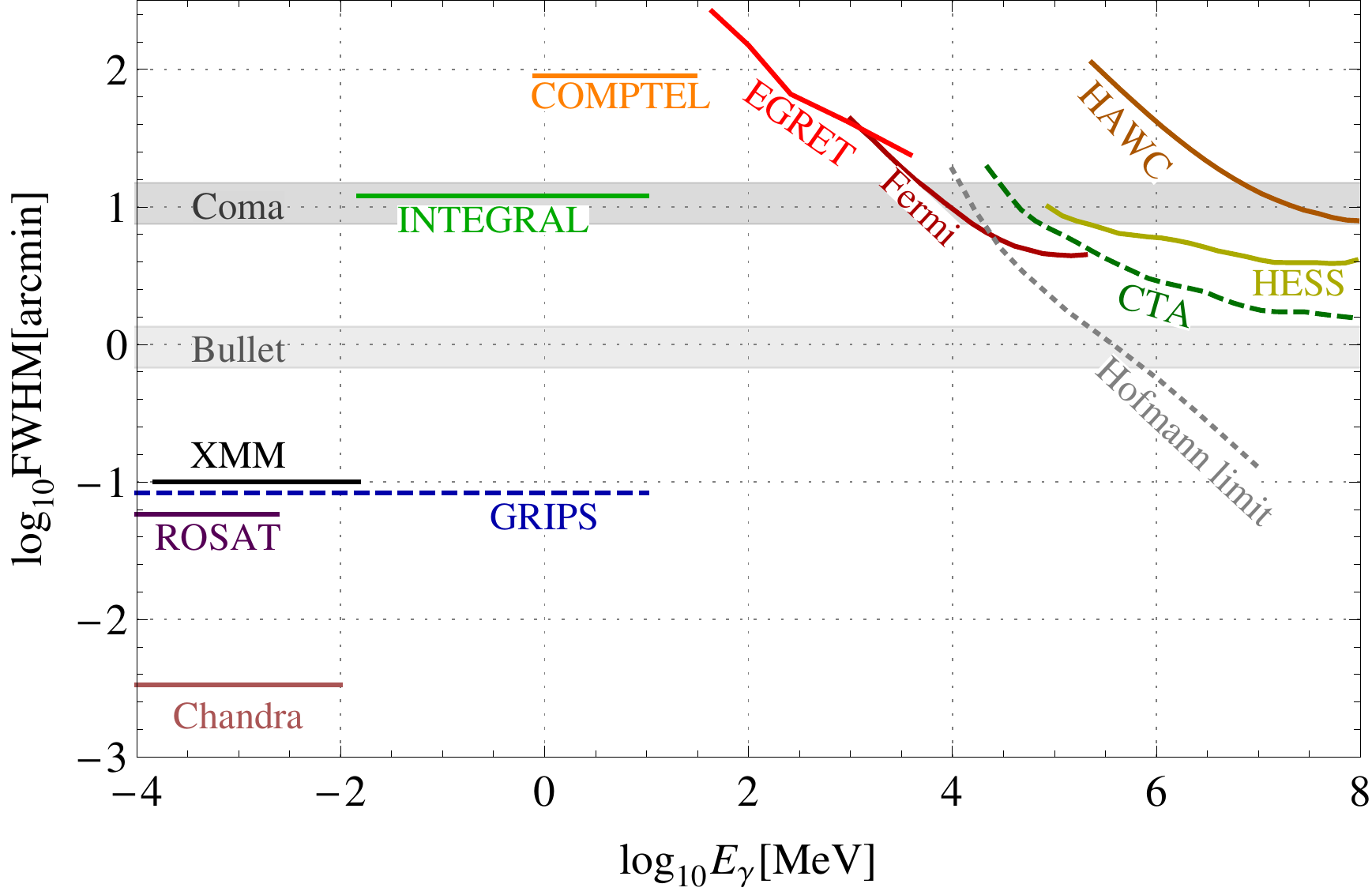}
\end{centering}
 \captionsetup{justification=justified, singlelinecheck=false}
 \caption{(Color online) Angular resolution as a function of photon energy $E_\gamma$ for current (solid) and planned (dashed) X-ray and gamma-ray observatories. The light and dark gray bands are between $1/2$ and $1/4$ of the DM separation (as quoted in Table~\ref{tab:clusterinfo}) in the Bullet and Coma Clusters, respectively. Our methods have reduced power inside the band, but apply fully below. Instrumental resolution is quantified as the estimated FWHM of X-ray satellites~\cite{xraycomp}, GRIPS~\cite{2012ExA....34..551G}, INTEGRAL-IBIS~\cite{Gros:2003wb}, COMPTEL~\cite{1993ApJS...86..657S}, EGRET~\cite{Thompson:1993zz}, and gamma-ray telescopes~\cite{Funk:2012ca}. The Hofmann limit (dotted) is the theoretical lower bound on the resolution of an air Cherenkov telescope~\cite{Hofmann:2006wf}.}
\label{fig:instruments}
\end{figure}

In the X-ray regime, galaxy clusters have been used as potential sources of photons from decaying DM, leading to limits competitive with those of other astrophysical targets~\cite{Boyarsky:2006kc,Boyarsky:2006zi}, or, in the case of the 3.55 keV line~\cite{0004-637X-789-1-13}, a potential signal consistent with previous upper limits. Soft X-ray observatories also have excellent angular resolution, more than sufficient to resolve the gross substructure of galaxy clusters, as shown in \Fig{fig:instruments}. For hard X-rays and soft gamma rays ($10 \keV \lesssim E_\gamma \lesssim 10 \GeV$), observations out of the galactic plane set stringent limits on models of decaying and annihilating dark matter (see~\cite{Essig:2013goa} for a comprehensive review).  However, observatories in this energy regime have yet to achieve the required angular resolving power to discern features of galaxy clusters (see \Fig{fig:instruments}). The same lack of angular resolution will make it difficult for e.g.~Fermi-LAT to check the Galactic excess at $1$--$10\GeV$ first reported in~\cite{Goodenough:2009gk} with merging cluster targets.
More energetic gamma-rays ($\gtrsim 100 \GeV$) can be sufficiently resolved by Fermi-LAT and air Cherenkov telescopes, as shown in \Fig{fig:instruments}.

Cherenkov telescope observations by MAGIC in the Perseus Cluster~\cite{2012A&A...541A..99A}, VERITAS in the Coma Cluster~\cite{2012ApJ...757..123A}, and HESS in the Fornax Cluster~\cite{2012ApJ...750..123A} have set limits on the annihilation cross section at the $\langle \sigma v \rangle \sim 10^{-21}-10^{-22} \cm^3 \second^{-1}$ level in the TeV mass range, while Fermi-LAT observations in nearby clusters~\cite{Ackermann:2010rg,Huang:2011xr,2012JCAP...07..017A} set limits down to $10^{-24} \cm^3 \second^{-1}$ in the 10 GeV range. These limits conservatively assume a smooth NFW profile, while baryonic contraction \cite{Blumenthal:1985qy,Gnedin:2011uj}, DM substructure \cite{Ghigna:1998vn,2011PhRvD..84l3509P,2012MNRAS.425.2169G} and Sommerfeld enhancement may substantially boost the DM flux to yield sensitivity to thermal cross sections of $3 \cdot 10^{-26} \cm^3 \second^{-1}$. For example, baryonic contraction is believed to boost the annihilation flux in Fornax by up to a factor of 2--6~\cite{2012JCAP...07..017A}, while the ``clumpiness" of DM may boost the potential annihilation flux by up to a factor of $10^3$ in a Coma-like cluster~\cite{2012MNRAS.419.1721G}. Future Cherenkov telescopes such as the Cherenkov Telescope Array (CTA) may be able to reach sensitivity to thermal annihilation cross sections in galaxy clusters with sufficient observation time and a high substructure boost factor~\cite{2013APh....43..189D,2011PhRvD..84l3509P}. Clusters are also promising targets for DM decays into gamma rays, as considered in Refs.~\cite{Dugger:2010ys,Huang:2011xr,Ke:2011xw}.  

We note here that for our methods to be applied exactly as described in \Sec{sec:statproc}, the total number of counts per bin must be large enough that Gaussian statistics is a good approximation. Though the number of counts per bin depends on many experimental details (including energy resolution and binning, exposure time, and distance to source), the Gaussian approximation is typically justified for X-ray telescopes (where signals are typically on top of a large continuum background from the ICM) and for air Cherenkov telescopes (which see a large atmospheric background). It may not be applicable for space-based gamma-ray observatories such as Fermi-LAT. In the case of small statistics, our methods can still be applied but their power will now depend on the number of counts per bin, making a model-independent prediction of their utility problematic. A full analysis of the small-statistics case is beyond the scope of our paper.

Finally, the third criterion and crux for our methods is prior knowledge about the spatial distribution of signal and background. In galaxy clusters, weak gravitational lensing maps provide substantial information about the morphology of the dark matter distribution (and thus the signal), while the spatial distributions of the dominant backgrounds are extended, mostly measurable in (X-ray) side energy bands, and, crucially, partially non-overlapping with the signal in merging galaxy clusters. This is in contrast to the Galactic Center, which has large uncertainties on the spatial distribution of both signal and background. As we will show in \Sec{sec:results}, spatial information provides a powerful handle on even modest photon excesses from clusters.

If an excess flux $\Phi_\gamma$ of photons were coming from dark matter decays, their spatial distribution in terms of the coordinates right ascension $\alpha$ and declination $\delta$ on the celestial sphere would have to be correlated with the integral along the line of sight of the DM density distribution: $\Phi_\gamma^{\text{decay}}(\alpha,\delta) \propto \int_\text{los} \rho_\text{DM}(\alpha,\delta,l)\,dl$. The right-hand side of this equation is very nearly the surface mass density $\kappa$ measured in weak gravitational lensing:  $\kappa(\alpha,\delta) \propto \int_\text{los} \rho_\text{M}(\alpha,\delta,l)\,dl$, where the integral is over the total matter contribution $\rho_\text{M} = \rho_\text{DM} + \rho_\text{VM}$, which is dominated by dark matter so that $\rho_\text{M} \simeq \rho_\text{DM}$.\footnote{One has to take into account a convolution with the resolution of both the observing instrument and the weak lensing map measurements to match $\Phi_\gamma^{\text{decay}}$ with $\kappa$.} Weak lensing maps such as those depicted in \Fig{fig:maps} thus provide an excellent model of the expected spatial distribution of a DM decay photon signal. For the Bullet Cluster, we used weak lensing data described in Ref.~\cite{Clowe:2006eq} publicly available at Ref.~\cite{data:bullet_lens}; the lensing map of the Coma Cluster is the one used in Ref.~\cite{Okabe:2013oza}.\footnote{We thank Nobuhiro Okabe for sharing Coma lensing data.}

The spatial distribution of photons coming from annihilations is proportional to the line-of-sight integral of the square of the dark matter distribution: $\Phi_\gamma^{\text{ann}}(\alpha,\delta) \propto \int_\text{los} \rho_\text{DM}^2(\alpha,\delta,l)\,dl$. There is no other direct measure of this quantity, although a reasonable proxy for this observable can be constructed based on the weak lensing map. The DM halo density of any structure is expected to fall off as the cube of the distance from the center of the halo $\rho_\text{DM}(r) \propto r^{-3}$ at relatively large radius, implying that $\int_\text{los} \rho_\text{DM}(r)\,dl \propto r^{-2}$ and $\int_\text{los} \rho_\text{DM}(r)^2\,dl \propto r^{-5}$ if $\rho_\text{DM}(r)$ is smooth. Hence the annihilation flux from the smooth DM halo should be tightly correlated with the function $\Phi_\gamma^{\text{ann}}(\alpha,\delta) \propto \kappa(\alpha,\delta)^{5/2}$, i.e.~a more concentrated version of the weak lensing map. However, the dark matter distribution is not expected to be smooth; in particular, it may be more ``cuspy" than $\kappa^{5/2}$ and/or ``clumpy", meaning that the quantity $\langle \rho_\text{DM}^2 \rangle / \langle \rho_\text{DM} \rangle^2$ is probably much larger than unity on resolvable scales, increasing the DM annihilation flux. Baryonic infall may make the DM halo core more cuspy~\cite{2012JCAP...07..017A}. 

Self-bound DM subhalos are also expected to significantly boost the annihilation flux.
Simulations of cold DM halos on the cluster scale~\cite{2011JCAP...12..011S,2012MNRAS.419.1721G,2012MNRAS.425.2169G} with resolutions down to $\sim 10^7 M_\odot$ show that these resolved substructures contribute roughly twice the annihilation flux of the NFW-like core. However, substructure  may exist down to scales of $10^{-6}$ to $10^{-12} M_\odot$ depending on the nature of the DM and perturbation spectrum, in which case the estimate of the flux from the unresolved substructure ranges from $\sim 50$ to $\sim 1000$ times the core flux.  The spatial distribution of the annihilation flux from substructure is expected to be approximately uniform throughout the cluster up to some cutoff radius, after which it falls off as $r^{-2}$, similar to the expected flux from DM decays~\cite{2012MNRAS.419.1721G}. 
In this paper, we construct templates for the core and substructure distributions from the weak lensing map. We then allow the relative contribution of core and substructure to vary in accordance with our fitting procedure. (The details are addressed in Sec.~\ref{sec:results_fit}.) We present our results as \emph{estimates} of the strength of our procedure as applied to annihilating DM, and hope that future simulations will shed more light on the expected substructure in merging clusters. 

For the spatial distribution of the ICM plasma, we use public data from X-ray observatories: XMM-Newton for the Bullet Cluster~\cite{data:xmm_0112980201}, and ROSAT for the Coma Cluster~\cite{data:rosat_RH800242A04}. The flux of X-rays from the ICM plasma is dominated by thermal bremsstrahlung with only a weak dependence on temperature; hence the intensity of X-rays is a measure of the (squared) density of the plasma. The data amounts to a direct measurement of the background in the X-ray energy range, and we also use it as a proxy for the distribution of other features correlated with the gas emission. In the gamma ray energy range, only the (spatially uniform) diffuse extragalactic background has so far been observed in clusters. (Other gamma-ray backgrounds, originating within the Galaxy, may also contribute to the observed background; for our purposes, we will assume that these are subdominant to the spatially uniform part.) However, the dominant non-DM source of gamma rays from clusters is expected to be produced by scattering of cosmic rays on the ICM. The cosmic rays (CR) are well-confined by the ICM's magnetic field with a long cooling time, resulting in a flux proportional to $n_\text{CR} n_\text{ICM} \simeq n_\text{ICM}^2$ and therefore similar in spatial distribution to the X-rays~\cite{Pinzke:2010st}.

\begin{figure*}[t]
\includegraphics[width=0.22\textwidth]{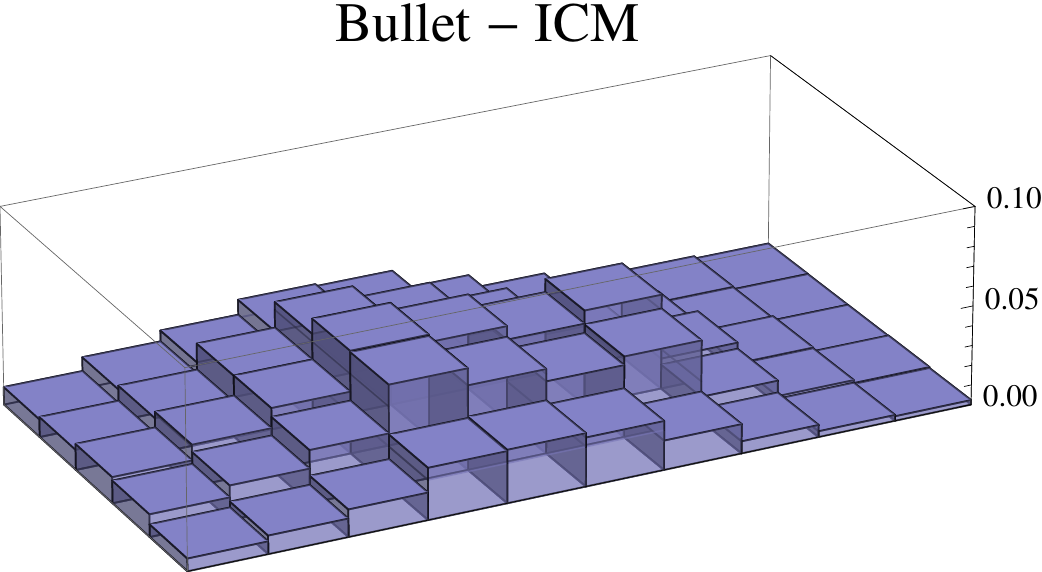}
\includegraphics[width=0.22\textwidth]{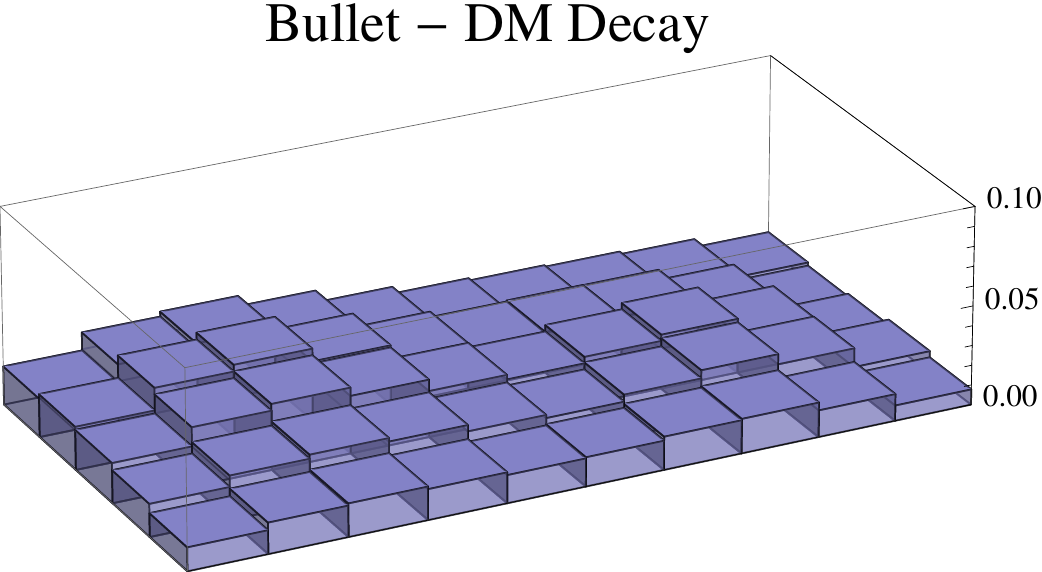}
\includegraphics[width=0.22\textwidth]{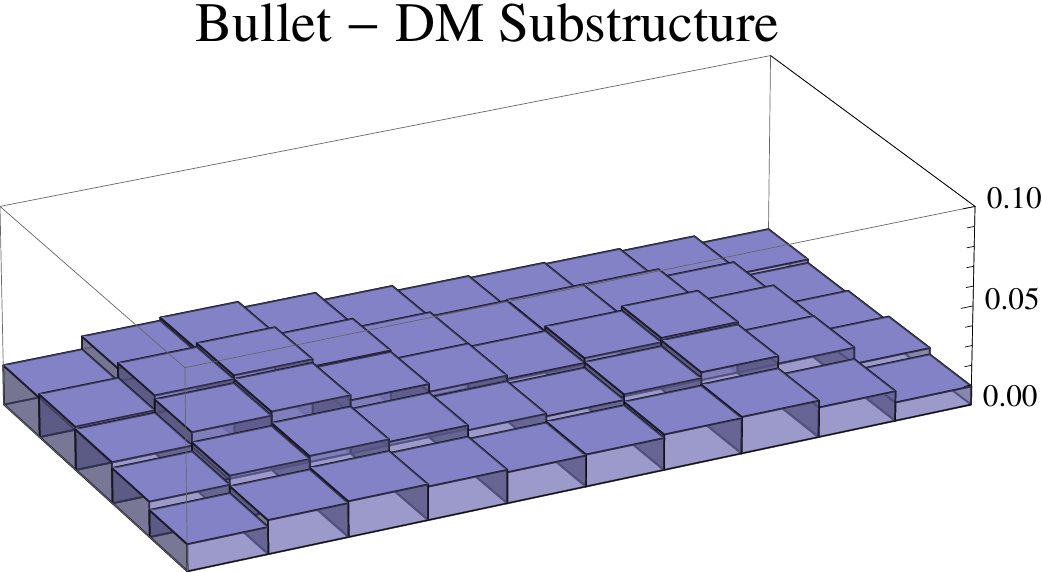}
\includegraphics[width=0.22\textwidth]{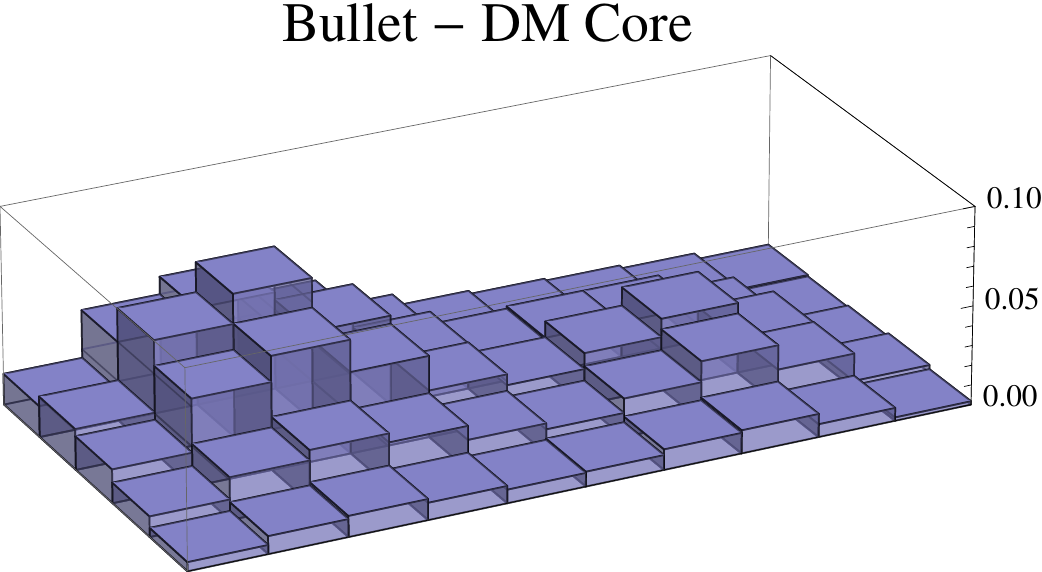}	
\\
\includegraphics[width=0.22\textwidth]{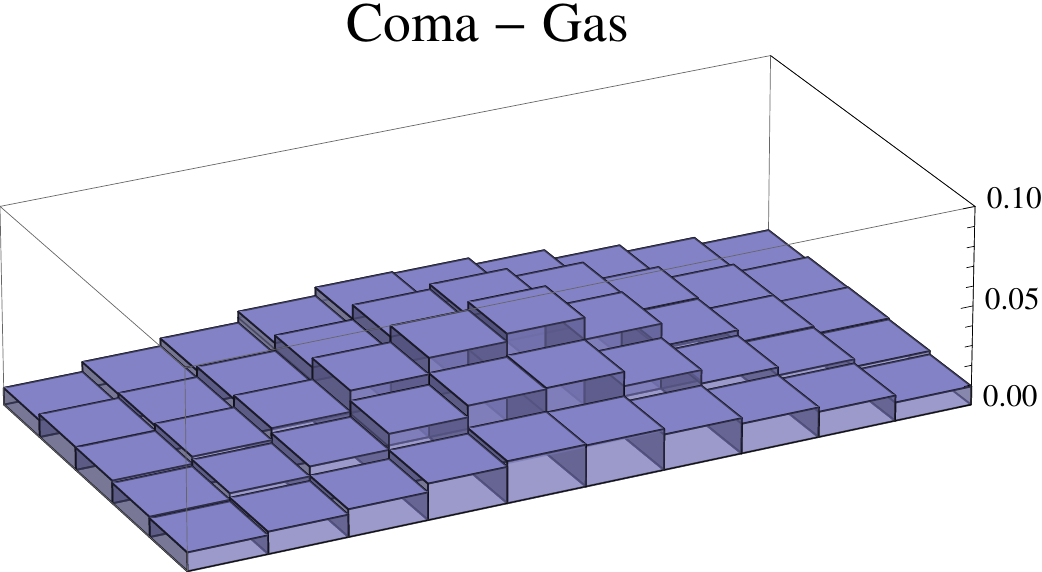}
\includegraphics[width=0.22\textwidth]{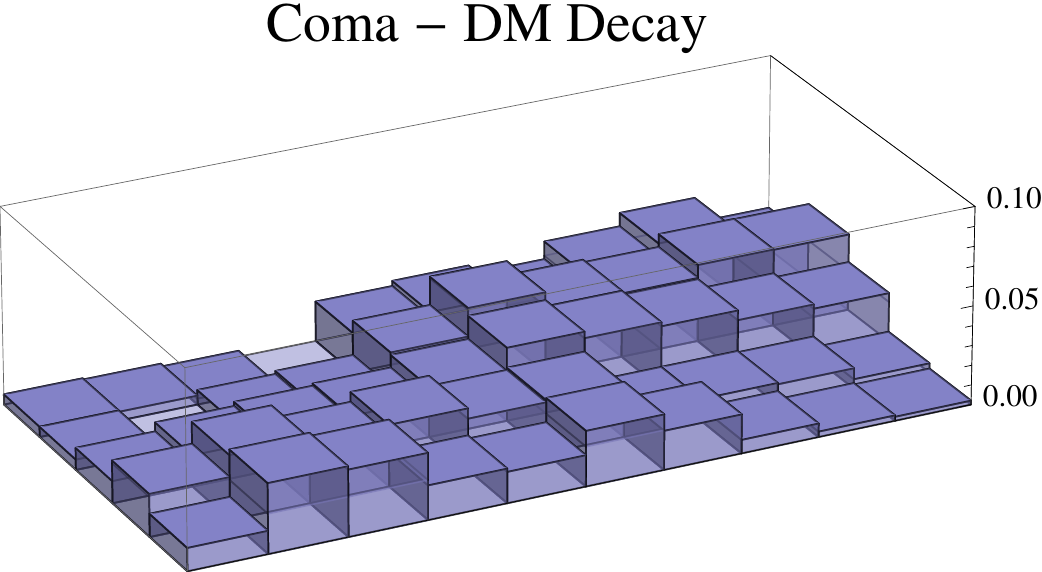}
\includegraphics[width=0.22\textwidth]{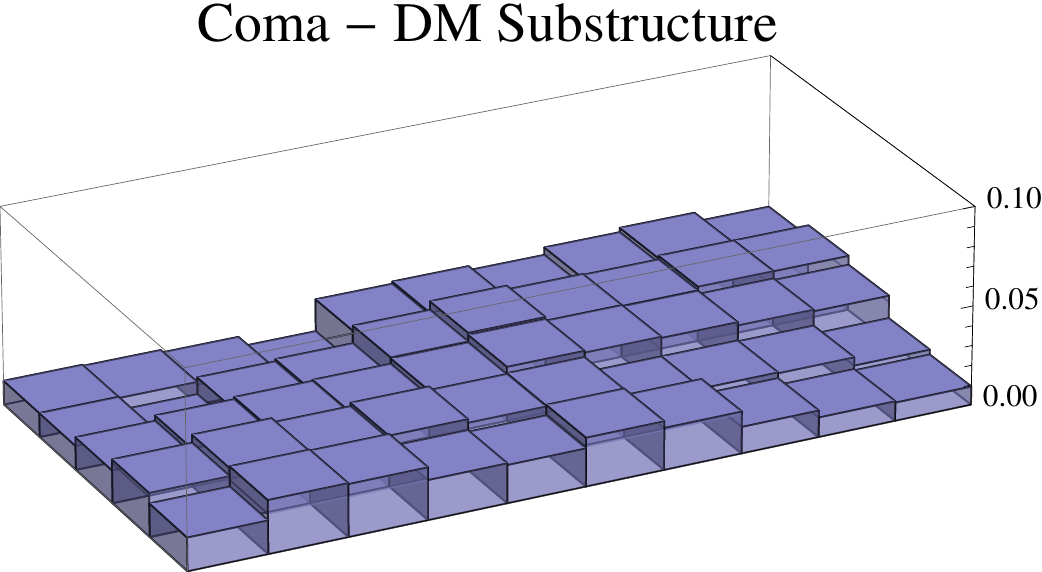}
\includegraphics[width=0.22\textwidth]{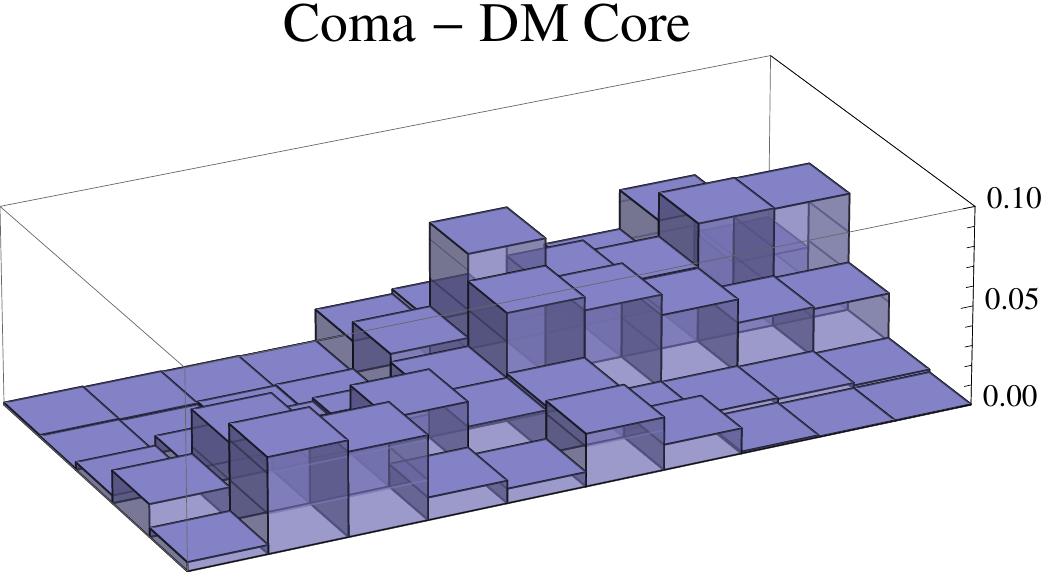}	
\caption{(Color online) Normalized spatial distributions of signals originating from gas, decaying DM, and annihilating DM (substructure and core components) in the Bullet (top) and Coma (bottom) Clusters.}\label{fig:bullet_coma_dists}
\end{figure*}

The normalized spatial distributions that we will use in this paper are displayed in \Fig{fig:bullet_coma_dists}. The ICM and DM decay distributions are measured data (X-ray flux and weak lensing convergence, respectively) as discussed above; the DM core and DM substructure distributions are approximations derived by combining the weak lensing map and phenomenological models from simulations. The exact constructions are discussed further in \Sec{sec:results_fit}.

\section{Results}
\label{sec:results}

Having developed our statistical techniques in Sec.~\ref{sec:statproc} and the expected spatial distributions of signal and background in \Sec{sec:application}, we now move on to quantify the power of these methods in realistic situations.\footnote{Even though we illustrate our methods on real X-ray data, we refrain from doing a careful analysis; we leave this to the experts with a better understanding of background modeling, point source subtraction, and instrumental response/calibration.}  We will discuss the application of (A.) the spatial test procedure to an X-ray line, (B.) the template fitting procedure to annihilating WIMP DM in gamma rays, and (C.) boosting a small X-ray excess using spatial reweighting.

\subsection{Spatial test: Sterile neutrino decay}
\label{sec:results_test}

We now quantify the power of the hypothesis test outlined in Sec.~\ref{sec:statproc_test}. Suppose an indirect detection experiment sees an anomalously large number of photons, and that this excess has a significance of $s$ sigma. We would like to know whether this excess of photons has a spatial distribution that is consistent either with a DM origin (in which it should be correlated in some way with the weak lensing map), or with an unmodeled component of the ICM. 
Specifically, we want to establish how well one can distinguish an excess of $s$ sigma significance with spatial distribution $f_i$ from one with spatial distribution $g_i$, on top of a large background with spatial distribution $b_i$. (As in Sec.~\ref{sec:statproc_test}, spatial bins are labeled by $i$ and the distributions are normalized so that $\sum_i f_i,g_i,b_i = 1$.) As a demonstrative example, we will use the $3.55\keV$ line~\cite{0004-637X-789-1-13} to show how the method works in practice.

For this particular anomaly, the main background is composed of the X-ray emission of the ICM, of which the spatial distribution $b_i$ can be measured accurately in side energy bands.
The most conservative hypothesis is that the observed excess is an unmodeled emission line of the ICM, in which case it should spatially follow the main background, i.e.~$f_i \simeq b_i$.
An exciting possibility is that the excess is due to a dark matter decay such as that of a sterile neutrino into an X-ray photon and a neutrino, in which the signal should be strongly correlated with the weak lensing convergence $\kappa_i$.
Therefore, our alternative hypothesis is that the excess has a spatial distribution $g_i \propto \kappa_i$.
In Sec.~\ref{sec:statproc_test}, we showed that the optimal way to discriminate between these two possibilities is to compute the quantity $\Lambda = (2N^{3/2})^{-1}\sum_i (g_i-f_i) x_i^2 / b_i^2$, where $x_i$ is the observed number of counts in each spatial bin. Since the expected value of $\Lambda$ is different for the two possibilities we are considering, the value of $\Lambda$ computed from the data can tell us which distribution is a better match.
If the ICM is the source of the excess (e.g.~an ICM emission line), we would expect $\Mean{\Lambda} \simeq 0$; if the excess originates from decaying DM, then $\Mean{\Lambda}\simeq T^2 s$, where $T^2=\sum_i (g_i-f_i)^2/b_i$ is determined by the separation of the spatial distributions of the ICM and DM. The standard deviation of $\Lambda$ under either hypothesis is $T$, so the two hypotheses can be distinguished at $\tilde s = T s$ sigma. 

Suppose an X-ray telescope with effective area of about $400\text{ cm}^2$ had an exposure of a few times $10^5\text{ s}$ to the Coma Cluster region depicted in \Fig{fig:comamap}.\footnote{For calculations in this section, we use ROSAT data \cite{data:rosat_RH800242A04} which does not cover the $3.55\keV$ energy band. XMM-Newton observations of Coma exist but require combining several exposures. Since we do not expect the spatial distribution of the X-ray background to change significantly from $2$--$3.55\keV$, we chose the ROSAT data for simplicity. Combining the XMM-Newton exposures adds a technical complication but does not affect the power of our test procedure.}  
Such an exposure would yield a very large number of background ICM photons in the energy band around $3.55\keV$ ($N \gtrsim 10^5$, depending on the energy resolution of the instrument), as well as potentially a few hundred ``excess'' photons from the unidentified emission line reported in~\cite{0004-637X-789-1-13}.
For definiteness, we assume that this experiment led to an $s = 5$ sigma detection of the emission line at $3.55\keV$ in the energy spectrum, not unlike what was reported in~\cite{0004-637X-789-1-13}.\footnote{We use the $3.55\keV$ line only as a specific case study for illustrating the implementation and power of our methods, and acknowledge the existing tension between different studies of the excess, as discussed in \Sec{sec:intro}.}
Spatial binning into square bins of $4'$, the FWHM angular resolution of the weak lensing map, as depicted in \Fig{fig:comamap} (for a total number of spatial bins $k=50$) satisfies the necessary criteria for our procedure to be effective.
The number of counts $x_i$ in each bin is sufficiently large, $N b_i \gg 1$, that they are appropriately described by Gaussian random variables with $\Var{x_i} \simeq N b_i$.
Furthermore, the angular resolution of a typical X-ray telescope (on the order of $10''$) is much smaller than $4'$, so we can treat the counts in each spatial bin as independent.

\begin{figure}
\begin{centering}
\includegraphics[width=0.45\textwidth]{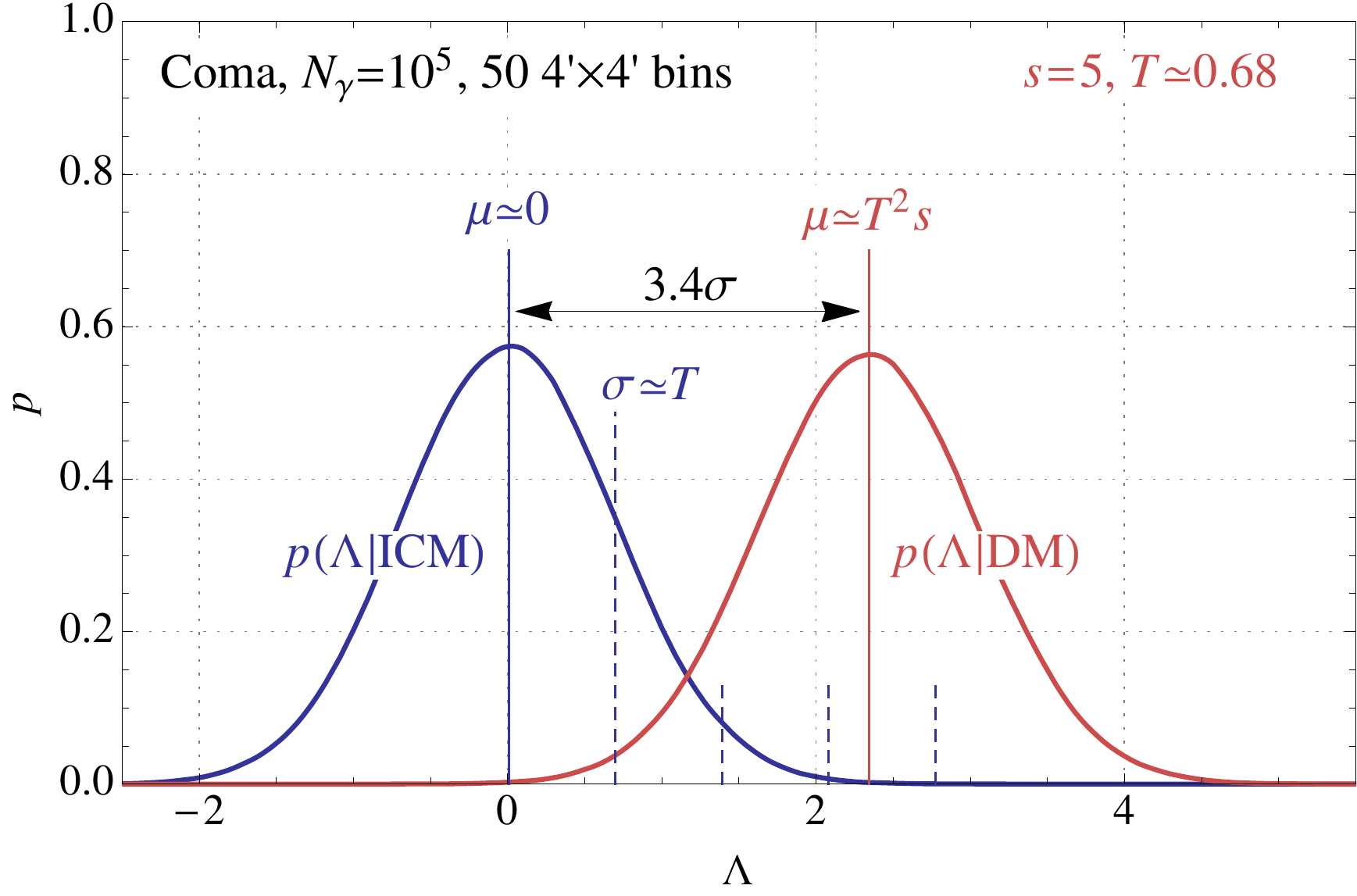}
\end{centering}
 \captionsetup{justification=justified, singlelinecheck=false}
 \caption{(Color online) Probability density functions of $\Lambda$ if an $s=5$ sigma excess is an unmodeled emission line in the ICM (blue, left) or a DM decay line (red, right).  An ICM background of $10^5$ photons originating from the Coma Cluster was injected with an $s=5$ sigma statistical excess following either the ICM or the cluster mass density profile; the photons were binned into square bins of $4'$ as in \Fig{fig:comamap}. The discrimination power of this test is $\tilde{s} = T s \pm 1\approx 3.4 \pm 1$; in other words, if the excess is truly from DM, an ICM interpretation of the excess is expected to be excluded at $3.4\pm1$ sigma (and vice versa).}
\label{fig:pLambda}
\end{figure}

The power of the proposed test is depicted in \Fig{fig:pLambda}.
Even with this relatively coarse binning, the ICM and DM distributions of the Coma Cluster are well-separated, with discrimination factor $T\approx 0.68$. 
This causes the probability density functions (pdfs) $p(\Lambda|\text{ICM})$ (blue, left) and $p(\Lambda|\text{DM})$ (red, right) to differ substantially: the expected value of $\Lambda$ under the DM hypothesis is $T s \approx 3.4$ standard deviations removed from the expected value of $\Lambda$ given the ICM hypothesis. 
For the case at hand, the above hypothesis test on the $s=5$ sigma excess could rule out an additional ICM line contribution in favor of a DM decay scenario with an expected significance of $T s \pm 1 \approx 3.4 \pm 1$ sigma.
Conversely, since the variances of $p(\Lambda|ICM)$ and $p(\Lambda|DM)$ are very nearly equal, a DM decay hypothesis could be ruled out in favor of an ordinary ICM component at the same expected significance.
If the excess was less prominent, the power of the hypothesis test decreases also, but always according to the relation $\tilde{s} \simeq T s$: e.g.~an $s=3$ sigma excess could be tested for an ICM contribution versus DM decay at $\tilde{s} \approx 2.0 \pm 1$ sigma. Hence, even moderately-sized excesses can be tested for consistency with a true signal (DM) or background (ICM) with high significance due to the spatial separation of DM and gas in merging galaxy clusters.

A large discrimination factor $T$ is thus all that is needed to distinguish a DM decay signal from an ICM source based on its spatial distribution. In Table \ref{tab:Tfactors}, we list the discrimination factor $T$ for several scenarios.  The left column is for distinguishing a DM decay signal from an ICM source, as discussed above. The right column lists the $T$ factors to differentiate DM decay from a gamma-ray source.  We took the baseline gamma ray background $b_i$ to be uniform, and the alternative spatial distribution $f_i$ of a potential excess to be that of cosmic rays scattering off the ICM (and, therefore, well-approximated by the X-ray distribution; see Sec.~\ref{sec:application}). For both the Bullet and Coma Clusters, $T$ is large ($\gtrsim 0.5$) if the instrumental angular resolution $\delta \theta$ is sufficient to resolve the separation of the DM and gas in the cluster.

\begin{table}[t]
\newcolumntype{Y}{>{\centering\arraybackslash}X}
\renewcommand{\arraystretch}{1.2}
	\begin{tabularx}{0.45\textwidth}{X c | YY}
		\multicolumn{2}{c|}{} & X-ray & gamma ray \\ \hline
		DM decay signal & $g_i$ & \multicolumn{2}{c}{lensing map $\kappa$ }\\
		alternative & $f_i$  & $ n_\text{ICM}^2$  & $ n_\text{CR} n_{ICM}$  \\
		background & $b_i$ & $n_\text{ICM}^2$ & uniform \\\hline\hline
		 & $\delta\theta$ & \multicolumn{2}{c}{$T$} \\
		\hline
		\multirow{3}{*}{Bullet Cluster} & $12''$ & 0.79 & 0.67 \\
								& $30''$ & 0.74 & 0.62 \\
								& $2.5'$ & 0.12 & 0.11 \\ \hline
		Coma Cluster					& $4'$ & 0.68 & 0.59
	\end{tabularx}
	\captionsetup{justification=justified, singlelinecheck=false}
	\caption{List of discrimination factors $T$ in our benchmark merging clusters for various angular resolutions $\delta\theta$ and choices of background (distributed like the ICM for X-rays, and uniform for the diffuse gamma ray background). An $s$-sigma excess of photons can be tested for consistency with decaying DM compared to an ICM-based source (e.g.~X-ray lines or cosmic ray-ICM scattering producing gamma rays) at $\tilde s = T s$ sigma. The necessity of sufficient angular resolution is evident for the Bullet Cluster $T$-values; a $2.5'$ resolution is barely sufficient to resolve the two subclusters, leading to a tiny discrimination factor. On the other hand, once the structure is resolved, improving the angular resolution does not increase $T$ significantly.}\label{tab:Tfactors}
\end{table}


\subsection{Spatial template fit: WIMP annihilation}
\label{sec:results_fit}

To demonstrate our methods in the case of annihilating DM, we turn to the fitting procedure for multiple spatial distributions described in Sec.~\ref{sec:statproc_fit}. While the spatial distribution of the decay products of DM can be measured directly by weak lensing, the distribution of photons produced in annihilations depends on the integral $\int_\text{los}\rho^2\,dl$ which can have a drastically different shape than the surface mass density. For a smooth NFW profile, the surface mass density $\kappa$ falls off like $r^{-2}$ at large $r$, while the annihilation signal falls off like $r^{-5}$. However, substructure at scales smaller than what weak lensing can resolve is expected to contribute substantially to the annihilation signal; in fact, simulations indicate that it may dominate the total flux, as we discussed in \Sec{sec:application}. In the absence of a definitive prediction for the distribution of the annihilation signal in a merging cluster, we will treat the annihilation signal as a mixture of an NFW-like ``core'' and a flatter ``substructure'' component, both of which we will estimate from the measured surface mass density. Based on the large-$r$ behavior for the NFW profile, we trade $r$ for $\kappa^{-1/2}$ and estimate the core component as $f^\text{core}_i \propto \kappa^{5/2}_i$. This distribution has the qualitatively correct feature that DM peaks become much sharper. For the substructure,~\cite{2012MNRAS.419.1721G} find that the distribution of the annihilation signal from substructure can be described by $(1+(4r/r_{200})^2)^{-1}$, i.e.~roughly uniform inside $r_{200}/4$ and falling off like $r^{-2}$ outside. Since the distribution in terms of $r$ is not very useful for a non-spherically-symmetric merging cluster, we again trade $r$ for $\kappa^{-1/2}$ and model the substructure as $f^\text{sub}_i\propto (1+16\kappa_{200}/\kappa)^{-1}$ where $\kappa_{200}$ corresponds to the surface mass density at $r_{200}$. Since our chosen region of the Coma Cluster is comparable in size to $r_{200}\approx 1\text{ Mpc}$, we pick $\kappa_{200}$ to be the smallest value along the boundary of the region. 
We stress that unlike the spatial profile of DM decay (which is directly measured by gravitational lensing), there is a large systematic uncertainty in the spatial profile of a DM annihilation signal for both the core and substructure components. The templates we construct from the weak lensing map are simply guesses that are consistent with the long-range behavior of DM profiles. However, we think they can be regarded as useful proxies: any linear combination of DM core and substructure components should be qualitatively very different from the ICM flux in a merging cluster. Lacking any observational evidence or simulations of merging clusters to suggest a better approach, we will continue using these constructed templates with the caveat that our results for annihilating DM will be only estimates.

We will demonstrate the discriminating power of the fitting procedure in the case of DM annihilating to gamma rays in the Coma Cluster. The basic idea is to find the best-fit combination of spatial templates to reproduce the excess and determine the extent to which various hypotheses (i.e., particular combinations of templates) are excluded.
Using the formalism introduced in Sec.~\ref{sec:statproc_fit}, we consider fitting a potential excess to the distributions $f^\text{core}$, $f^\text{sub}$, and $f^\text{ICM} \propto f^\text{X-ray}$. The fit has two free parameters, $\theta_\text{core}$ and $\theta_\text{sub}$, which are (respectively) the fractional contribution of the DM core and DM substructure to the excess. The ICM contribution is constrained to be $\theta_\text{ICM}=1-\theta_\text{core}-\theta_\text{sub}$. We take the background distribution $b_i=\text{uniform}$ for gamma rays and the cosmic ray induced distribution to be the same as the X-ray distribution, as discussed in Sec.~\ref{sec:application}. The ability of the fit to distinguish between the three scenarios is encoded by the inverse covariance matrix of $\theta_1=\theta_\text{core}$ and $\theta_2=\theta_\text{sub}$, calculated via \Eq{eq:covariancematrix}:
\begin{align}
	V^{-1}_{ab} &= s^2 \sum_i \frac{(f^a_i-f^n_i)(f^b_i-f^n_i)}{b_i} = s^2 \begin{pmatrix}
		0.811 & 0.259 \\
		0.259 & 0.189
	\end{pmatrix} \label{eq:coma_ann_icov}
\end{align}
From this we can determine the $1\sigma$ error bars on each $\theta_a$ independent of the best-fit values: $\sigma_\text{core}\simeq 1.5s^{-1}$, $\sigma_\text{sub}\simeq 3.1s^{-1}$, and $\sigma_\text{ICM}\simeq 2.4s^{-1}$. However, the exclusion of the ICM-only hypothesis depends on the best-fit parameters via \Eq{eq:gen_exclusion}; when the substructure boost is large, $\theta_\text{sub} \to 1$ and $\tilde s \to 0.43 s$. The inverse covariance matrix also defines the error ellipse in the $\theta_\text{core}$--$\theta_\text{sub}$ plane.

\begin{figure}[t]
\includegraphics[width=0.45\textwidth]{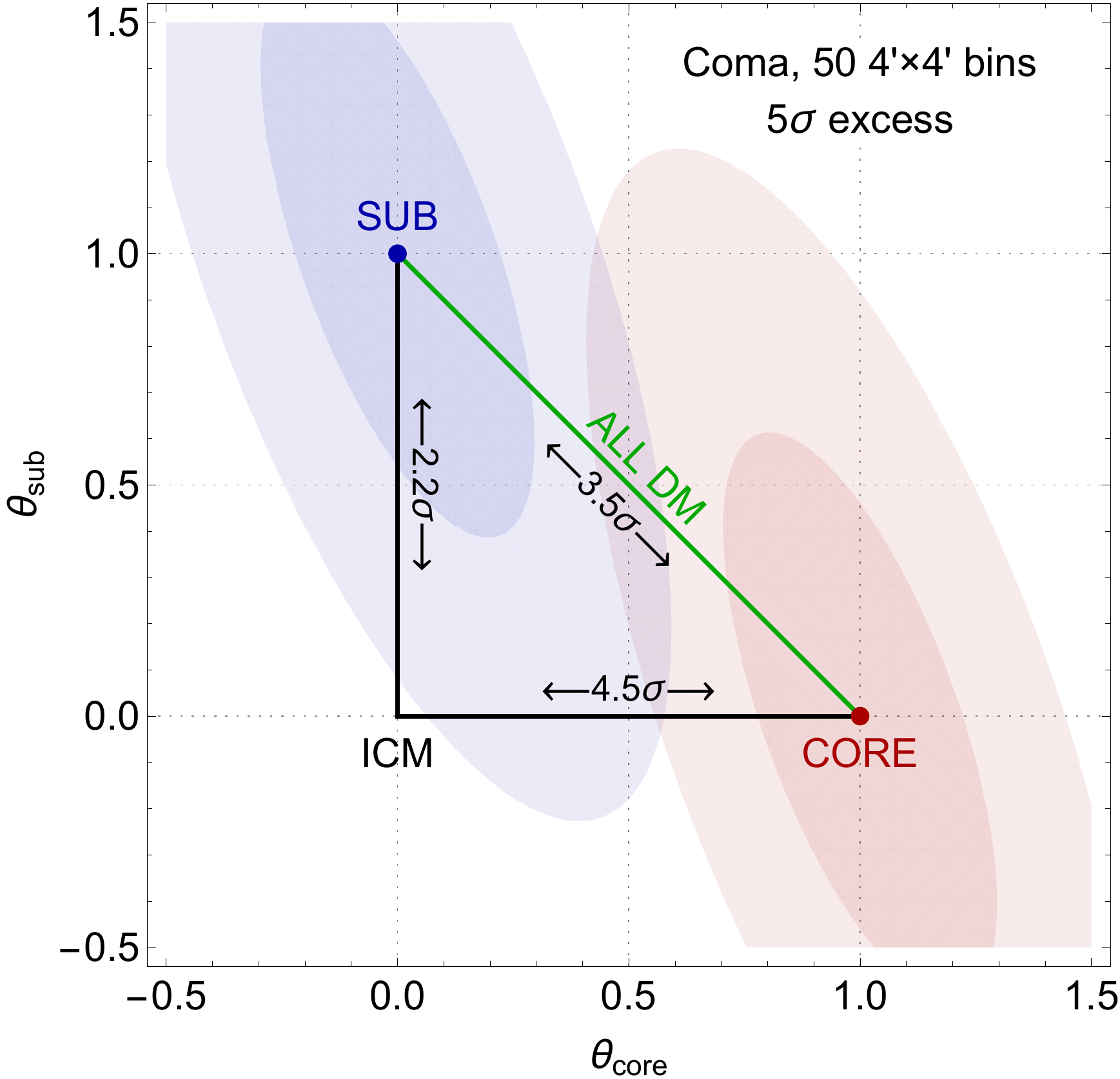}
\caption{(Color online) The expected 1- and 2-sigma confidence regions for fitting the spatial profile of a 5-sigma excess in the Coma Cluster. $\theta_\text{core}$ and $\theta_\text{sub}$ are the fractions of the excess originating from DM annihilation in the NFW-like core and clumpy substructure, respectively; the remainder of the excess is assumed to originate from cosmic ray--ICM scattering. The red ellipses (bottom right) correspond to the errors on the fit parameters for a perfect fit to a core-shaped DM annihilation flux distribution, while the blue ellipses (top left) are for a perfect fit to a flux dominated DM substructure.
The triangle corresponds to the region of hypotheses that we are testing (although statistical fluctuations may cause the best-fit point to lie outside the triangle), and the green side (hypotenuse) is the set of DM-only hypotheses. In these coordinates, the shape of the confidence region (i.e.~size of error bars) is independent of the best-fit point, but the exclusion of the ICM-only hypothesis does depend on the best-fit point, even if it lies along the DM-only line. 
}\label{fig:triangle}
\end{figure}

In \Fig{fig:triangle}, we depict potential outcomes of our fitting procedure for an $s=5$ sigma excess in the Coma Cluster. A spatial fit to an observed excess would yield a best-fit value anywhere in the $\theta_\text{core}$--$\theta_\text{sub}$ plane with an error ellipse determined by the covariance matrix of \Eq{eq:covariancematrix}. Note that the size and shape of the error ellipse are independent of the best-fit point. The red ovals represent 1- and 2-sigma contours for a best-fit value of $\theta_\text{core} = 1$ and $\theta_\text{sub} = \theta_\text{ICM} = 0$, i.e.~consistent with annihilation produced mainly by an NFW-like core distribution. In this case, an explanation of the excess via cosmic rays scattering off the ICM would be excluded at $4.5$ sigma. The blue ovals are for a best-fit value of $\theta_\text{sub} = 1$ and $\theta_\text{core}=\theta_\text{ICM} = 0$, consistent with a substructure-dominated signal, which would rule out a cosmic ray explanation at $2.2$ sigma.

\subsection{Spatial reweighting: dark matter decay}
\label{sec:results_weight}

We now wish to apply the spatial reweighting technique described in Sec.~\ref{sec:statproc_weight} to physical situations. Methods A and B in the previous sections were designed to test for a DM origin of an already-existing statistical excess. This procedure is optimized for ``DM bump hunting'', i.e.~to amplify putative DM-induced photon excesses in the spectrum. The reweighting procedure also constitutes a test for DM origin of an excess, though a less powerful one than the one we devised in \Sec{sec:results_test}.  We shall mostly restrict ourselves here to the case of potential photon excesses from DM decay, as the spatial distribution from an annihilation signal is less certain.

As detailed in Sec.~\ref{sec:statproc_weight}, we can reweight each photon with a factor dependent on its spatial position in the sky. The weight factors are chosen to amplify photons from DM-rich regions of a cluster and diminish those from background-rich regions. As long as the total number of observed counts $B_i$ in each spatial bin $i$ is large enough to be approximately Gaussian, and as long as the dominant error is statistical, we showed that the optimal weights are
\begin{align}
	w_{*i} = c\frac{S_i}{B_i}, \label{eq:wfactors}
\end{align}
where $S_i$ is proportional to the expected number of signal photons and $c$ is an arbitrary normalization factor. In this work, we choose $c=k/(\sum_{i=1}^k S_i/B_i)$ so that the typical value of a weight is $1$.

We also showed that the expected boost in statistical significance of a true DM signal is given by
\begin{align}
	R\equiv \frac{\tilde s}{s} = \frac{\sqrt{\sum_i S_i^2/B_i}}{\sum_i S_i / \sqrt{\sum_i B_i}}, \label{eq:Rvalue}
\end{align}
where $s\ (\tilde s)$ is the significance of the signal before (after) reweighting. Because $R$ (like $w_{*i}$) is insensitive to rescalings of $S_i$ and $B_i$, it is a purely geometric property of a particular cluster, depending only on the spatial distributions of DM and background. However, to the extent that the shape of the background energy spectrum varies with position, e.g.~with temperature variations of the ICM in the X-ray energy band, $R$ will vary with energy.
For DM decaying to X-rays, the energy dependence is quite weak, varying from $R(1\text{ keV})\approx 1.3$ to  $R(5\text{ keV})\approx 1.2$ in the Bullet Cluster. For DM decaying to gamma rays, the background is uniform and independent of energy. We report values of $R$ for both the Bullet and Coma Clusters in Table~\ref{tab:Rvalues}.

\begin{table}[t]
	\newcolumntype{Y}{>{\centering\arraybackslash}X}
	\renewcommand{\arraystretch}{1.2}
	\begin{tabularx}{0.45\textwidth}{Xc|D{.}{.}{-1}D{.}{.}{-1}}
		& & \multicolumn{1}{Y}{X-ray} & \multicolumn{1}{Y}{gamma ray} \\ \hline
		DM decay signal & $S_i$ & \multicolumn{2}{c}{lensing map $\kappa$} \\
		background & $B_i$ & \multicolumn{1}{Y}{$n_{ICM}^2$} & \multicolumn{1}{Y}{uniform} \\ \hline\hline
		\multicolumn{2}{l|}{} & \multicolumn{2}{c}{$R$} \\ \hline
		\multicolumn{2}{l|}{Bullet Cluster} & 1.24\footnotemark[1] & 1.06 \\
		\multicolumn{2}{l|}{Coma Cluster} & 1.21 & 1.16
	\end{tabularx}
	\footnotetext[1]{average value; $R(E)$ is weakly energy dependent, see text}
	\captionsetup{justification=justified, singlelinecheck=false}
	\caption{Values of the spatial reweighting significance boost $R$ for DM decaying to either X-rays or gamma rays, in our two representative clusters, calculated according to \Eq{eq:Rvalue}. A true decaying DM signal with unweighted significance $s$ will be boosted by reweighting to a significance $\tilde s = R s$. The regions and bins used are those of \Fig{fig:maps}. The bin size is $30''$ for the Bullet Cluster and $4'$ for the Coma Cluster.}\label{tab:Rvalues}
\end{table}

The reweighting technique is not effective for DM annihilating to gamma rays if the flux from the substructure dominates the flux from the smooth core, since in that case the spatial distributions of the signal and background are both very nearly uniform within the cluster and the spatial distribution provides no extra information. We find $R\lesssim1.05$ in both the Bullet and Coma Clusters. The hypothesis test (A) and fitting procedure (B) are still effective in this scenario because the excess is already in hand, and we are only comparing its shape to an alternative (e.g.~cosmic ray--ICM collisions) which is non-uniform. Here, the alternative is a fluctuation of the background (with uniform spatial distribution), which is practically indistinguishable from DM annihilation dominated by substructure flux (also nearly spatially uniform).

\begin{figure}[t]
	\includegraphics[width=0.45\textwidth]{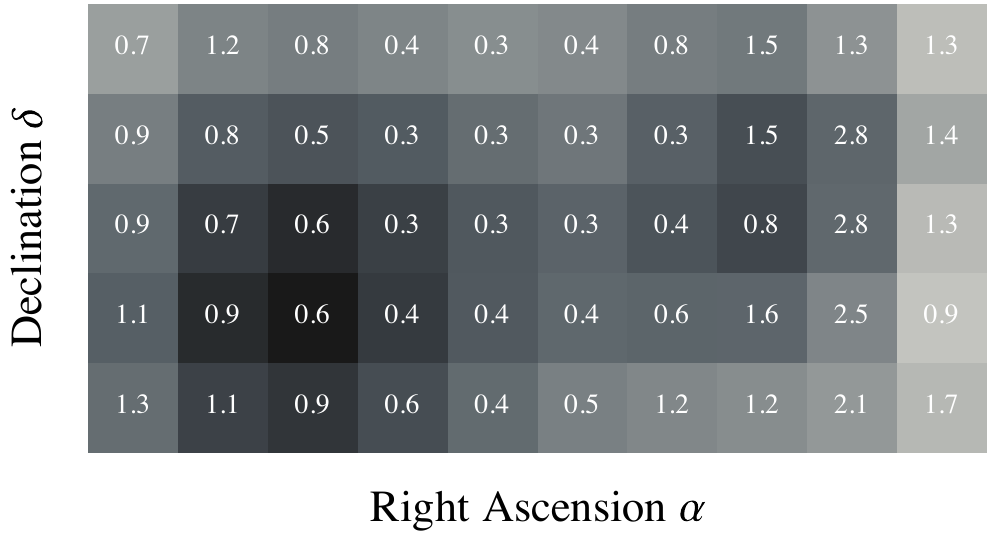}
	\caption{Distribution of decaying DM signal photons in the Bullet Cluster (in grayscale) in $30''$ bins overlaid with the weight factors $w_i$ calculated according to Eq.~\ref{eq:wfactors}. The region and binning are identical to those of \Fig{fig:bulletmap}. The background is taken to be the X-ray photons in the range $3.75$--$4.75 \text{ keV}$, i.e.~the weight factors used at the energy of our injected signal (cf.~\Fig{fig:spectrum}). Note that the bins with the largest weights are not those with the highest DM content (used in Ref.~\cite{Boyarsky:2006kc}), but those on the outskirts of the cluster where the DM has moved beyond the ICM and is correspondingly more pure.}\label{fig:weights}
\end{figure}

\begin{figure*}[t]
\centering
	\subfloat[]{\includegraphics[width=0.45\textwidth]{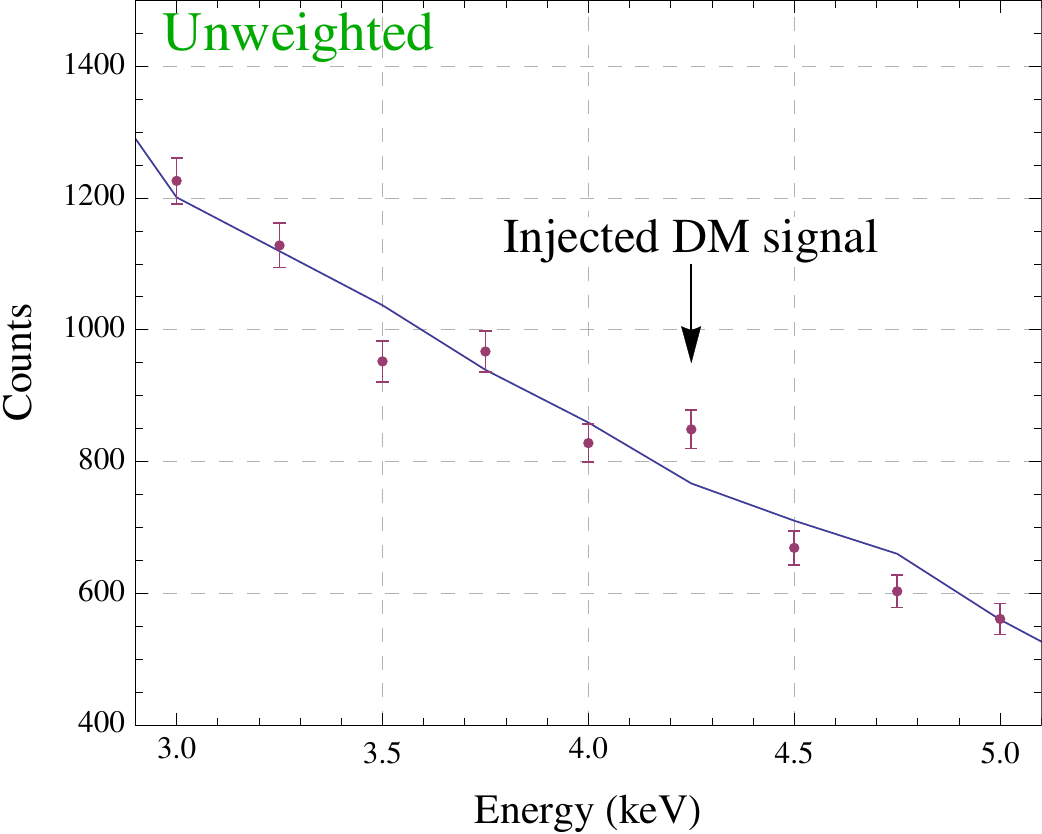}}
	\hfill
	\subfloat[]{\includegraphics[width=0.45\textwidth]{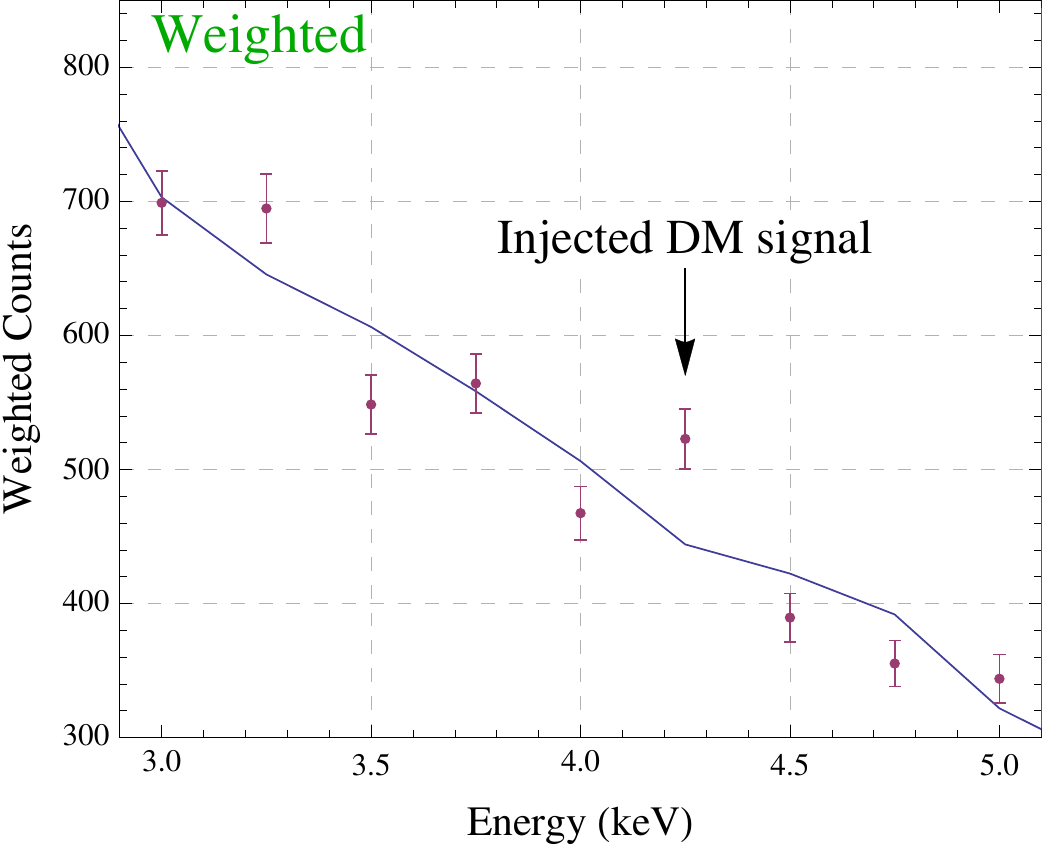}}
	\caption{(Color online) A sample energy spectrum of the Bullet Cluster with 30'' binning, before (left) and after (right) reweighting the photons based on their spatial distribution. The blue solid line is the estimated background contribution, calculated from the sidebands (average of two bins on each side). A fictitious 2.5$\sigma$ line excess with the spatial distribution of decaying DM has been injected at $E=4.25\text{ keV}$. The $R$-factor at this energy is roughly $1.2$, and the significance increases accordingly to $2.5R\sigma\approx3\sigma$ after reweighting. The weights are calculated using $1\keV$-wide bins to reduce their variance.
}\label{fig:spectrum}
\end{figure*}

Spatial reweighting can enhance DM spectral features before they are prominent enough to be considered for the hypothesis test of Sec.~\ref{sec:results_test} or the fitting procedure of Sec.~\ref{sec:results_fit}. As a concrete example, we inject a $2.5\sigma$ excess, spatially distributed like decaying DM, into the X-ray spectrum of the Bullet Cluster at $E=4.25\keV$. We use the $30''$ spatial bins shown in \Fig{fig:bulletmap}. We then calculate the optimal weight factors according to \Eq{eq:wfactors} with $B_i$ given by the total number of counts in a $1\keV$-wide bin. (Although $R$ is only weakly energy dependent, the weight factors themselves vary more strongly; the optimal weights calculated at low energies are not effective for excesses at higher energies. However, the wide energy binning we used is sufficient to enhance the excess.) The weight factors are displayed in \Fig{fig:weights}. The unweighted and reweighted spectra are shown in \Fig{fig:spectrum}; the injected DM excess at $4.25\keV$, which was only $2.5\sigma$ above background in the unweighted case, is visibly enhanced to a $3\sigma$ excess, in accordance with our calculation of $R\approx 1.2$. The blue line is a simplistic estimate of the background contribution, calculated with by averaging side energy bands, meant only to aid in estimating the significance of the injected signal. Other 1-2$\sigma$ excesses visible in the unweighted spectrum, e.g.~at $3.75\keV$, are in fact suppressed by the reweighting.

Spatial reweighting can also be used to strengthen exclusion limits by an amount proportional to $R$. A background-subtracted $k$-sigma exclusion limit on some DM parameter $\alpha$, where the received flux in the detector is $\Phi_\text{DM}\propto \alpha^n$, is set by enforcing that $\Phi_\text{DM}$ produce no more than a $k$ sigma excess above the observed background. Since the significance of such a signal would be boosted by $R$ after reweighting, the limit on $\alpha$ will be multiplied by a factor of $R^{-1/n}$. For example, the flux from decaying sterile neutrino DM is linearly proportional to the parameter $\sin^2 2\theta$, and so reweighting can make those limits stronger by a factor of $R$. In Ref.~\cite{Boyarsky:2006kc}, the regions selected for their method correspond to a non-optimal choice of weight factors in our method. In fact, the $R$-value calculated using these regions is less than $1$; selecting only the DM peaks gives a \emph{weaker} limit on $\sin^2 2\theta$ than cutting out no region at all. This can be understood from \Fig{fig:weights} by noting that the weights in DM-rich regions are often smaller than one, while the regions on the outskirts of the cluster (where there is some DM but almost no gas) are heavily weighted. By reweighting the photons optimally, the limit could be improved by a factor of $R\approx 1.2$ instead.

A similar method (with different weight factors) is applicable for strengthening conservative, non-background-subtracted limits; see Appendix~\ref{sec:app_conslim}.

\section{Discussion \& future outlook} \label{sec:discussion}

We have developed three data analysis methods for indirect detection of dark matter in galaxy clusters,  where the extended dark matter spatial distribution can be measured via weak gravitational lensing. 
All three methods A, B, and C rely on this spatial information to enhance the discrimination power between dark matter signals and astrophysical backgrounds.
\begin{itemize}
\item {\bf Method A} is a spatial test on a putative statistical excess, and assesses the compatibility of the excess with two known morphologies. It is especially powerful to determine whether an excess is consistent with a \emph{dark matter decay}, as shown in \Sec{sec:results_test}.
\item {\bf Method B} is a spatial fit of any number of spatial templates to a putative statistical excess. For two templates, it is equivalent to Method A, but it can be extended to three or more templates. It is most useful for distinguishing \emph{dark matter annihilation} signals (which is expected to have a smooth halo component, and an extended subhalo component in galaxy clusters, with unknown ratio) from background, as explained in \Sec{sec:results_fit}.
\item {\bf Method C} is a spatial reweighting of photons, designed to enlarge any statistical excess with known spatial distribution. It applies to both decays and annihilations, although it is more promising for the former, as discussed in \Sec{sec:results_weight}.
\end{itemize}

Our methods have several important advantages:
\begin{itemize}
\item {\bf Robustness:} In contrast to methods used in the Galactic Center and dwarf spheroidals, each of our methods are based on spatial distributions for signal and background which are both \emph{measured} and \emph{separated}. For decaying DM, the spatial distribution is directly measured by weak lensing; for annihilations, we combine the weak lensing map with general results from simulations. Method B accounts for some of the uncertainty in the annihilation template. The background distributions are also well-known and substantially different from the signal distributions due to the cluster merger.
\item {\bf Universality:} Our results are insensitive to details of spatial binning (as long as the gross substructure of the merging cluster is resolved), energy range, exposure time, etc., and only depend on the \emph{intrinsic spatial distributions} of signal and background.
\item {\bf Optimality:} For the questions they try to answer, Methods A and C are provably optimal, and use all of the spatial information available. Method B generalizes Method A and also appears to use all of the spatial information.
\item {\bf Wide range of applicability:} Our methods apply to any type of emission for which the spatial distributions of signal and background are known or can be reliably estimated. We showed how weak lensing maps of galaxy clusters can be used to test potential decaying or annihilating dark matter signals in the well-motivated X-ray and gamma-ray energy bands. This type of analysis can be done offline and with existing data. The statistical methods presented here may also find a use outside the realm of dark matter detection, in areas where a weak signal is to be extracted and tested for consistency with an independently measured property of the expected signal.
\end{itemize}

We have also showed that these methods are promising in realistic scenarios, including current indirect detection anomalies. Method A can be applied to the $3.55 \keV$ line in already existing observations of the Coma Cluster. A longer observation of Coma leading to a $5\sigma$ observation of the line could distinguish a decaying sterile neutrino from ICM emission at more than $3\sigma$. Fermi-LAT, HESS, and (in the future) CTA have the angular resolution to test future excesses in high-energy gamma rays in the Coma Cluster.

The main downside of our method is that galaxy clusters are fainter targets in terms of DM-induced photon flux than the Galactic Center (and perhaps also dwarf spheroidals). As a target for discovery of an anomalous excess, the latter is thus more interesting; to elucidate the DM nature of an excess, we have argued that merging galaxy clusters are the most compelling targets. In this sense, indirect searches of DM in clusters are complementary to those in the Galactic Center and dwarf spheroidals.

Our methods may become even better as more merging galaxy clusters are discovered. The well-known and very nearby Coma Cluster was only recently lensed and confirmed to have undergone a significant merger leading to DM--ICM separation,
and galaxy clusters with major and minor mergers continue to be found \cite{Harvey:2015hha}. Although Coma is likely the closest cluster with a substantial merging event, and gravitational lensing becomes prohibitively difficult for closer targets, future lensing studies may discover interesting DM morphologies in other nearby clusters.
New instruments with good angular resolution may close the hard X-ray/soft gamma ray gap (see \Fig{fig:instruments}) and allow observations of merging clusters in interesting new energy ranges.

The data analysis methods outlined in this paper make indirect detection in galaxy clusters more attractive than before. In the near term, they can be applied to existing anomalies such as the $3.55\keV$ line, and provide extra motivation for constructing weak gravitational lensing maps of nearby galaxy clusters. In the future, they motivate dedicating substantial instrumental exposure time to galaxy clusters, as well as pushing the boundaries on angular resolution and effective area of future X-ray and gamma-ray observatories.

\acknowledgments{
We thank Masha Baryakhtar, Maru\v{s}a Brada\v{c}, Kiel Howe, Xinlu Huang, David E.~Kaplan, Jeremy Mardon, Ondrej Urban, and Yue Zhao for discussions. We are especially grateful to Nobuhiro Okabe for providing us with weak lensing data of the Coma Cluster. 
This research has made use of data and software provided by the High Energy Astrophysics Science Archive Research Center (HEASARC), which is a service of the Astrophysics Science Division at NASA/GSFC and the High Energy Astrophysics Division of the Smithsonian Astrophysical Observatory. Bullet Cluster X-ray data is based on observations obtained with XMM-Newton, an ESA science mission with instruments and contributions directly funded by ESA Member States and NASA.
SR acknowledges the support of NSF grant PHY-1417295.
This work was supported in part by NSF grant PHY-1316706 and DOE Early Career Award DE-SC0012012.
}

\appendix
\section{Conservative limit reweighting}
\label{sec:app_conslim}
If no excess is seen, a limit on the DM flux can be set. When the background is well-modeled, a background-subtracted limit may be a reasonable approach; spatial reweighting can improve such a limit only by a factor of $R$ as described in Sec.~\ref{sec:statproc_weight} and quantified in Sec.~\ref{sec:results_weight}. However, if the background is not well-understood, the most conservative way to set a limit is to constrain the DM flux to be less than the total observed flux (plus uncertainty):
\begin{align} \Phi_{\rm DM} < \Phi_{\rm obs} + s\sigma_{\Phi_{\rm obs}}, \end{align}
where $s$ is the number of sigma corresponding to the confidence level of the limit (typically $s=2$, for a 95\% confidence limit).

With spatial information, we can reweight the photons to set a stronger limit on the DM hypothesis. In terms of the weighted variables $\hat S$ and $\hat B$, the limit is set by $K(\alpha) \langle\hat S\rangle < \langle \hat B \rangle + s\sqrt{\Var \hat B}$ where $K$ gives the observed counts per unit $S_i$ in terms of the DM parameter $\alpha$. For example, in the case of DM decay, $\alpha$ is some coupling in the model and $K(\alpha)$ gives the number of observed counts per unit surface mass density on the sky, taking $S_i=\kappa_i$.

Since the strongest limit is set by minimizing the limiting value of $K(\alpha)$, we choose to maximize the target function
\begin{align} \Sigma_{\rm CL}(\{w_i\}) &\equiv \frac{\langle\hat S\rangle}{\langle \hat B \rangle + s\sqrt{\Var \hat B}}\nonumber \\
	&= \frac{\sum_i w_i S_i}{\sum_i w_i B_i + s\sqrt{\sum_i w_i^2 B_i}}. \end{align}
Again, the target function is unchanged by rescaling $w_i$ so we impose $\sum_{i=1}^k w_i = k$. There is no closed form solution for the optimal weight $w_{*i}$, but it satisfies the equation
\begin{align} w_{*j} = c\left(\frac{\sum_i w_{*i} B_i + s \sqrt{\sum_i w_{*i}^2 B_i}}{\sum_i w_{*i} S_i}\frac{S_j}{B_j} - 1\right) \end{align}
which can be efficiently solved by iteration if the weights are restricted to be nonnegative. These weights are insensitive to rescalings of $S_i$ but not to those of $B_i$; as the total number of counts increases, the optimal weighting for setting a conservative limit changes, eventually approaching a Kronecker $\delta$ at the location of greatest $S_i/B_i$.

The boost in limit-setting power is given by $R_{\rm CL}\equiv \Sigma_{\rm CL}(\{w_{*i}\})/\Sigma_{\rm CL}(\{1\})$. Typically $R_{\rm CL}\sim\mathcal{O}(\rm{few})$. For the same exposure and binning of the Bullet Cluster discussed in Section~\ref{sec:results_weight}, $R_{\rm CL}\approx 3$ for $s=2$ and grows slowly with exposure time.

\bibliographystyle{apsrev4-1-etal}
\bibliography{bullet}

\begin{thebibliography}{79}%
\makeatletter
\providecommand \@ifxundefined [1]{%
 \@ifx{#1\undefined}
}%
\providecommand \@ifnum [1]{%
 \ifnum #1\expandafter \@firstoftwo
 \else \expandafter \@secondoftwo
 \fi
}%
\providecommand \@ifx [1]{%
 \ifx #1\expandafter \@firstoftwo
 \else \expandafter \@secondoftwo
 \fi
}%
\providecommand \natexlab [1]{#1}%
\providecommand \enquote  [1]{``#1''}%
\providecommand \bibnamefont  [1]{#1}%
\providecommand \bibfnamefont [1]{#1}%
\providecommand \citenamefont [1]{#1}%
\providecommand \href@noop [0]{\@secondoftwo}%
\providecommand \href [0]{\begingroup \@sanitize@url \@href}%
\providecommand \@href[1]{\@@startlink{#1}\@@href}%
\providecommand \@@href[1]{\endgroup#1\@@endlink}%
\providecommand \@sanitize@url [0]{\catcode `\\12\catcode `\$12\catcode
  `\&12\catcode `\#12\catcode `\^12\catcode `\_12\catcode `\%12\relax}%
\providecommand \@@startlink[1]{}%
\providecommand \@@endlink[0]{}%
\providecommand \url  [0]{\begingroup\@sanitize@url \@url }%
\providecommand \@url [1]{\endgroup\@href {#1}{\urlprefix }}%
\providecommand \urlprefix  [0]{URL }%
\providecommand \Eprint [0]{\href }%
\providecommand \doibase [0]{http://dx.doi.org/}%
\providecommand \selectlanguage [0]{\@gobble}%
\providecommand \bibinfo  [0]{\@secondoftwo}%
\providecommand \bibfield  [0]{\@secondoftwo}%
\providecommand \translation [1]{[#1]}%
\providecommand \BibitemOpen [0]{}%
\providecommand \bibitemStop [0]{}%
\providecommand \bibitemNoStop [0]{.\EOS\space}%
\providecommand \EOS [0]{\spacefactor3000\relax}%
\providecommand \BibitemShut  [1]{\csname bibitem#1\endcsname}%
\let\auto@bib@innerbib\@empty
\bibitem [{\citenamefont {Bertone}\ \emph {et~al.}(2005)\citenamefont
  {Bertone}, \citenamefont {Hooper},\ and\ \citenamefont
  {Silk}}]{Bertone:2004pz}%
  \BibitemOpen
  \bibfield  {author} {\bibinfo {author} {\bibfnamefont {G.}~\bibnamefont
  {Bertone}}, \bibinfo {author} {\bibfnamefont {D.}~\bibnamefont {Hooper}}, \
  and\ \bibinfo {author} {\bibfnamefont {J.}~\bibnamefont {Silk}},\ }\href
  {\doibase 10.1016/j.physrep.2004.08.031} {\bibfield  {journal} {\bibinfo
  {journal} {Phys.Rept.}\ }\textbf {\bibinfo {volume} {405}},\ \bibinfo {pages}
  {279} (\bibinfo {year} {2005})},\ \Eprint
  {http://arxiv.org/abs/hep-ph/0404175} {arXiv:hep-ph/0404175 [hep-ph]}
  \BibitemShut {NoStop}%
\bibitem [{\citenamefont {Gelmini}(2015)}]{Gelmini:2015zpa}%
  \BibitemOpen
  \bibfield  {author} {\bibinfo {author} {\bibfnamefont {G.~B.}\ \bibnamefont
  {Gelmini}},\ }\href@noop {} {\  (\bibinfo {year} {2015})},\ \Eprint
  {http://arxiv.org/abs/1502.01320} {arXiv:1502.01320 [hep-ph]} \BibitemShut
  {NoStop}%
\bibitem [{\citenamefont {Essig}\ \emph {et~al.}(2013)\citenamefont {Essig},
  \citenamefont {Kuflik}, \citenamefont {McDermott}, \citenamefont {Volansky},\
  and\ \citenamefont {Zurek}}]{Essig:2013goa}%
  \BibitemOpen
  \bibfield  {author} {\bibinfo {author} {\bibfnamefont {R.}~\bibnamefont
  {Essig}}, \bibinfo {author} {\bibfnamefont {E.}~\bibnamefont {Kuflik}},
  \bibinfo {author} {\bibfnamefont {S.~D.}\ \bibnamefont {McDermott}}, \bibinfo
  {author} {\bibfnamefont {T.}~\bibnamefont {Volansky}}, \ and\ \bibinfo
  {author} {\bibfnamefont {K.~M.}\ \bibnamefont {Zurek}},\ }\href {\doibase
  10.1007/JHEP11(2013)193} {\bibfield  {journal} {\bibinfo  {journal} {JHEP}\
  }\textbf {\bibinfo {volume} {1311}},\ \bibinfo {pages} {193} (\bibinfo {year}
  {2013})},\ \Eprint {http://arxiv.org/abs/1309.4091} {arXiv:1309.4091
  [hep-ph]} \BibitemShut {NoStop}%
\bibitem [{\citenamefont {Funk}(2013)}]{Funk:2013gxa}%
  \BibitemOpen
  \bibfield  {author} {\bibinfo {author} {\bibfnamefont {S.}~\bibnamefont
  {Funk}},\ }\href@noop {} {\  (\bibinfo {year} {2013})},\ \Eprint
  {http://arxiv.org/abs/1310.2695} {arXiv:1310.2695 [astro-ph.HE]} \BibitemShut
  {NoStop}%
\bibitem [{\citenamefont {Askew}\ \emph {et~al.}(2014)\citenamefont {Askew},
  \citenamefont {Chauhan}, \citenamefont {Penning}, \citenamefont {Shepherd},\
  and\ \citenamefont {Tripathi}}]{Askew:2014kqa}%
  \BibitemOpen
  \bibfield  {author} {\bibinfo {author} {\bibfnamefont {A.}~\bibnamefont
  {Askew}}, \bibinfo {author} {\bibfnamefont {S.}~\bibnamefont {Chauhan}},
  \bibinfo {author} {\bibfnamefont {B.}~\bibnamefont {Penning}}, \bibinfo
  {author} {\bibfnamefont {W.}~\bibnamefont {Shepherd}}, \ and\ \bibinfo
  {author} {\bibfnamefont {M.}~\bibnamefont {Tripathi}},\ }\href {\doibase
  10.1142/S0217751X14300415} {\bibfield  {journal} {\bibinfo  {journal}
  {Int.J.Mod.Phys.}\ }\textbf {\bibinfo {volume} {A29}},\ \bibinfo {pages}
  {1430041} (\bibinfo {year} {2014})},\ \Eprint
  {http://arxiv.org/abs/1406.5662} {arXiv:1406.5662 [hep-ph]} \BibitemShut
  {NoStop}%
\bibitem [{\citenamefont {Cushman}\ \emph {et~al.}(2013)\citenamefont
  {Cushman}, \citenamefont {Galbiati}, \citenamefont {McKinsey}, \citenamefont
  {Robertson}, \citenamefont {Tait} \emph {et~al.}}]{Cushman:2013zza}%
  \BibitemOpen
  \bibfield  {author} {\bibinfo {author} {\bibfnamefont {P.}~\bibnamefont
  {Cushman}}, \bibinfo {author} {\bibfnamefont {C.}~\bibnamefont {Galbiati}},
  \bibinfo {author} {\bibfnamefont {D.}~\bibnamefont {McKinsey}}, \bibinfo
  {author} {\bibfnamefont {H.}~\bibnamefont {Robertson}}, \bibinfo {author}
  {\bibfnamefont {T.}~\bibnamefont {Tait}},  \emph {et~al.},\ }\href@noop {} {\
   (\bibinfo {year} {2013})},\ \Eprint {http://arxiv.org/abs/1310.8327}
  {arXiv:1310.8327 [hep-ex]} \BibitemShut {NoStop}%
\bibitem [{\citenamefont {Barger}\ \emph {et~al.}(2002)\citenamefont {Barger},
  \citenamefont {Halzen}, \citenamefont {Hooper},\ and\ \citenamefont
  {Kao}}]{Barger:2001ur}%
  \BibitemOpen
  \bibfield  {author} {\bibinfo {author} {\bibfnamefont {V.~D.}\ \bibnamefont
  {Barger}}, \bibinfo {author} {\bibfnamefont {F.}~\bibnamefont {Halzen}},
  \bibinfo {author} {\bibfnamefont {D.}~\bibnamefont {Hooper}}, \ and\ \bibinfo
  {author} {\bibfnamefont {C.}~\bibnamefont {Kao}},\ }\href {\doibase
  10.1103/PhysRevD.65.075022} {\bibfield  {journal} {\bibinfo  {journal}
  {Phys.Rev.}\ }\textbf {\bibinfo {volume} {D65}},\ \bibinfo {pages} {075022}
  (\bibinfo {year} {2002})},\ \Eprint {http://arxiv.org/abs/hep-ph/0105182}
  {arXiv:hep-ph/0105182 [hep-ph]} \BibitemShut {NoStop}%
\bibitem [{\citenamefont {Cirelli}\ \emph {et~al.}(2009)\citenamefont
  {Cirelli}, \citenamefont {Kadastik}, \citenamefont {Raidal},\ and\
  \citenamefont {Strumia}}]{Cirelli:2008pk}%
  \BibitemOpen
  \bibfield  {author} {\bibinfo {author} {\bibfnamefont {M.}~\bibnamefont
  {Cirelli}}, \bibinfo {author} {\bibfnamefont {M.}~\bibnamefont {Kadastik}},
  \bibinfo {author} {\bibfnamefont {M.}~\bibnamefont {Raidal}}, \ and\ \bibinfo
  {author} {\bibfnamefont {A.}~\bibnamefont {Strumia}},\ }\href {\doibase
  10.1016/j.nuclphysb.2013.05.002, 10.1016/j.nuclphysb.2008.11.031} {\bibfield
  {journal} {\bibinfo  {journal} {Nucl.Phys.}\ }\textbf {\bibinfo {volume}
  {B813}},\ \bibinfo {pages} {1} (\bibinfo {year} {2009})},\ \Eprint
  {http://arxiv.org/abs/0809.2409} {arXiv:0809.2409 [hep-ph]} \BibitemShut
  {NoStop}%
\bibitem [{\citenamefont {{Prantzos}}\ \emph {et~al.}(2011)\citenamefont
  {{Prantzos}}, \citenamefont {{Boehm}}, \citenamefont {{Bykov}}, \citenamefont
  {{Diehl}}, \citenamefont {{Ferri{\`e}re}}, \citenamefont {{Guessoum}},
  \citenamefont {{Jean}}, \citenamefont {{Knoedlseder}}, \citenamefont
  {{Marcowith}}, \citenamefont {{Moskalenko}}, \citenamefont {{Strong}},\ and\
  \citenamefont {{Weidenspointner}}}]{2011RvMP...83.1001P}%
  \BibitemOpen
  \bibfield  {author} {\bibinfo {author} {\bibfnamefont {N.}~\bibnamefont
  {{Prantzos}}}, \bibinfo {author} {\bibfnamefont {C.}~\bibnamefont {{Boehm}}},
  \bibinfo {author} {\bibfnamefont {A.~M.}\ \bibnamefont {{Bykov}}}, \bibinfo
  {author} {\bibfnamefont {R.}~\bibnamefont {{Diehl}}}, \bibinfo {author}
  {\bibfnamefont {K.}~\bibnamefont {{Ferri{\`e}re}}}, \bibinfo {author}
  {\bibfnamefont {N.}~\bibnamefont {{Guessoum}}}, \bibinfo {author}
  {\bibfnamefont {P.}~\bibnamefont {{Jean}}}, \bibinfo {author} {\bibfnamefont
  {J.}~\bibnamefont {{Knoedlseder}}}, \bibinfo {author} {\bibfnamefont
  {A.}~\bibnamefont {{Marcowith}}},  \emph {et~al.},\ }\href {\doibase
  10.1103/RevModPhys.83.1001} {\bibfield  {journal} {\bibinfo  {journal}
  {Reviews of Modern Physics}\ }\textbf {\bibinfo {volume} {83}},\ \bibinfo
  {pages} {1001} (\bibinfo {year} {2011})},\ \Eprint
  {http://arxiv.org/abs/1009.4620} {arXiv:1009.4620 [astro-ph.HE]} \BibitemShut
  {NoStop}%
\bibitem [{\citenamefont {Goodenough}\ and\ \citenamefont
  {Hooper}(2009)}]{Goodenough:2009gk}%
  \BibitemOpen
  \bibfield  {author} {\bibinfo {author} {\bibfnamefont {L.}~\bibnamefont
  {Goodenough}}\ and\ \bibinfo {author} {\bibfnamefont {D.}~\bibnamefont
  {Hooper}},\ }\href@noop {} {\  (\bibinfo {year} {2009})},\ \Eprint
  {http://arxiv.org/abs/0910.2998} {arXiv:0910.2998 [hep-ph]} \BibitemShut
  {NoStop}%
\bibitem [{\citenamefont {Hooper}\ and\ \citenamefont
  {Goodenough}(2011)}]{Hooper:2010mq}%
  \BibitemOpen
  \bibfield  {author} {\bibinfo {author} {\bibfnamefont {D.}~\bibnamefont
  {Hooper}}\ and\ \bibinfo {author} {\bibfnamefont {L.}~\bibnamefont
  {Goodenough}},\ }\href {\doibase 10.1016/j.physletb.2011.02.029} {\bibfield
  {journal} {\bibinfo  {journal} {Phys.Lett.}\ }\textbf {\bibinfo {volume}
  {B697}},\ \bibinfo {pages} {412} (\bibinfo {year} {2011})},\ \Eprint
  {http://arxiv.org/abs/1010.2752} {arXiv:1010.2752 [hep-ph]} \BibitemShut
  {NoStop}%
\bibitem [{\citenamefont {Hooper}\ and\ \citenamefont
  {Linden}(2011)}]{Hooper:2011ti}%
  \BibitemOpen
  \bibfield  {author} {\bibinfo {author} {\bibfnamefont {D.}~\bibnamefont
  {Hooper}}\ and\ \bibinfo {author} {\bibfnamefont {T.}~\bibnamefont
  {Linden}},\ }\href {\doibase 10.1103/PhysRevD.84.123005} {\bibfield
  {journal} {\bibinfo  {journal} {Phys.Rev.}\ }\textbf {\bibinfo {volume}
  {D84}},\ \bibinfo {pages} {123005} (\bibinfo {year} {2011})},\ \Eprint
  {http://arxiv.org/abs/1110.0006} {arXiv:1110.0006 [astro-ph.HE]} \BibitemShut
  {NoStop}%
\bibitem [{\citenamefont {Daylan}\ \emph {et~al.}(2014)\citenamefont {Daylan},
  \citenamefont {Finkbeiner}, \citenamefont {Hooper}, \citenamefont {Linden},
  \citenamefont {Portillo} \emph {et~al.}}]{Daylan:2014rsa}%
  \BibitemOpen
  \bibfield  {author} {\bibinfo {author} {\bibfnamefont {T.}~\bibnamefont
  {Daylan}}, \bibinfo {author} {\bibfnamefont {D.~P.}\ \bibnamefont
  {Finkbeiner}}, \bibinfo {author} {\bibfnamefont {D.}~\bibnamefont {Hooper}},
  \bibinfo {author} {\bibfnamefont {T.}~\bibnamefont {Linden}}, \bibinfo
  {author} {\bibfnamefont {S.~K.~N.}\ \bibnamefont {Portillo}},  \emph
  {et~al.},\ }\href@noop {} {\  (\bibinfo {year} {2014})},\ \Eprint
  {http://arxiv.org/abs/1402.6703} {arXiv:1402.6703 [astro-ph.HE]} \BibitemShut
  {NoStop}%
\bibitem [{\citenamefont {Abazajian}\ and\ \citenamefont
  {Kaplinghat}(2012)}]{Abazajian:2012pn}%
  \BibitemOpen
  \bibfield  {author} {\bibinfo {author} {\bibfnamefont {K.~N.}\ \bibnamefont
  {Abazajian}}\ and\ \bibinfo {author} {\bibfnamefont {M.}~\bibnamefont
  {Kaplinghat}},\ }\href {\doibase 10.1103/PhysRevD.86.083511} {\bibfield
  {journal} {\bibinfo  {journal} {Phys.Rev.}\ }\textbf {\bibinfo {volume}
  {D86}},\ \bibinfo {pages} {083511} (\bibinfo {year} {2012})},\ \Eprint
  {http://arxiv.org/abs/1207.6047} {arXiv:1207.6047 [astro-ph.HE]} \BibitemShut
  {NoStop}%
\bibitem [{\citenamefont {Macias}\ and\ \citenamefont
  {Gordon}(2014)}]{Macias:2013vya}%
  \BibitemOpen
  \bibfield  {author} {\bibinfo {author} {\bibfnamefont {O.}~\bibnamefont
  {Macias}}\ and\ \bibinfo {author} {\bibfnamefont {C.}~\bibnamefont
  {Gordon}},\ }\href {\doibase 10.1103/PhysRevD.89.063515} {\bibfield
  {journal} {\bibinfo  {journal} {Phys.Rev.}\ }\textbf {\bibinfo {volume}
  {D89}},\ \bibinfo {pages} {063515} (\bibinfo {year} {2014})},\ \Eprint
  {http://arxiv.org/abs/1312.6671} {arXiv:1312.6671 [astro-ph.HE]} \BibitemShut
  {NoStop}%
\bibitem [{\citenamefont {Abazajian}\ \emph {et~al.}(2014)\citenamefont
  {Abazajian}, \citenamefont {Canac}, \citenamefont {Horiuchi},\ and\
  \citenamefont {Kaplinghat}}]{Abazajian:2014fta}%
  \BibitemOpen
  \bibfield  {author} {\bibinfo {author} {\bibfnamefont {K.~N.}\ \bibnamefont
  {Abazajian}}, \bibinfo {author} {\bibfnamefont {N.}~\bibnamefont {Canac}},
  \bibinfo {author} {\bibfnamefont {S.}~\bibnamefont {Horiuchi}}, \ and\
  \bibinfo {author} {\bibfnamefont {M.}~\bibnamefont {Kaplinghat}},\ }\href
  {\doibase 10.1103/PhysRevD.90.023526} {\bibfield  {journal} {\bibinfo
  {journal} {Phys.Rev.}\ }\textbf {\bibinfo {volume} {D90}},\ \bibinfo {pages}
  {023526} (\bibinfo {year} {2014})},\ \Eprint {http://arxiv.org/abs/1402.4090}
  {arXiv:1402.4090 [astro-ph.HE]} \BibitemShut {NoStop}%
\bibitem [{\citenamefont {Bulbul}\ \emph {et~al.}(2014)\citenamefont {Bulbul},
  \citenamefont {Markevitch}, \citenamefont {Foster}, \citenamefont {Smith},
  \citenamefont {Loewenstein},\ and\ \citenamefont
  {Randall}}]{0004-637X-789-1-13}%
  \BibitemOpen
  \bibfield  {author} {\bibinfo {author} {\bibfnamefont {E.}~\bibnamefont
  {Bulbul}}, \bibinfo {author} {\bibfnamefont {M.}~\bibnamefont {Markevitch}},
  \bibinfo {author} {\bibfnamefont {A.}~\bibnamefont {Foster}}, \bibinfo
  {author} {\bibfnamefont {R.~K.}\ \bibnamefont {Smith}}, \bibinfo {author}
  {\bibfnamefont {M.}~\bibnamefont {Loewenstein}}, \ and\ \bibinfo {author}
  {\bibfnamefont {S.~W.}\ \bibnamefont {Randall}},\ }\href
  {http://stacks.iop.org/0004-637X/789/i=1/a=13} {\bibfield  {journal}
  {\bibinfo  {journal} {The Astrophysical Journal}\ }\textbf {\bibinfo {volume}
  {789}},\ \bibinfo {pages} {13} (\bibinfo {year} {2014})}\BibitemShut
  {NoStop}%
\bibitem [{\citenamefont {Boyarsky}\ \emph
  {et~al.}(2014{\natexlab{a}})\citenamefont {Boyarsky}, \citenamefont
  {Ruchayskiy}, \citenamefont {Iakubovskyi},\ and\ \citenamefont
  {Franse}}]{Boyarsky:2014jta}%
  \BibitemOpen
  \bibfield  {author} {\bibinfo {author} {\bibfnamefont {A.}~\bibnamefont
  {Boyarsky}}, \bibinfo {author} {\bibfnamefont {O.}~\bibnamefont
  {Ruchayskiy}}, \bibinfo {author} {\bibfnamefont {D.}~\bibnamefont
  {Iakubovskyi}}, \ and\ \bibinfo {author} {\bibfnamefont {J.}~\bibnamefont
  {Franse}},\ }\href@noop {} {\  (\bibinfo {year} {2014}{\natexlab{a}})},\
  \Eprint {http://arxiv.org/abs/1402.4119} {arXiv:1402.4119 [astro-ph.CO]}
  \BibitemShut {NoStop}%
\bibitem [{\citenamefont {Boyarsky}\ \emph
  {et~al.}(2014{\natexlab{b}})\citenamefont {Boyarsky}, \citenamefont {Franse},
  \citenamefont {Iakubovskyi},\ and\ \citenamefont
  {Ruchayskiy}}]{Boyarsky:2014ska}%
  \BibitemOpen
  \bibfield  {author} {\bibinfo {author} {\bibfnamefont {A.}~\bibnamefont
  {Boyarsky}}, \bibinfo {author} {\bibfnamefont {J.}~\bibnamefont {Franse}},
  \bibinfo {author} {\bibfnamefont {D.}~\bibnamefont {Iakubovskyi}}, \ and\
  \bibinfo {author} {\bibfnamefont {O.}~\bibnamefont {Ruchayskiy}},\
  }\href@noop {} {\  (\bibinfo {year} {2014}{\natexlab{b}})},\ \Eprint
  {http://arxiv.org/abs/1408.2503} {arXiv:1408.2503 [astro-ph.CO]} \BibitemShut
  {NoStop}%
\bibitem [{\citenamefont {Jeltema}\ and\ \citenamefont
  {Profumo}(2014)}]{Jeltema:2014qfa}%
  \BibitemOpen
  \bibfield  {author} {\bibinfo {author} {\bibfnamefont {T.~E.}\ \bibnamefont
  {Jeltema}}\ and\ \bibinfo {author} {\bibfnamefont {S.}~\bibnamefont
  {Profumo}},\ }\href@noop {} {\  (\bibinfo {year} {2014})},\ \Eprint
  {http://arxiv.org/abs/1408.1699} {arXiv:1408.1699 [astro-ph.HE]} \BibitemShut
  {NoStop}%
\bibitem [{\citenamefont {Springel}\ \emph {et~al.}(2008)\citenamefont
  {Springel}, \citenamefont {Wang}, \citenamefont {Vogelsberger}, \citenamefont
  {Ludlow}, \citenamefont {Jenkins} \emph {et~al.}}]{Springel:2008cc}%
  \BibitemOpen
  \bibfield  {author} {\bibinfo {author} {\bibfnamefont {V.}~\bibnamefont
  {Springel}}, \bibinfo {author} {\bibfnamefont {J.}~\bibnamefont {Wang}},
  \bibinfo {author} {\bibfnamefont {M.}~\bibnamefont {Vogelsberger}}, \bibinfo
  {author} {\bibfnamefont {A.}~\bibnamefont {Ludlow}}, \bibinfo {author}
  {\bibfnamefont {A.}~\bibnamefont {Jenkins}},  \emph {et~al.},\ }\href
  {\doibase 10.1111/j.1365-2966.2008.14066.x} {\bibfield  {journal} {\bibinfo
  {journal} {Mon.Not.Roy.Astron.Soc.}\ }\textbf {\bibinfo {volume} {391}},\
  \bibinfo {pages} {1685} (\bibinfo {year} {2008})},\ \Eprint
  {http://arxiv.org/abs/0809.0898} {arXiv:0809.0898 [astro-ph]} \BibitemShut
  {NoStop}%
\bibitem [{\citenamefont {Clowe}\ \emph
  {et~al.}(2006{\natexlab{a}})\citenamefont {Clowe}, \citenamefont {Bradac},
  \citenamefont {Gonzalez}, \citenamefont {Markevitch}, \citenamefont {Randall}
  \emph {et~al.}}]{Clowe:2006eq}%
  \BibitemOpen
  \bibfield  {author} {\bibinfo {author} {\bibfnamefont {D.}~\bibnamefont
  {Clowe}}, \bibinfo {author} {\bibfnamefont {M.}~\bibnamefont {Bradac}},
  \bibinfo {author} {\bibfnamefont {A.~H.}\ \bibnamefont {Gonzalez}}, \bibinfo
  {author} {\bibfnamefont {M.}~\bibnamefont {Markevitch}}, \bibinfo {author}
  {\bibfnamefont {S.~W.}\ \bibnamefont {Randall}},  \emph {et~al.},\ }\href
  {\doibase 10.1086/508162} {\bibfield  {journal} {\bibinfo  {journal}
  {Astrophys.J.}\ }\textbf {\bibinfo {volume} {648}},\ \bibinfo {pages} {L109}
  (\bibinfo {year} {2006}{\natexlab{a}})},\ \Eprint
  {http://arxiv.org/abs/astro-ph/0608407} {arXiv:astro-ph/0608407 [astro-ph]}
  \BibitemShut {NoStop}%
\bibitem [{\citenamefont {{Gabici}}\ and\ \citenamefont
  {{Blasi}}(2003)}]{2003ApJ...583..695G}%
  \BibitemOpen
  \bibfield  {author} {\bibinfo {author} {\bibfnamefont {S.}~\bibnamefont
  {{Gabici}}}\ and\ \bibinfo {author} {\bibfnamefont {P.}~\bibnamefont
  {{Blasi}}},\ }\href {\doibase 10.1086/345429} {\bibfield  {journal} {\bibinfo
   {journal} {\apj}\ }\textbf {\bibinfo {volume} {583}},\ \bibinfo {pages}
  {695} (\bibinfo {year} {2003})},\ \Eprint
  {http://arxiv.org/abs/astro-ph/0207523} {astro-ph/0207523} \BibitemShut
  {NoStop}%
\bibitem [{\citenamefont {{Fujita}}\ \emph {et~al.}(2003)\citenamefont
  {{Fujita}}, \citenamefont {{Takizawa}},\ and\ \citenamefont
  {{Sarazin}}}]{2003ApJ...584..190F}%
  \BibitemOpen
  \bibfield  {author} {\bibinfo {author} {\bibfnamefont {Y.}~\bibnamefont
  {{Fujita}}}, \bibinfo {author} {\bibfnamefont {M.}~\bibnamefont
  {{Takizawa}}}, \ and\ \bibinfo {author} {\bibfnamefont {C.~L.}\ \bibnamefont
  {{Sarazin}}},\ }\href {\doibase 10.1086/345599} {\bibfield  {journal}
  {\bibinfo  {journal} {\apj}\ }\textbf {\bibinfo {volume} {584}},\ \bibinfo
  {pages} {190} (\bibinfo {year} {2003})},\ \Eprint
  {http://arxiv.org/abs/astro-ph/0210320} {astro-ph/0210320} \BibitemShut
  {NoStop}%
\bibitem [{\citenamefont {{Honda}}\ \emph {et~al.}(1996)\citenamefont
  {{Honda}}, \citenamefont {{Hirayama}}, \citenamefont {{Watanabe}},
  \citenamefont {{Kunieda}}, \citenamefont {{Tawara}}, \citenamefont
  {{Yamashita}}, \citenamefont {{Ohashi}}, \citenamefont {{Hughes}},\ and\
  \citenamefont {{Henry}}}]{1996ApJ...473L..71H}%
  \BibitemOpen
  \bibfield  {author} {\bibinfo {author} {\bibfnamefont {H.}~\bibnamefont
  {{Honda}}}, \bibinfo {author} {\bibfnamefont {M.}~\bibnamefont {{Hirayama}}},
  \bibinfo {author} {\bibfnamefont {M.}~\bibnamefont {{Watanabe}}}, \bibinfo
  {author} {\bibfnamefont {H.}~\bibnamefont {{Kunieda}}}, \bibinfo {author}
  {\bibfnamefont {Y.}~\bibnamefont {{Tawara}}}, \bibinfo {author}
  {\bibfnamefont {K.}~\bibnamefont {{Yamashita}}}, \bibinfo {author}
  {\bibfnamefont {T.}~\bibnamefont {{Ohashi}}}, \bibinfo {author}
  {\bibfnamefont {J.~P.}\ \bibnamefont {{Hughes}}}, \ and\ \bibinfo {author}
  {\bibfnamefont {J.~P.}\ \bibnamefont {{Henry}}},\ }\href {\doibase
  10.1086/310400} {\bibfield  {journal} {\bibinfo  {journal} {Astrophys.J.}\
  }\textbf {\bibinfo {volume} {473}},\ \bibinfo {pages} {L71} (\bibinfo {year}
  {1996})}\BibitemShut {NoStop}%
\bibitem [{\citenamefont {Okabe}\ \emph {et~al.}(2014)\citenamefont {Okabe},
  \citenamefont {Futamase}, \citenamefont {Kajisawa},\ and\ \citenamefont
  {Kuroshima}}]{Okabe:2013oza}%
  \BibitemOpen
  \bibfield  {author} {\bibinfo {author} {\bibfnamefont {N.}~\bibnamefont
  {Okabe}}, \bibinfo {author} {\bibfnamefont {T.}~\bibnamefont {Futamase}},
  \bibinfo {author} {\bibfnamefont {M.}~\bibnamefont {Kajisawa}}, \ and\
  \bibinfo {author} {\bibfnamefont {R.}~\bibnamefont {Kuroshima}},\ }\href
  {\doibase 10.1088/0004-637X/784/2/90} {\bibfield  {journal} {\bibinfo
  {journal} {Astrophys.J.}\ }\textbf {\bibinfo {volume} {784}},\ \bibinfo
  {pages} {90} (\bibinfo {year} {2014})},\ \Eprint
  {http://arxiv.org/abs/1304.2399} {arXiv:1304.2399 [astro-ph.CO]} \BibitemShut
  {NoStop}%
\bibitem [{\citenamefont {{Gavazzi}}\ \emph {et~al.}(2009)\citenamefont
  {{Gavazzi}}, \citenamefont {{Adami}}, \citenamefont {{Durret}}, \citenamefont
  {{Cuillandre}}, \citenamefont {{Ilbert}}, \citenamefont {{Mazure}},
  \citenamefont {{Pell{\'o}}},\ and\ \citenamefont
  {{Ulmer}}}]{2009A&A...498L..33G}%
  \BibitemOpen
  \bibfield  {author} {\bibinfo {author} {\bibfnamefont {R.}~\bibnamefont
  {{Gavazzi}}}, \bibinfo {author} {\bibfnamefont {C.}~\bibnamefont {{Adami}}},
  \bibinfo {author} {\bibfnamefont {F.}~\bibnamefont {{Durret}}}, \bibinfo
  {author} {\bibfnamefont {J.-C.}\ \bibnamefont {{Cuillandre}}}, \bibinfo
  {author} {\bibfnamefont {O.}~\bibnamefont {{Ilbert}}}, \bibinfo {author}
  {\bibfnamefont {A.}~\bibnamefont {{Mazure}}}, \bibinfo {author}
  {\bibfnamefont {R.}~\bibnamefont {{Pell{\'o}}}}, \ and\ \bibinfo {author}
  {\bibfnamefont {M.~P.}\ \bibnamefont {{Ulmer}}},\ }\href {\doibase
  10.1051/0004-6361/200911841} {\bibfield  {journal} {\bibinfo  {journal}
  {Astron.Astrophys.}\ }\textbf {\bibinfo {volume} {498}},\ \bibinfo {pages}
  {L33} (\bibinfo {year} {2009})},\ \Eprint {http://arxiv.org/abs/0904.0220}
  {arXiv:0904.0220 [astro-ph.CO]} \BibitemShut {NoStop}%
\bibitem [{\citenamefont {Clowe}\ \emph
  {et~al.}(2006{\natexlab{b}})\citenamefont {Clowe}, \citenamefont {Bradac},
  \citenamefont {Gonzalez}, \citenamefont {Markevitch}, \citenamefont {Randall}
  \emph {et~al.}}]{data:bullet_lens}%
  \BibitemOpen
  \bibfield  {author} {\bibinfo {author} {\bibfnamefont {D.}~\bibnamefont
  {Clowe}}, \bibinfo {author} {\bibfnamefont {M.}~\bibnamefont {Bradac}},
  \bibinfo {author} {\bibfnamefont {A.~H.}\ \bibnamefont {Gonzalez}}, \bibinfo
  {author} {\bibfnamefont {M.}~\bibnamefont {Markevitch}}, \bibinfo {author}
  {\bibfnamefont {S.~W.}\ \bibnamefont {Randall}},  \emph {et~al.},\
  }\href@noop {} {} (\bibinfo {year} {2006}{\natexlab{b}}),\ \bibinfo {note}
  {\url{http://flamingos.astro.ufl.edu/1e0657/public.html}}\BibitemShut
  {NoStop}%
\bibitem [{\citenamefont {{XMM-Newton}}(2000)}]{data:xmm_0112980201}%
  \BibitemOpen
  \bibfield  {author} {\bibinfo {author} {\bibnamefont {{XMM-Newton}}},\
  }\href@noop {} {} (\bibinfo {year} {2000}),\ \bibinfo {note} {target:
  RXJ0658-55 (Bullet Cluster); date: 2000-10-21 18:44:57; obs.~ID:
  0112980201}\BibitemShut {NoStop}%
\bibitem [{\citenamefont {{ROSAT}}(1994)}]{data:rosat_RH800242A04}%
  \BibitemOpen
  \bibfield  {author} {\bibinfo {author} {\bibnamefont {{ROSAT}}},\ }\href@noop
  {} {} (\bibinfo {year} {1994}),\ \bibinfo {note} {target: Coma Cluster; date:
  1994-06-08 05:00:17; seq.~ID: RH800242A04}\BibitemShut {NoStop}%
\bibitem [{\citenamefont {Kubo}\ \emph {et~al.}(2007)\citenamefont {Kubo},
  \citenamefont {Stebbins}, \citenamefont {Annis}, \citenamefont
  {Dell'Antonio}, \citenamefont {Lin} \emph {et~al.}}]{Kubo:2007wt}%
  \BibitemOpen
  \bibfield  {author} {\bibinfo {author} {\bibfnamefont {J.~M.}\ \bibnamefont
  {Kubo}}, \bibinfo {author} {\bibfnamefont {A.}~\bibnamefont {Stebbins}},
  \bibinfo {author} {\bibfnamefont {J.}~\bibnamefont {Annis}}, \bibinfo
  {author} {\bibfnamefont {I.~P.}\ \bibnamefont {Dell'Antonio}}, \bibinfo
  {author} {\bibfnamefont {H.}~\bibnamefont {Lin}},  \emph {et~al.},\ }\href
  {\doibase 10.1086/523101} {\bibfield  {journal} {\bibinfo  {journal}
  {Astrophys.J.}\ }\textbf {\bibinfo {volume} {671}},\ \bibinfo {pages} {1466}
  (\bibinfo {year} {2007})},\ \Eprint {http://arxiv.org/abs/0709.0506}
  {arXiv:0709.0506 [astro-ph]} \BibitemShut {NoStop}%
\bibitem [{\citenamefont {{Clowe}}\ \emph {et~al.}(2004)\citenamefont
  {{Clowe}}, \citenamefont {{Gonzalez}},\ and\ \citenamefont
  {{Markevitch}}}]{2004ApJ...604..596C}%
  \BibitemOpen
  \bibfield  {author} {\bibinfo {author} {\bibfnamefont {D.}~\bibnamefont
  {{Clowe}}}, \bibinfo {author} {\bibfnamefont {A.}~\bibnamefont {{Gonzalez}}},
  \ and\ \bibinfo {author} {\bibfnamefont {M.}~\bibnamefont {{Markevitch}}},\
  }\href {\doibase 10.1086/381970} {\bibfield  {journal} {\bibinfo  {journal}
  {\apj}\ }\textbf {\bibinfo {volume} {604}},\ \bibinfo {pages} {596} (\bibinfo
  {year} {2004})},\ \Eprint {http://arxiv.org/abs/astro-ph/0312273}
  {astro-ph/0312273} \BibitemShut {NoStop}%
\bibitem [{\citenamefont {Durret}\ \emph {et~al.}(2013)\citenamefont {Durret},
  \citenamefont {Perrot}, \citenamefont {Neto}, \citenamefont {Adami},
  \citenamefont {Bertin} \emph {et~al.}}]{Durret:2013mna}%
  \BibitemOpen
  \bibfield  {author} {\bibinfo {author} {\bibfnamefont {F.}~\bibnamefont
  {Durret}}, \bibinfo {author} {\bibfnamefont {C.}~\bibnamefont {Perrot}},
  \bibinfo {author} {\bibfnamefont {G.~B.~L.}\ \bibnamefont {Neto}}, \bibinfo
  {author} {\bibfnamefont {C.}~\bibnamefont {Adami}}, \bibinfo {author}
  {\bibfnamefont {E.}~\bibnamefont {Bertin}},  \emph {et~al.},\ }\href@noop {}
  {\  (\bibinfo {year} {2013})},\ \Eprint {http://arxiv.org/abs/1310.7493}
  {arXiv:1310.7493 [astro-ph.CO]} \BibitemShut {NoStop}%
\bibitem [{\citenamefont {Machado}\ and\ \citenamefont
  {Neto}(2013)}]{Machado:2013jq}%
  \BibitemOpen
  \bibfield  {author} {\bibinfo {author} {\bibfnamefont {R.~E.}\ \bibnamefont
  {Machado}}\ and\ \bibinfo {author} {\bibfnamefont {G.~B.~L.}\ \bibnamefont
  {Neto}},\ }\href@noop {} {\  (\bibinfo {year} {2013})},\ \Eprint
  {http://arxiv.org/abs/1301.4434} {arXiv:1301.4434 [astro-ph.CO]} \BibitemShut
  {NoStop}%
\bibitem [{\citenamefont {Jee}\ \emph {et~al.}(2014)\citenamefont {Jee},
  \citenamefont {Hoekstra}, \citenamefont {Mahdavi},\ and\ \citenamefont
  {Babul}}]{Jee:2014hja}%
  \BibitemOpen
  \bibfield  {author} {\bibinfo {author} {\bibfnamefont {M.~J.}\ \bibnamefont
  {Jee}}, \bibinfo {author} {\bibfnamefont {H.}~\bibnamefont {Hoekstra}},
  \bibinfo {author} {\bibfnamefont {A.}~\bibnamefont {Mahdavi}}, \ and\
  \bibinfo {author} {\bibfnamefont {A.}~\bibnamefont {Babul}},\ }\href
  {\doibase 10.1088/0004-637X/783/2/78} {\bibfield  {journal} {\bibinfo
  {journal} {Astrophys.J.}\ }\textbf {\bibinfo {volume} {783}},\ \bibinfo
  {pages} {78} (\bibinfo {year} {2014})},\ \Eprint
  {http://arxiv.org/abs/1401.3356} {arXiv:1401.3356 [astro-ph.CO]} \BibitemShut
  {NoStop}%
\bibitem [{\citenamefont {{Clowe}}\ \emph {et~al.}(2012)\citenamefont
  {{Clowe}}, \citenamefont {{Markevitch}}, \citenamefont {{Brada{\v c}}},
  \citenamefont {{Gonzalez}}, \citenamefont {{Chung}}, \citenamefont
  {{Massey}},\ and\ \citenamefont {{Zaritsky}}}]{2012ApJ...758..128C}%
  \BibitemOpen
  \bibfield  {author} {\bibinfo {author} {\bibfnamefont {D.}~\bibnamefont
  {{Clowe}}}, \bibinfo {author} {\bibfnamefont {M.}~\bibnamefont
  {{Markevitch}}}, \bibinfo {author} {\bibfnamefont {M.}~\bibnamefont
  {{Brada{\v c}}}}, \bibinfo {author} {\bibfnamefont {A.~H.}\ \bibnamefont
  {{Gonzalez}}}, \bibinfo {author} {\bibfnamefont {S.~M.}\ \bibnamefont
  {{Chung}}}, \bibinfo {author} {\bibfnamefont {R.}~\bibnamefont {{Massey}}}, \
  and\ \bibinfo {author} {\bibfnamefont {D.}~\bibnamefont {{Zaritsky}}},\
  }\href {\doibase 10.1088/0004-637X/758/2/128} {\bibfield  {journal} {\bibinfo
   {journal} {\apj}\ }\textbf {\bibinfo {volume} {758}},\ \bibinfo {eid} {128}
  (\bibinfo {year} {2012})},\ \Eprint {http://arxiv.org/abs/1209.2143}
  {arXiv:1209.2143 [astro-ph.CO]} \BibitemShut {NoStop}%
\bibitem [{\citenamefont {Okabe}\ \emph {et~al.}(2011)\citenamefont {Okabe},
  \citenamefont {Bourdin}, \citenamefont {Mazzotta},\ and\ \citenamefont
  {Maurogordato}}]{Okabe:2011dt}%
  \BibitemOpen
  \bibfield  {author} {\bibinfo {author} {\bibfnamefont {N.}~\bibnamefont
  {Okabe}}, \bibinfo {author} {\bibfnamefont {H.}~\bibnamefont {Bourdin}},
  \bibinfo {author} {\bibfnamefont {P.}~\bibnamefont {Mazzotta}}, \ and\
  \bibinfo {author} {\bibfnamefont {S.}~\bibnamefont {Maurogordato}},\ }\href
  {\doibase 10.1088/0004-637X/741/2/116} {\bibfield  {journal} {\bibinfo
  {journal} {Astrophys.J.}\ }\textbf {\bibinfo {volume} {741}},\ \bibinfo
  {pages} {116} (\bibinfo {year} {2011})},\ \Eprint
  {http://arxiv.org/abs/1107.0004} {arXiv:1107.0004 [astro-ph.CO]} \BibitemShut
  {NoStop}%
\bibitem [{\citenamefont {Ragozzine}\ \emph {et~al.}(2012)\citenamefont
  {Ragozzine}, \citenamefont {Clowe}, \citenamefont {Markevitch}, \citenamefont
  {Gonzalez},\ and\ \citenamefont {Bradac}}]{Ragozzine:2011qt}%
  \BibitemOpen
  \bibfield  {author} {\bibinfo {author} {\bibfnamefont {B.}~\bibnamefont
  {Ragozzine}}, \bibinfo {author} {\bibfnamefont {D.}~\bibnamefont {Clowe}},
  \bibinfo {author} {\bibfnamefont {M.}~\bibnamefont {Markevitch}}, \bibinfo
  {author} {\bibfnamefont {A.~H.}\ \bibnamefont {Gonzalez}}, \ and\ \bibinfo
  {author} {\bibfnamefont {M.}~\bibnamefont {Bradac}},\ }\href {\doibase
  10.1088/0004-637X/744/2/94} {\bibfield  {journal} {\bibinfo  {journal}
  {Astrophys.J.}\ }\textbf {\bibinfo {volume} {744}},\ \bibinfo {pages} {94}
  (\bibinfo {year} {2012})},\ \Eprint {http://arxiv.org/abs/1111.4983}
  {arXiv:1111.4983 [astro-ph.CO]} \BibitemShut {NoStop}%
\bibitem [{\citenamefont {{Babyk}}\ \emph {et~al.}(2012)\citenamefont
  {{Babyk}}, \citenamefont {{Melnyk}},\ and\ \citenamefont
  {{Elyiv}}}]{2012AASP....2...56B}%
  \BibitemOpen
  \bibfield  {author} {\bibinfo {author} {\bibfnamefont {I.}~\bibnamefont
  {{Babyk}}}, \bibinfo {author} {\bibfnamefont {O.}~\bibnamefont {{Melnyk}}}, \
  and\ \bibinfo {author} {\bibfnamefont {A.}~\bibnamefont {{Elyiv}}},\
  }\href@noop {} {\bibfield  {journal} {\bibinfo  {journal} {Advances in
  Astronomy and Space Physics}\ }\textbf {\bibinfo {volume} {2}},\ \bibinfo
  {pages} {56} (\bibinfo {year} {2012})}\BibitemShut {NoStop}%
\bibitem [{\citenamefont {Merten}\ \emph {et~al.}(2011)\citenamefont {Merten},
  \citenamefont {Coe}, \citenamefont {Dupke}, \citenamefont {Massey},
  \citenamefont {Zitrin} \emph {et~al.}}]{Merten:2011wj}%
  \BibitemOpen
  \bibfield  {author} {\bibinfo {author} {\bibfnamefont {J.}~\bibnamefont
  {Merten}}, \bibinfo {author} {\bibfnamefont {D.}~\bibnamefont {Coe}},
  \bibinfo {author} {\bibfnamefont {R.}~\bibnamefont {Dupke}}, \bibinfo
  {author} {\bibfnamefont {R.}~\bibnamefont {Massey}}, \bibinfo {author}
  {\bibfnamefont {A.}~\bibnamefont {Zitrin}},  \emph {et~al.},\ }\href
  {\doibase 10.1111/j.1365-2966.2011.19266.x} {\bibfield  {journal} {\bibinfo
  {journal} {Mon.Not.Roy.Astron.Soc.}\ }\textbf {\bibinfo {volume} {417}},\
  \bibinfo {pages} {333} (\bibinfo {year} {2011})},\ \Eprint
  {http://arxiv.org/abs/1103.2772} {arXiv:1103.2772 [astro-ph.CO]} \BibitemShut
  {NoStop}%
\bibitem [{\citenamefont {{Dawson}}\ \emph {et~al.}(2012)\citenamefont
  {{Dawson}}, \citenamefont {{Wittman}}, \citenamefont {{Jee}}, \citenamefont
  {{Gee}}, \citenamefont {{Hughes}}, \citenamefont {{Tyson}}, \citenamefont
  {{Schmidt}}, \citenamefont {{Thorman}}, \citenamefont {{Brada{\v c}}},
  \citenamefont {{Miyazaki}}, \citenamefont {{Lemaux}}, \citenamefont
  {{Utsumi}},\ and\ \citenamefont {{Margoniner}}}]{2012ApJ...747L..42D}%
  \BibitemOpen
  \bibfield  {author} {\bibinfo {author} {\bibfnamefont {W.~A.}\ \bibnamefont
  {{Dawson}}}, \bibinfo {author} {\bibfnamefont {D.}~\bibnamefont {{Wittman}}},
  \bibinfo {author} {\bibfnamefont {M.~J.}\ \bibnamefont {{Jee}}}, \bibinfo
  {author} {\bibfnamefont {P.}~\bibnamefont {{Gee}}}, \bibinfo {author}
  {\bibfnamefont {J.~P.}\ \bibnamefont {{Hughes}}}, \bibinfo {author}
  {\bibfnamefont {J.~A.}\ \bibnamefont {{Tyson}}}, \bibinfo {author}
  {\bibfnamefont {S.}~\bibnamefont {{Schmidt}}}, \bibinfo {author}
  {\bibfnamefont {P.}~\bibnamefont {{Thorman}}}, \bibinfo {author}
  {\bibfnamefont {M.}~\bibnamefont {{Brada{\v c}}}},  \emph {et~al.},\ }\href
  {\doibase 10.1088/2041-8205/747/2/L42} {\bibfield  {journal} {\bibinfo
  {journal} {\apj}\ }\textbf {\bibinfo {volume} {747}},\ \bibinfo {eid} {L42}
  (\bibinfo {year} {2012})},\ \Eprint {http://arxiv.org/abs/1110.4391}
  {arXiv:1110.4391 [astro-ph.CO]} \BibitemShut {NoStop}%
\bibitem [{\citenamefont {Bradac}\ \emph {et~al.}(2008)\citenamefont {Bradac},
  \citenamefont {Allen}, \citenamefont {Treu}, \citenamefont {Ebeling},
  \citenamefont {Massey} \emph {et~al.}}]{Bradac:2008eu}%
  \BibitemOpen
  \bibfield  {author} {\bibinfo {author} {\bibfnamefont {M.}~\bibnamefont
  {Bradac}}, \bibinfo {author} {\bibfnamefont {S.~W.}\ \bibnamefont {Allen}},
  \bibinfo {author} {\bibfnamefont {T.}~\bibnamefont {Treu}}, \bibinfo {author}
  {\bibfnamefont {H.}~\bibnamefont {Ebeling}}, \bibinfo {author} {\bibfnamefont
  {R.}~\bibnamefont {Massey}},  \emph {et~al.},\ }\href@noop {} {\  (\bibinfo
  {year} {2008})},\ \Eprint {http://arxiv.org/abs/0806.2320} {arXiv:0806.2320
  [astro-ph]} \BibitemShut {NoStop}%
\bibitem [{\citenamefont {Boyarsky}\ \emph {et~al.}(2008)\citenamefont
  {Boyarsky}, \citenamefont {Ruchayskiy},\ and\ \citenamefont
  {Markevitch}}]{Boyarsky:2006kc}%
  \BibitemOpen
  \bibfield  {author} {\bibinfo {author} {\bibfnamefont {A.}~\bibnamefont
  {Boyarsky}}, \bibinfo {author} {\bibfnamefont {O.}~\bibnamefont
  {Ruchayskiy}}, \ and\ \bibinfo {author} {\bibfnamefont {M.}~\bibnamefont
  {Markevitch}},\ }\href {\doibase 10.1086/524397} {\bibfield  {journal}
  {\bibinfo  {journal} {Astrophys.J.}\ }\textbf {\bibinfo {volume} {673}},\
  \bibinfo {pages} {752} (\bibinfo {year} {2008})},\ \Eprint
  {http://arxiv.org/abs/astro-ph/0611168} {arXiv:astro-ph/0611168 [astro-ph]}
  \BibitemShut {NoStop}%
\bibitem [{\citenamefont {Weniger}(2012)}]{Weniger:2012tx}%
  \BibitemOpen
  \bibfield  {author} {\bibinfo {author} {\bibfnamefont {C.}~\bibnamefont
  {Weniger}},\ }\href {\doibase 10.1088/1475-7516/2012/08/007} {\bibfield
  {journal} {\bibinfo  {journal} {JCAP}\ }\textbf {\bibinfo {volume} {1208}},\
  \bibinfo {pages} {007} (\bibinfo {year} {2012})},\ \Eprint
  {http://arxiv.org/abs/1204.2797} {arXiv:1204.2797 [hep-ph]} \BibitemShut
  {NoStop}%
\bibitem [{\citenamefont {Boyarsky}\ \emph {et~al.}(2010)\citenamefont
  {Boyarsky}, \citenamefont {Ruchayskiy}, \citenamefont {Iakubovskyi},
  \citenamefont {Walker}, \citenamefont {Riemer-Sorensen} \emph
  {et~al.}}]{2010MNRAS.407.1188B}%
  \BibitemOpen
  \bibfield  {author} {\bibinfo {author} {\bibfnamefont {A.}~\bibnamefont
  {Boyarsky}}, \bibinfo {author} {\bibfnamefont {O.}~\bibnamefont
  {Ruchayskiy}}, \bibinfo {author} {\bibfnamefont {D.}~\bibnamefont
  {Iakubovskyi}}, \bibinfo {author} {\bibfnamefont {M.~G.}\ \bibnamefont
  {Walker}}, \bibinfo {author} {\bibfnamefont {S.}~\bibnamefont
  {Riemer-Sorensen}},  \emph {et~al.},\ }\href {\doibase
  10.1111/j.1365-2966.2010.17004.x} {\bibfield  {journal} {\bibinfo  {journal}
  {Mon.Not.Roy.Astron.Soc.}\ }\textbf {\bibinfo {volume} {407}},\ \bibinfo
  {pages} {1188} (\bibinfo {year} {2010})},\ \Eprint
  {http://arxiv.org/abs/1001.0644} {arXiv:1001.0644 [astro-ph.CO]} \BibitemShut
  {NoStop}%
\bibitem [{\citenamefont {Kusenko}\ and\ \citenamefont
  {Loewenstein}(2010)}]{2010arXiv1001.4055K}%
  \BibitemOpen
  \bibfield  {author} {\bibinfo {author} {\bibfnamefont {A.}~\bibnamefont
  {Kusenko}}\ and\ \bibinfo {author} {\bibfnamefont {M.}~\bibnamefont
  {Loewenstein}},\ }\href@noop {} {\  (\bibinfo {year} {2010})},\ \Eprint
  {http://arxiv.org/abs/1001.4055} {arXiv:1001.4055 [astro-ph.CO]} \BibitemShut
  {NoStop}%
\bibitem [{\citenamefont {Carlson}\ \emph {et~al.}(2014)\citenamefont
  {Carlson}, \citenamefont {Jeltema},\ and\ \citenamefont
  {Profumo}}]{Carlson:2014lla}%
  \BibitemOpen
  \bibfield  {author} {\bibinfo {author} {\bibfnamefont {E.}~\bibnamefont
  {Carlson}}, \bibinfo {author} {\bibfnamefont {T.}~\bibnamefont {Jeltema}}, \
  and\ \bibinfo {author} {\bibfnamefont {S.}~\bibnamefont {Profumo}},\
  }\href@noop {} {\  (\bibinfo {year} {2014})},\ \Eprint
  {http://arxiv.org/abs/1411.1758} {arXiv:1411.1758 [astro-ph.HE]} \BibitemShut
  {NoStop}%
\bibitem [{\citenamefont {Maurin}\ \emph {et~al.}(2012)\citenamefont {Maurin},
  \citenamefont {Combet}, \citenamefont {Nezri},\ and\ \citenamefont
  {Pointecouteau}}]{Maurin:2012tv}%
  \BibitemOpen
  \bibfield  {author} {\bibinfo {author} {\bibfnamefont {D.}~\bibnamefont
  {Maurin}}, \bibinfo {author} {\bibfnamefont {C.}~\bibnamefont {Combet}},
  \bibinfo {author} {\bibfnamefont {E.}~\bibnamefont {Nezri}}, \ and\ \bibinfo
  {author} {\bibfnamefont {E.}~\bibnamefont {Pointecouteau}},\ }\href {\doibase
  10.1051/0004-6361/201218986} {\bibfield  {journal} {\bibinfo  {journal}
  {Astron.Astrophys.}\ }\textbf {\bibinfo {volume} {547}},\ \bibinfo {pages}
  {A16} (\bibinfo {year} {2012})},\ \Eprint {http://arxiv.org/abs/1203.1166}
  {arXiv:1203.1166 [astro-ph.HE]} \BibitemShut {NoStop}%
\bibitem [{\citenamefont {Geringer-Sameth}\ \emph {et~al.}(2014)\citenamefont
  {Geringer-Sameth}, \citenamefont {Koushiappas},\ and\ \citenamefont
  {Walker}}]{Geringer-Sameth:2014qqa}%
  \BibitemOpen
  \bibfield  {author} {\bibinfo {author} {\bibfnamefont {A.}~\bibnamefont
  {Geringer-Sameth}}, \bibinfo {author} {\bibfnamefont {S.~M.}\ \bibnamefont
  {Koushiappas}}, \ and\ \bibinfo {author} {\bibfnamefont {M.~G.}\ \bibnamefont
  {Walker}},\ }\href@noop {} {\  (\bibinfo {year} {2014})},\ \Eprint
  {http://arxiv.org/abs/1410.2242} {arXiv:1410.2242 [astro-ph.CO]} \BibitemShut
  {NoStop}%
\bibitem [{\citenamefont {Boyarsky}\ \emph {et~al.}(2009)\citenamefont
  {Boyarsky}, \citenamefont {Ruchayskiy},\ and\ \citenamefont
  {Iakubovskyi}}]{Boyarsky:2008ju}%
  \BibitemOpen
  \bibfield  {author} {\bibinfo {author} {\bibfnamefont {A.}~\bibnamefont
  {Boyarsky}}, \bibinfo {author} {\bibfnamefont {O.}~\bibnamefont
  {Ruchayskiy}}, \ and\ \bibinfo {author} {\bibfnamefont {D.}~\bibnamefont
  {Iakubovskyi}},\ }\href {\doibase 10.1088/1475-7516/2009/03/005} {\bibfield
  {journal} {\bibinfo  {journal} {J.Cosmol.Astropart.Phys.}\ }\textbf {\bibinfo
  {volume} {0903}},\ \bibinfo {pages} {005} (\bibinfo {year} {2009})},\ \Eprint
  {http://arxiv.org/abs/0808.3902} {arXiv:0808.3902 [hep-ph]} \BibitemShut
  {NoStop}%
\bibitem [{\citenamefont {Viel}\ \emph {et~al.}(2013)\citenamefont {Viel},
  \citenamefont {Becker}, \citenamefont {Bolton},\ and\ \citenamefont
  {Haehnelt}}]{Viel:2013fqw}%
  \BibitemOpen
  \bibfield  {author} {\bibinfo {author} {\bibfnamefont {M.}~\bibnamefont
  {Viel}}, \bibinfo {author} {\bibfnamefont {G.~D.}\ \bibnamefont {Becker}},
  \bibinfo {author} {\bibfnamefont {J.~S.}\ \bibnamefont {Bolton}}, \ and\
  \bibinfo {author} {\bibfnamefont {M.~G.}\ \bibnamefont {Haehnelt}},\ }\href
  {\doibase 10.1103/PhysRevD.88.043502} {\bibfield  {journal} {\bibinfo
  {journal} {Phys.Rev.}\ }\textbf {\bibinfo {volume} {D88}},\ \bibinfo {pages}
  {043502} (\bibinfo {year} {2013})},\ \Eprint {http://arxiv.org/abs/1306.2314}
  {arXiv:1306.2314 [astro-ph.CO]} \BibitemShut {NoStop}%
\bibitem [{\citenamefont {Marsh}\ and\ \citenamefont
  {Silk}(2013)}]{Marsh:2013ywa}%
  \BibitemOpen
  \bibfield  {author} {\bibinfo {author} {\bibfnamefont {D.~J.~E.}\
  \bibnamefont {Marsh}}\ and\ \bibinfo {author} {\bibfnamefont
  {J.}~\bibnamefont {Silk}},\ }\href {\doibase 10.1093/mnras/stt2079}
  {\bibfield  {journal} {\bibinfo  {journal} {Mon.Not.Roy.Astron.Soc.}\
  }\textbf {\bibinfo {volume} {437}},\ \bibinfo {pages} {2652} (\bibinfo {year}
  {2013})},\ \Eprint {http://arxiv.org/abs/1307.1705} {arXiv:1307.1705
  [astro-ph.CO]} \BibitemShut {NoStop}%
\bibitem [{\citenamefont {Voit}(2005)}]{Voit:2004ah}%
  \BibitemOpen
  \bibfield  {author} {\bibinfo {author} {\bibfnamefont {G.~M.}\ \bibnamefont
  {Voit}},\ }\href {\doibase 10.1103/RevModPhys.77.207} {\bibfield  {journal}
  {\bibinfo  {journal} {Rev.Mod.Phys.}\ }\textbf {\bibinfo {volume} {77}},\
  \bibinfo {pages} {207} (\bibinfo {year} {2005})},\ \Eprint
  {http://arxiv.org/abs/astro-ph/0410173} {arXiv:astro-ph/0410173 [astro-ph]}
  \BibitemShut {NoStop}%
\bibitem [{\citenamefont {Boyarsky}\ \emph {et~al.}(2006)\citenamefont
  {Boyarsky}, \citenamefont {Neronov}, \citenamefont {Ruchayskiy},\ and\
  \citenamefont {Shaposhnikov}}]{Boyarsky:2006zi}%
  \BibitemOpen
  \bibfield  {author} {\bibinfo {author} {\bibfnamefont {A.}~\bibnamefont
  {Boyarsky}}, \bibinfo {author} {\bibfnamefont {A.}~\bibnamefont {Neronov}},
  \bibinfo {author} {\bibfnamefont {O.}~\bibnamefont {Ruchayskiy}}, \ and\
  \bibinfo {author} {\bibfnamefont {M.}~\bibnamefont {Shaposhnikov}},\ }\href
  {\doibase 10.1103/PhysRevD.74.103506} {\bibfield  {journal} {\bibinfo
  {journal} {Phys.Rev.}\ }\textbf {\bibinfo {volume} {D74}},\ \bibinfo {pages}
  {103506} (\bibinfo {year} {2006})},\ \Eprint
  {http://arxiv.org/abs/astro-ph/0603368} {arXiv:astro-ph/0603368 [astro-ph]}
  \BibitemShut {NoStop}%
\bibitem [{\citenamefont {{Aleksi{\'c}}}\ \emph {et~al.}(2012)\citenamefont
  {{Aleksi{\'c}}}, \citenamefont {{Alvarez}}, \citenamefont {{Antonelli}},
  \citenamefont {{Antoranz}}, \citenamefont {{Asensio}}, \citenamefont
  {{Backes}}, \citenamefont {{Barres de Almeida}}, \citenamefont {{Barrio}},
  \citenamefont {{Bastieri}}, \citenamefont {{Becerra Gonz{\'a}lez}},
  \citenamefont {{Bednarek}}, \citenamefont {{Berdyugin}}, \citenamefont
  {{Berger}}, \citenamefont {{Bernardini}}, \citenamefont {{Biland}},
  \citenamefont {{Blanch}}, \citenamefont {{Bock}}, \citenamefont {{Boller}},
  \citenamefont {{Bonnoli}}, \citenamefont {{Borla Tridon}}, \citenamefont
  {{Braun}}, \citenamefont {{Bretz}}, \citenamefont {{Ca{\~n}ellas}},
  \citenamefont {{Carmona}}, \citenamefont {{Carosi}}, \citenamefont {{Colin}},
  \citenamefont {{Colombo}}, \citenamefont {{Contreras}}, \citenamefont
  {{Cortina}}, \citenamefont {{Cossio}}, \citenamefont {{Covino}},
  \citenamefont {{Dazzi}}, \citenamefont {{de Angelis}}, \citenamefont {{de
  Caneva}}, \citenamefont {{de Cea Del Pozo}}, \citenamefont {{de Lotto}},
  \citenamefont {{Delgado Mendez}}, \citenamefont {{Diago Ortega}},
  \citenamefont {{Doert}}, \citenamefont {{Dom{\'{\i}}nguez}}, \citenamefont
  {{Dominis Prester}}, \citenamefont {{Dorner}}, \citenamefont {{Doro}},
  \citenamefont {{Eisenacher}}, \citenamefont {{Elsaesser}}, \citenamefont
  {{Ferenc}}, \citenamefont {{Fonseca}}, \citenamefont {{Font}}, \citenamefont
  {{Fruck}}, \citenamefont {{Garc{\'{\i}}a L{\'o}pez}}, \citenamefont
  {{Garczarczyk}}, \citenamefont {{Garrido}}, \citenamefont {{Giavitto}},
  \citenamefont {{Godinovi{\'c}}}, \citenamefont {{Gozzini}}, \citenamefont
  {{Hadasch}}, \citenamefont {{H{\"a}fner}}, \citenamefont {{Herrero}},
  \citenamefont {{Hildebrand}}, \citenamefont {{H{\"o}hne-M{\"o}nch}},
  \citenamefont {{Hose}}, \citenamefont {{Hrupec}}, \citenamefont {{Jogler}},
  \citenamefont {{Kellermann}}, \citenamefont {{Klepser}}, \citenamefont
  {{Kr{\"a}henb{\"u}hl}}, \citenamefont {{Krause}}, \citenamefont {{Kushida}},
  \citenamefont {{La Barbera}}, \citenamefont {{Lelas}}, \citenamefont
  {{Leonardo}}, \citenamefont {{Lewandowska}}, \citenamefont {{Lindfors}},
  \citenamefont {{Lombardi}}, \citenamefont {{L{\'o}pez}}, \citenamefont
  {{L{\'o}pez}}, \citenamefont {{L{\'o}pez-Oramas}}, \citenamefont {{Lorenz}},
  \citenamefont {{Makariev}}, \citenamefont {{Maneva}}, \citenamefont
  {{Mankuzhiyil}}, \citenamefont {{Mannheim}}, \citenamefont {{Maraschi}},
  \citenamefont {{Mariotti}}, \citenamefont {{Mart{\'{\i}}nez}}, \citenamefont
  {{Mazin}}, \citenamefont {{Meucci}}, \citenamefont {{Miranda}}, \citenamefont
  {{Mirzoyan}}, \citenamefont {{Mold{\'o}n}}, \citenamefont {{Moralejo}},
  \citenamefont {{Munar-Adrover}}, \citenamefont {{Niedzwiecki}}, \citenamefont
  {{Nieto}}, \citenamefont {{Nilsson}}, \citenamefont {{Nowak}}, \citenamefont
  {{Orito}}, \citenamefont {{Paiano}}, \citenamefont {{Paneque}}, \citenamefont
  {{Paoletti}}, \citenamefont {{Pardo}}, \citenamefont {{Paredes}},
  \citenamefont {{Partini}}, \citenamefont {{Perez-Torres}}, \citenamefont
  {{Persic}}, \citenamefont {{Peruzzo}}, \citenamefont {{Pilia}}, \citenamefont
  {{Pochon}}, \citenamefont {{Prada}}, \citenamefont {{Prada Moroni}},
  \citenamefont {{Prandini}}, \citenamefont {{Puerto Gimenez}}, \citenamefont
  {{Puljak}}, \citenamefont {{Reichardt}}, \citenamefont {{Reinthal}},
  \citenamefont {{Rhode}}, \citenamefont {{Rib{\'o}}}, \citenamefont {{Rico}},
  \citenamefont {{R{\"u}gamer}}, \citenamefont {{Saggion}}, \citenamefont
  {{Saito}}, \citenamefont {{Saito}}, \citenamefont {{Salvati}}, \citenamefont
  {{Satalecka}}, \citenamefont {{Scalzotto}}, \citenamefont {{Scapin}},
  \citenamefont {{Schultz}}, \citenamefont {{Schweizer}}, \citenamefont
  {{Shayduk}}, \citenamefont {{Shore}}, \citenamefont {{Sillanp{\"a}{\"a}}},
  \citenamefont {{Sitarek}}, \citenamefont {{Snidaric}}, \citenamefont
  {{Sobczynska}}, \citenamefont {{Spanier}}, \citenamefont {{Spiro}},
  \citenamefont {{Stamatescu}}, \citenamefont {{Stamerra}}, \citenamefont
  {{Steinke}}, \citenamefont {{Storz}}, \citenamefont {{Strah}}, \citenamefont
  {{Sun}}, \citenamefont {{Suri{\'c}}}, \citenamefont {{Takalo}}, \citenamefont
  {{Takami}}, \citenamefont {{Tavecchio}}, \citenamefont {{Temnikov}},
  \citenamefont {{Terzi{\'c}}}, \citenamefont {{Tescaro}}, \citenamefont
  {{Teshima}}, \citenamefont {{Tibolla}}, \citenamefont {{Torres}},
  \citenamefont {{Treves}}, \citenamefont {{Uellenbeck}}, \citenamefont
  {{Vankov}}, \citenamefont {{Vogler}}, \citenamefont {{Wagner}}, \citenamefont
  {{Weitzel}}, \citenamefont {{Zabalza}}, \citenamefont {{Zandanel}},
  \citenamefont {{Zanin}}, \citenamefont {{MAGIC Collaboration}}, \citenamefont
  {{Pfrommer}},\ and\ \citenamefont {{Pinzke}}}]{2012A&A...541A..99A}%
  \BibitemOpen
  \bibfield  {author} {\bibinfo {author} {\bibfnamefont {J.}~\bibnamefont
  {{Aleksi{\'c}}}}, \bibinfo {author} {\bibfnamefont {E.~A.}\ \bibnamefont
  {{Alvarez}}}, \bibinfo {author} {\bibfnamefont {L.~A.}\ \bibnamefont
  {{Antonelli}}}, \bibinfo {author} {\bibfnamefont {P.}~\bibnamefont
  {{Antoranz}}}, \bibinfo {author} {\bibfnamefont {M.}~\bibnamefont
  {{Asensio}}}, \bibinfo {author} {\bibfnamefont {M.}~\bibnamefont {{Backes}}},
  \bibinfo {author} {\bibfnamefont {U.}~\bibnamefont {{Barres de Almeida}}},
  \bibinfo {author} {\bibfnamefont {J.~A.}\ \bibnamefont {{Barrio}}}, \bibinfo
  {author} {\bibfnamefont {D.}~\bibnamefont {{Bastieri}}},  \emph {et~al.},\
  }\href {\doibase 10.1051/0004-6361/201118502} {\bibfield  {journal} {\bibinfo
   {journal} {Astron.Astrophys.}\ }\textbf {\bibinfo {volume} {541}},\ \bibinfo
  {eid} {A99} (\bibinfo {year} {2012})},\ \Eprint
  {http://arxiv.org/abs/1111.5544} {arXiv:1111.5544 [astro-ph.HE]} \BibitemShut
  {NoStop}%
\bibitem [{\citenamefont {{Arlen}}\ \emph {et~al.}(2012)\citenamefont
  {{Arlen}}, \citenamefont {{Aune}}, \citenamefont {{Beilicke}}, \citenamefont
  {{Benbow}}, \citenamefont {{Bouvier}}, \citenamefont {{Buckley}},
  \citenamefont {{Bugaev}}, \citenamefont {{Byrum}}, \citenamefont {{Cannon}},
  \citenamefont {{Cesarini}}, \citenamefont {{Ciupik}}, \citenamefont
  {{Collins-Hughes}}, \citenamefont {{Connolly}}, \citenamefont {{Cui}},
  \citenamefont {{Dickherber}}, \citenamefont {{Dumm}}, \citenamefont
  {{Falcone}}, \citenamefont {{Federici}}, \citenamefont {{Feng}},
  \citenamefont {{Finley}}, \citenamefont {{Finnegan}}, \citenamefont
  {{Fortson}}, \citenamefont {{Furniss}}, \citenamefont {{Galante}},
  \citenamefont {{Gall}}, \citenamefont {{Godambe}}, \citenamefont {{Griffin}},
  \citenamefont {{Grube}}, \citenamefont {{Gyuk}}, \citenamefont {{Holder}},
  \citenamefont {{Huan}}, \citenamefont {{Hughes}}, \citenamefont {{Humensky}},
  \citenamefont {{Imran}}, \citenamefont {{Kaaret}}, \citenamefont
  {{Karlsson}}, \citenamefont {{Kertzman}}, \citenamefont {{Khassen}},
  \citenamefont {{Kieda}}, \citenamefont {{Krawczynski}}, \citenamefont
  {{Krennrich}}, \citenamefont {{Lee}}, \citenamefont {{Madhavan}},
  \citenamefont {{Maier}}, \citenamefont {{Majumdar}}, \citenamefont
  {{McArthur}}, \citenamefont {{McCann}}, \citenamefont {{Moriarty}},
  \citenamefont {{Mukherjee}}, \citenamefont {{Nelson}}, \citenamefont
  {{O'Faol{\'a}in de Bhr{\'o}ithe}}, \citenamefont {{Ong}}, \citenamefont
  {{Orr}}, \citenamefont {{Otte}}, \citenamefont {{Park}}, \citenamefont
  {{Perkins}}, \citenamefont {{Pohl}}, \citenamefont {{Prokoph}}, \citenamefont
  {{Quinn}}, \citenamefont {{Ragan}}, \citenamefont {{Reyes}}, \citenamefont
  {{Reynolds}}, \citenamefont {{Roache}}, \citenamefont {{Ruppel}},
  \citenamefont {{Saxon}}, \citenamefont {{Schroedter}}, \citenamefont
  {{Sembroski}}, \citenamefont {{Skole}}, \citenamefont {{Smith}},
  \citenamefont {{Telezhinsky}}, \citenamefont {{Te{\v s}i{\'c}}},
  \citenamefont {{Theiling}}, \citenamefont {{Thibadeau}}, \citenamefont
  {{Tsurusaki}}, \citenamefont {{Varlotta}}, \citenamefont {{Vivier}},
  \citenamefont {{Wakely}}, \citenamefont {{Ward}}, \citenamefont
  {{Weinstein}}, \citenamefont {{Welsing}}, \citenamefont {{Williams}},
  \citenamefont {{Zitzer}}, \citenamefont {{Pfrommer}},\ and\ \citenamefont
  {{Pinzke}}}]{2012ApJ...757..123A}%
  \BibitemOpen
  \bibfield  {author} {\bibinfo {author} {\bibfnamefont {T.}~\bibnamefont
  {{Arlen}}}, \bibinfo {author} {\bibfnamefont {T.}~\bibnamefont {{Aune}}},
  \bibinfo {author} {\bibfnamefont {M.}~\bibnamefont {{Beilicke}}}, \bibinfo
  {author} {\bibfnamefont {W.}~\bibnamefont {{Benbow}}}, \bibinfo {author}
  {\bibfnamefont {A.}~\bibnamefont {{Bouvier}}}, \bibinfo {author}
  {\bibfnamefont {J.~H.}\ \bibnamefont {{Buckley}}}, \bibinfo {author}
  {\bibfnamefont {V.}~\bibnamefont {{Bugaev}}}, \bibinfo {author}
  {\bibfnamefont {K.}~\bibnamefont {{Byrum}}}, \bibinfo {author} {\bibfnamefont
  {A.}~\bibnamefont {{Cannon}}},  \emph {et~al.},\ }\href {\doibase
  10.1088/0004-637X/757/2/123} {\bibfield  {journal} {\bibinfo  {journal}
  {\apj}\ }\textbf {\bibinfo {volume} {757}},\ \bibinfo {eid} {123} (\bibinfo
  {year} {2012})},\ \Eprint {http://arxiv.org/abs/1208.0676} {arXiv:1208.0676
  [astro-ph.HE]} \BibitemShut {NoStop}%
\bibitem [{\citenamefont {{Abramowski}}\ \emph {et~al.}(2012)\citenamefont
  {{Abramowski}}, \citenamefont {{Acero}}, \citenamefont {{Aharonian}},
  \citenamefont {{Akhperjanian}}, \citenamefont {{Anton}}, \citenamefont
  {{Balzer}}, \citenamefont {{Barnacka}}, \citenamefont {{Barres de Almeida}},
  \citenamefont {{Becherini}}, \citenamefont {{Becker}}, \citenamefont
  {{Behera}}, \citenamefont {{Bernl{\"o}hr}}, \citenamefont {{Birsin}},
  \citenamefont {{Biteau}}, \citenamefont {{Bochow}}, \citenamefont
  {{Boisson}}, \citenamefont {{Bolmont}}, \citenamefont {{Bordas}},
  \citenamefont {{Brucker}}, \citenamefont {{Brun}}, \citenamefont {{Brun}},
  \citenamefont {{Bulik}}, \citenamefont {{B{\"u}sching}}, \citenamefont
  {{Carrigan}}, \citenamefont {{Casanova}}, \citenamefont {{Cerruti}},
  \citenamefont {{Chadwick}}, \citenamefont {{Charbonnier}}, \citenamefont
  {{Chaves}}, \citenamefont {{Cheesebrough}}, \citenamefont {{Clapson}},
  \citenamefont {{Coignet}}, \citenamefont {{Cologna}}, \citenamefont
  {{Conrad}}, \citenamefont {{Dalton}}, \citenamefont {{Daniel}}, \citenamefont
  {{Davids}}, \citenamefont {{Degrange}}, \citenamefont {{Deil}}, \citenamefont
  {{Dickinson}}, \citenamefont {{Djannati-Ata{\"i}}}, \citenamefont
  {{Domainko}}, \citenamefont {{Drury}}, \citenamefont {{Dubus}}, \citenamefont
  {{Dutson}}, \citenamefont {{Dyks}}, \citenamefont {{Dyrda}}, \citenamefont
  {{Egberts}}, \citenamefont {{Eger}}, \citenamefont {{Espigat}}, \citenamefont
  {{Fallon}}, \citenamefont {{Farnier}}, \citenamefont {{Fegan}}, \citenamefont
  {{Feinstein}}, \citenamefont {{Fernandes}}, \citenamefont {{Fiasson}},
  \citenamefont {{Fontaine}}, \citenamefont {{F{\"o}rster}}, \citenamefont
  {{F{\"u}{\ss}ling}}, \citenamefont {{Gallant}}, \citenamefont {{Gast}},
  \citenamefont {{G{\'e}rard}}, \citenamefont {{Gerbig}}, \citenamefont
  {{Giebels}}, \citenamefont {{Glicenstein}}, \citenamefont {{Gl{\"u}ck}},
  \citenamefont {{Goret}}, \citenamefont {{G{\"o}ring}}, \citenamefont
  {{H{\"a}ffner}}, \citenamefont {{Hague}}, \citenamefont {{Hampf}},
  \citenamefont {{Hauser}}, \citenamefont {{Heinz}}, \citenamefont
  {{Heinzelmann}}, \citenamefont {{Henri}}, \citenamefont {{Hermann}},
  \citenamefont {{Hinton}}, \citenamefont {{Hoffmann}}, \citenamefont
  {{Hofmann}}, \citenamefont {{Hofverberg}}, \citenamefont {{Holler}},
  \citenamefont {{Horns}}, \citenamefont {{Jacholkowska}}, \citenamefont {{de
  Jager}}, \citenamefont {{Jahn}}, \citenamefont {{Jamrozy}}, \citenamefont
  {{Jung}}, \citenamefont {{Kastendieck}}, \citenamefont {{Katarzy{\'n}ski}},
  \citenamefont {{Katz}}, \citenamefont {{Kaufmann}}, \citenamefont {{Keogh}},
  \citenamefont {{Khangulyan}}, \citenamefont {{Kh{\'e}lifi}}, \citenamefont
  {{Klochkov}}, \citenamefont {{Klu{\'z}niak}}, \citenamefont {{Kneiske}},
  \citenamefont {{Komin}}, \citenamefont {{Kosack}}, \citenamefont
  {{Kossakowski}}, \citenamefont {{Laffon}}, \citenamefont {{Lamanna}},
  \citenamefont {{Lennarz}}, \citenamefont {{Lohse}}, \citenamefont
  {{Lopatin}}, \citenamefont {{Lu}}, \citenamefont {{Marandon}}, \citenamefont
  {{Marcowith}}, \citenamefont {{Masbou}}, \citenamefont {{Maurin}},
  \citenamefont {{Maxted}}, \citenamefont {{Mayer}}, \citenamefont {{McComb}},
  \citenamefont {{Medina}}, \citenamefont {{M{\'e}hault}}, \citenamefont
  {{Moderski}}, \citenamefont {{Moulin}}, \citenamefont {{Naumann}},
  \citenamefont {{Naumann-Godo}}, \citenamefont {{de Naurois}}, \citenamefont
  {{Nedbal}}, \citenamefont {{Nekrassov}}, \citenamefont {{Nguyen}},
  \citenamefont {{Nicholas}}, \citenamefont {{Niemiec}}, \citenamefont
  {{Nolan}}, \citenamefont {{Ohm}}, \citenamefont {{de O{\~n}a Wilhelmi}},
  \citenamefont {{Opitz}}, \citenamefont {{Ostrowski}}, \citenamefont {{Oya}},
  \citenamefont {{Panter}}, \citenamefont {{Paz Arribas}}, \citenamefont
  {{Pedaletti}}, \citenamefont {{Pelletier}}, \citenamefont {{Petrucci}},
  \citenamefont {{Pita}}, \citenamefont {{P{\"u}hlhofer}}, \citenamefont
  {{Punch}}, \citenamefont {{Quirrenbach}}, \citenamefont {{Raue}},
  \citenamefont {{Rayner}}, \citenamefont {{Reimer}}, \citenamefont {{Reimer}},
  \citenamefont {{Renaud}}, \citenamefont {{de los Reyes}}, \citenamefont
  {{Rieger}}, \citenamefont {{Ripken}}, \citenamefont {{Rob}}, \citenamefont
  {{Rosier-Lees}}, \citenamefont {{Rowell}}, \citenamefont {{Rudak}},
  \citenamefont {{Rulten}}, \citenamefont {{Ruppel}}, \citenamefont
  {{Sahakian}}, \citenamefont {{Sanchez}}, \citenamefont {{Santangelo}},
  \citenamefont {{Schlickeiser}}, \citenamefont {{Sch{\"o}ck}}, \citenamefont
  {{Schulz}}, \citenamefont {{Schwanke}}, \citenamefont {{Schwarzburg}},
  \citenamefont {{Schwemmer}}, \citenamefont {{Sheidaei}}, \citenamefont
  {{Skilton}}, \citenamefont {{Sol}}, \citenamefont {{Spengler}}, \citenamefont
  {{Stawarz}}, \citenamefont {{Steenkamp}}, \citenamefont {{Stegmann}},
  \citenamefont {{Stinzing}}, \citenamefont {{Stycz}}, \citenamefont
  {{Sushch}}, \citenamefont {{Szostek}}, \citenamefont {{Tavernet}},
  \citenamefont {{Terrier}}, \citenamefont {{Tluczykont}}, \citenamefont
  {{Valerius}}, \citenamefont {{van Eldik}}, \citenamefont {{Vasileiadis}},
  \citenamefont {{Venter}}, \citenamefont {{Vialle}}, \citenamefont {{Viana}},
  \citenamefont {{Vincent}}, \citenamefont {{V{\"o}lk}}, \citenamefont
  {{Volpe}}, \citenamefont {{Vorobiov}}, \citenamefont {{Vorster}},
  \citenamefont {{Wagner}}, \citenamefont {{Ward}}, \citenamefont {{White}},
  \citenamefont {{Wierzcholska}}, \citenamefont {{Zacharias}}, \citenamefont
  {{Zajczyk}}, \citenamefont {{Zdziarski}}, \citenamefont {{Zech}},
  \citenamefont {{Zechlin}},\ and\ \citenamefont
  {{H.~E.~S.~S.~Collaboration}}}]{2012ApJ...750..123A}%
  \BibitemOpen
  \bibfield  {author} {\bibinfo {author} {\bibfnamefont {A.}~\bibnamefont
  {{Abramowski}}}, \bibinfo {author} {\bibfnamefont {F.}~\bibnamefont
  {{Acero}}}, \bibinfo {author} {\bibfnamefont {F.}~\bibnamefont
  {{Aharonian}}}, \bibinfo {author} {\bibfnamefont {A.~G.}\ \bibnamefont
  {{Akhperjanian}}}, \bibinfo {author} {\bibfnamefont {G.}~\bibnamefont
  {{Anton}}}, \bibinfo {author} {\bibfnamefont {A.}~\bibnamefont {{Balzer}}},
  \bibinfo {author} {\bibfnamefont {A.}~\bibnamefont {{Barnacka}}}, \bibinfo
  {author} {\bibfnamefont {U.}~\bibnamefont {{Barres de Almeida}}}, \bibinfo
  {author} {\bibfnamefont {Y.}~\bibnamefont {{Becherini}}},  \emph {et~al.},\
  }\href {\doibase 10.1088/0004-637X/750/2/123} {\bibfield  {journal} {\bibinfo
   {journal} {\apj}\ }\textbf {\bibinfo {volume} {750}},\ \bibinfo {eid} {123}
  (\bibinfo {year} {2012})},\ \Eprint {http://arxiv.org/abs/1202.5494}
  {arXiv:1202.5494 [astro-ph.HE]} \BibitemShut {NoStop}%
\bibitem [{\citenamefont {Ackermann}\ \emph {et~al.}(2010)\citenamefont
  {Ackermann}, \citenamefont {Ajello}, \citenamefont {Allafort}, \citenamefont
  {Baldini}, \citenamefont {Ballet} \emph {et~al.}}]{Ackermann:2010rg}%
  \BibitemOpen
  \bibfield  {author} {\bibinfo {author} {\bibfnamefont {M.}~\bibnamefont
  {Ackermann}}, \bibinfo {author} {\bibfnamefont {M.}~\bibnamefont {Ajello}},
  \bibinfo {author} {\bibfnamefont {A.}~\bibnamefont {Allafort}}, \bibinfo
  {author} {\bibfnamefont {L.}~\bibnamefont {Baldini}}, \bibinfo {author}
  {\bibfnamefont {J.}~\bibnamefont {Ballet}},  \emph {et~al.},\ }\href
  {\doibase 10.1088/1475-7516/2010/05/025} {\bibfield  {journal} {\bibinfo
  {journal} {JCAP}\ }\textbf {\bibinfo {volume} {1005}},\ \bibinfo {pages}
  {025} (\bibinfo {year} {2010})},\ \Eprint {http://arxiv.org/abs/1002.2239}
  {arXiv:1002.2239 [astro-ph.CO]} \BibitemShut {NoStop}%
\bibitem [{\citenamefont {Huang}\ \emph {et~al.}(2012)\citenamefont {Huang},
  \citenamefont {Vertongen},\ and\ \citenamefont {Weniger}}]{Huang:2011xr}%
  \BibitemOpen
  \bibfield  {author} {\bibinfo {author} {\bibfnamefont {X.}~\bibnamefont
  {Huang}}, \bibinfo {author} {\bibfnamefont {G.}~\bibnamefont {Vertongen}}, \
  and\ \bibinfo {author} {\bibfnamefont {C.}~\bibnamefont {Weniger}},\ }\href
  {\doibase 10.1088/1475-7516/2012/01/042} {\bibfield  {journal} {\bibinfo
  {journal} {JCAP}\ }\textbf {\bibinfo {volume} {1201}},\ \bibinfo {pages}
  {042} (\bibinfo {year} {2012})},\ \Eprint {http://arxiv.org/abs/1110.1529}
  {arXiv:1110.1529 [hep-ph]} \BibitemShut {NoStop}%
\bibitem [{\citenamefont {{Ando}}\ and\ \citenamefont
  {{Nagai}}(2012)}]{2012JCAP...07..017A}%
  \BibitemOpen
  \bibfield  {author} {\bibinfo {author} {\bibfnamefont {S.}~\bibnamefont
  {{Ando}}}\ and\ \bibinfo {author} {\bibfnamefont {D.}~\bibnamefont
  {{Nagai}}},\ }\href {\doibase 10.1088/1475-7516/2012/07/017} {\bibfield
  {journal} {\bibinfo  {journal} {J.Cosmol.Astropart.Phys.}\ }\textbf {\bibinfo
  {volume} {7}},\ \bibinfo {eid} {017} (\bibinfo {year} {2012})},\ \Eprint
  {http://arxiv.org/abs/1201.0753} {arXiv:1201.0753 [astro-ph.HE]} \BibitemShut
  {NoStop}%
\bibitem [{\citenamefont {Blumenthal}\ \emph {et~al.}(1986)\citenamefont
  {Blumenthal}, \citenamefont {Faber}, \citenamefont {Flores},\ and\
  \citenamefont {Primack}}]{Blumenthal:1985qy}%
  \BibitemOpen
  \bibfield  {author} {\bibinfo {author} {\bibfnamefont {G.~R.}\ \bibnamefont
  {Blumenthal}}, \bibinfo {author} {\bibfnamefont {S.}~\bibnamefont {Faber}},
  \bibinfo {author} {\bibfnamefont {R.}~\bibnamefont {Flores}}, \ and\ \bibinfo
  {author} {\bibfnamefont {J.~R.}\ \bibnamefont {Primack}},\ }\href {\doibase
  10.1086/163867} {\bibfield  {journal} {\bibinfo  {journal} {Astrophys.J.}\
  }\textbf {\bibinfo {volume} {301}},\ \bibinfo {pages} {27} (\bibinfo {year}
  {1986})}\BibitemShut {NoStop}%
\bibitem [{\citenamefont {Gnedin}\ \emph {et~al.}(2011)\citenamefont {Gnedin},
  \citenamefont {Ceverino}, \citenamefont {Gnedin}, \citenamefont {Klypin},
  \citenamefont {Kravtsov} \emph {et~al.}}]{Gnedin:2011uj}%
  \BibitemOpen
  \bibfield  {author} {\bibinfo {author} {\bibfnamefont {O.~Y.}\ \bibnamefont
  {Gnedin}}, \bibinfo {author} {\bibfnamefont {D.}~\bibnamefont {Ceverino}},
  \bibinfo {author} {\bibfnamefont {N.~Y.}\ \bibnamefont {Gnedin}}, \bibinfo
  {author} {\bibfnamefont {A.~A.}\ \bibnamefont {Klypin}}, \bibinfo {author}
  {\bibfnamefont {A.~V.}\ \bibnamefont {Kravtsov}},  \emph {et~al.},\
  }\href@noop {} {\  (\bibinfo {year} {2011})},\ \Eprint
  {http://arxiv.org/abs/1108.5736} {arXiv:1108.5736 [astro-ph.CO]} \BibitemShut
  {NoStop}%
\bibitem [{\citenamefont {Ghigna}\ \emph {et~al.}(1998)\citenamefont {Ghigna},
  \citenamefont {Moore}, \citenamefont {Governato}, \citenamefont {Lake},
  \citenamefont {Quinn} \emph {et~al.}}]{Ghigna:1998vn}%
  \BibitemOpen
  \bibfield  {author} {\bibinfo {author} {\bibfnamefont {S.}~\bibnamefont
  {Ghigna}}, \bibinfo {author} {\bibfnamefont {B.}~\bibnamefont {Moore}},
  \bibinfo {author} {\bibfnamefont {F.}~\bibnamefont {Governato}}, \bibinfo
  {author} {\bibfnamefont {G.}~\bibnamefont {Lake}}, \bibinfo {author}
  {\bibfnamefont {T.~R.}\ \bibnamefont {Quinn}},  \emph {et~al.},\ }\href
  {\doibase 10.1046/j.1365-8711.1998.01918.x} {\bibfield  {journal} {\bibinfo
  {journal} {Mon.Not.Roy.Astron.Soc.}\ }\textbf {\bibinfo {volume} {300}},\
  \bibinfo {pages} {146} (\bibinfo {year} {1998})},\ \Eprint
  {http://arxiv.org/abs/astro-ph/9801192} {arXiv:astro-ph/9801192 [astro-ph]}
  \BibitemShut {NoStop}%
\bibitem [{\citenamefont {{Pinzke}}\ \emph {et~al.}(2011)\citenamefont
  {{Pinzke}}, \citenamefont {{Pfrommer}},\ and\ \citenamefont
  {{Bergstr{\"o}m}}}]{2011PhRvD..84l3509P}%
  \BibitemOpen
  \bibfield  {author} {\bibinfo {author} {\bibfnamefont {A.}~\bibnamefont
  {{Pinzke}}}, \bibinfo {author} {\bibfnamefont {C.}~\bibnamefont
  {{Pfrommer}}}, \ and\ \bibinfo {author} {\bibfnamefont {L.}~\bibnamefont
  {{Bergstr{\"o}m}}},\ }\href {\doibase 10.1103/PhysRevD.84.123509} {\bibfield
  {journal} {\bibinfo  {journal} {Phys.Rev.}\ }\textbf {\bibinfo {volume}
  {D84}},\ \bibinfo {eid} {123509} (\bibinfo {year} {2011})},\ \Eprint
  {http://arxiv.org/abs/1105.3240} {arXiv:1105.3240 [astro-ph.HE]} \BibitemShut
  {NoStop}%
\bibitem [{\citenamefont {{Gao}}\ \emph
  {et~al.}(2012{\natexlab{a}})\citenamefont {{Gao}}, \citenamefont {{Navarro}},
  \citenamefont {{Frenk}}, \citenamefont {{Jenkins}}, \citenamefont
  {{Springel}},\ and\ \citenamefont {{White}}}]{2012MNRAS.425.2169G}%
  \BibitemOpen
  \bibfield  {author} {\bibinfo {author} {\bibfnamefont {L.}~\bibnamefont
  {{Gao}}}, \bibinfo {author} {\bibfnamefont {J.~F.}\ \bibnamefont
  {{Navarro}}}, \bibinfo {author} {\bibfnamefont {C.~S.}\ \bibnamefont
  {{Frenk}}}, \bibinfo {author} {\bibfnamefont {A.}~\bibnamefont {{Jenkins}}},
  \bibinfo {author} {\bibfnamefont {V.}~\bibnamefont {{Springel}}}, \ and\
  \bibinfo {author} {\bibfnamefont {S.~D.~M.}\ \bibnamefont {{White}}},\ }\href
  {\doibase 10.1111/j.1365-2966.2012.21564.x} {\bibfield  {journal} {\bibinfo
  {journal} {Mon.Not.R.Astron.Soc.}\ }\textbf {\bibinfo {volume} {425}},\
  \bibinfo {pages} {2169} (\bibinfo {year} {2012}{\natexlab{a}})},\ \Eprint
  {http://arxiv.org/abs/1201.1940} {arXiv:1201.1940 [astro-ph.CO]} \BibitemShut
  {NoStop}%
\bibitem [{\citenamefont {{Gao}}\ \emph
  {et~al.}(2012{\natexlab{b}})\citenamefont {{Gao}}, \citenamefont {{Frenk}},
  \citenamefont {{Jenkins}}, \citenamefont {{Springel}},\ and\ \citenamefont
  {{White}}}]{2012MNRAS.419.1721G}%
  \BibitemOpen
  \bibfield  {author} {\bibinfo {author} {\bibfnamefont {L.}~\bibnamefont
  {{Gao}}}, \bibinfo {author} {\bibfnamefont {C.~S.}\ \bibnamefont {{Frenk}}},
  \bibinfo {author} {\bibfnamefont {A.}~\bibnamefont {{Jenkins}}}, \bibinfo
  {author} {\bibfnamefont {V.}~\bibnamefont {{Springel}}}, \ and\ \bibinfo
  {author} {\bibfnamefont {S.~D.~M.}\ \bibnamefont {{White}}},\ }\href
  {\doibase 10.1111/j.1365-2966.2011.19836.x} {\bibfield  {journal} {\bibinfo
  {journal} {Mon.Not.R.Astron.Soc.}\ }\textbf {\bibinfo {volume} {419}},\
  \bibinfo {pages} {1721} (\bibinfo {year} {2012}{\natexlab{b}})},\ \Eprint
  {http://arxiv.org/abs/1107.1916} {arXiv:1107.1916 [astro-ph.CO]} \BibitemShut
  {NoStop}%
\bibitem [{\citenamefont {{Doro}}\ \emph {et~al.}(2013)\citenamefont {{Doro}},
  \citenamefont {{Conrad}}, \citenamefont {{Emmanoulopoulos}}, \citenamefont
  {{S{\`a}nchez-Conde}}, \citenamefont {{Barrio}}, \citenamefont {{Birsin}},
  \citenamefont {{Bolmont}}, \citenamefont {{Brun}}, \citenamefont
  {{Colafrancesco}}, \citenamefont {{Connell}}, \citenamefont {{Contreras}},
  \citenamefont {{Daniel}}, \citenamefont {{Fornasa}}, \citenamefont {{Gaug}},
  \citenamefont {{Glicenstein}}, \citenamefont {{Gonz{\'a}lez-Mu{\~n}oz}},
  \citenamefont {{Hassan}}, \citenamefont {{Horns}}, \citenamefont
  {{Jacholkowska}}, \citenamefont {{Jahn}}, \citenamefont {{Mazini}},
  \citenamefont {{Mirabal}}, \citenamefont {{Moralejo}}, \citenamefont
  {{Moulin}}, \citenamefont {{Nieto}}, \citenamefont {{Ripken}}, \citenamefont
  {{Sandaker}}, \citenamefont {{Schwanke}}, \citenamefont {{Spengler}},
  \citenamefont {{Stamerra}}, \citenamefont {{Viana}}, \citenamefont
  {{Zechlin}}, \citenamefont {{Zimmer}},\ and\ \citenamefont {{CTA
  Consortium}}}]{2013APh....43..189D}%
  \BibitemOpen
  \bibfield  {author} {\bibinfo {author} {\bibfnamefont {M.}~\bibnamefont
  {{Doro}}}, \bibinfo {author} {\bibfnamefont {J.}~\bibnamefont {{Conrad}}},
  \bibinfo {author} {\bibfnamefont {D.}~\bibnamefont {{Emmanoulopoulos}}},
  \bibinfo {author} {\bibfnamefont {M.~A.}\ \bibnamefont
  {{S{\`a}nchez-Conde}}}, \bibinfo {author} {\bibfnamefont {J.~A.}\
  \bibnamefont {{Barrio}}}, \bibinfo {author} {\bibfnamefont {E.}~\bibnamefont
  {{Birsin}}}, \bibinfo {author} {\bibfnamefont {J.}~\bibnamefont {{Bolmont}}},
  \bibinfo {author} {\bibfnamefont {P.}~\bibnamefont {{Brun}}}, \bibinfo
  {author} {\bibfnamefont {S.}~\bibnamefont {{Colafrancesco}}},  \emph
  {et~al.},\ }\href {\doibase 10.1016/j.astropartphys.2012.08.002} {\bibfield
  {journal} {\bibinfo  {journal} {Astroparticle Physics}\ }\textbf {\bibinfo
  {volume} {43}},\ \bibinfo {pages} {189} (\bibinfo {year} {2013})},\ \Eprint
  {http://arxiv.org/abs/1208.5356} {arXiv:1208.5356 [astro-ph.IM]} \BibitemShut
  {NoStop}%
\bibitem [{\citenamefont {Dugger}\ \emph {et~al.}(2010)\citenamefont {Dugger},
  \citenamefont {Jeltema},\ and\ \citenamefont {Profumo}}]{Dugger:2010ys}%
  \BibitemOpen
  \bibfield  {author} {\bibinfo {author} {\bibfnamefont {L.}~\bibnamefont
  {Dugger}}, \bibinfo {author} {\bibfnamefont {T.~E.}\ \bibnamefont {Jeltema}},
  \ and\ \bibinfo {author} {\bibfnamefont {S.}~\bibnamefont {Profumo}},\ }\href
  {\doibase 10.1088/1475-7516/2010/12/015} {\bibfield  {journal} {\bibinfo
  {journal} {JCAP}\ }\textbf {\bibinfo {volume} {1012}},\ \bibinfo {pages}
  {015} (\bibinfo {year} {2010})},\ \Eprint {http://arxiv.org/abs/1009.5988}
  {arXiv:1009.5988 [astro-ph.HE]} \BibitemShut {NoStop}%
\bibitem [{\citenamefont {Ke}\ \emph {et~al.}(2011)\citenamefont {Ke},
  \citenamefont {Luo}, \citenamefont {Wang},\ and\ \citenamefont
  {Zhu}}]{Ke:2011xw}%
  \BibitemOpen
  \bibfield  {author} {\bibinfo {author} {\bibfnamefont {J.}~\bibnamefont
  {Ke}}, \bibinfo {author} {\bibfnamefont {M.}~\bibnamefont {Luo}}, \bibinfo
  {author} {\bibfnamefont {L.}~\bibnamefont {Wang}}, \ and\ \bibinfo {author}
  {\bibfnamefont {G.}~\bibnamefont {Zhu}},\ }\href {\doibase
  10.1016/j.physletb.2011.02.055} {\bibfield  {journal} {\bibinfo  {journal}
  {Phys.Lett.}\ }\textbf {\bibinfo {volume} {B698}},\ \bibinfo {pages} {44}
  (\bibinfo {year} {2011})},\ \Eprint {http://arxiv.org/abs/1101.5878}
  {arXiv:1101.5878 [hep-ph]} \BibitemShut {NoStop}%
\bibitem [{xra(2014)}]{xraycomp}%
  \BibitemOpen
  \href@noop {} {\emph {\bibinfo {title} {XMM-Newton Users Handbook (Issue
  2.12)}}},\ \bibinfo {organization} {ESA: XMM-Newton SOC} (\bibinfo {year}
  {2014})\BibitemShut {NoStop}%
\bibitem [{\citenamefont {{Greiner}}\ \emph {et~al.}(2012)\citenamefont
  {{Greiner}}, \citenamefont {{Mannheim}}, \citenamefont {{Aharonian}},
  \citenamefont {{Ajello}}, \citenamefont {{Balasz}}, \citenamefont
  {{Barbiellini}}, \citenamefont {{Bellazzini}}, \citenamefont {{Bishop}},
  \citenamefont {{Bisnovatij-Kogan}}, \citenamefont {{Boggs}}, \citenamefont
  {{Bykov}}, \citenamefont {{DiCocco}}, \citenamefont {{Diehl}}, \citenamefont
  {{Els{\"a}sser}}, \citenamefont {{Foley}}, \citenamefont {{Fransson}},
  \citenamefont {{Gehrels}}, \citenamefont {{Hanlon}}, \citenamefont
  {{Hartmann}}, \citenamefont {{Hermsen}}, \citenamefont {{Hillebrandt}},
  \citenamefont {{Hudec}}, \citenamefont {{Iyudin}}, \citenamefont {{Jose}},
  \citenamefont {{Kadler}}, \citenamefont {{Kanbach}}, \citenamefont
  {{Klamra}}, \citenamefont {{Kiener}}, \citenamefont {{Klose}}, \citenamefont
  {{Kreykenbohm}}, \citenamefont {{Kuiper}}, \citenamefont {{Kylafis}},
  \citenamefont {{Labanti}}, \citenamefont {{Langanke}}, \citenamefont
  {{Langer}}, \citenamefont {{Larsson}}, \citenamefont {{Leibundgut}},
  \citenamefont {{Laux}}, \citenamefont {{Longo}}, \citenamefont {{Maeda}},
  \citenamefont {{Marcinkowski}}, \citenamefont {{Marisaldi}}, \citenamefont
  {{McBreen}}, \citenamefont {{McBreen}}, \citenamefont {{Meszaros}},
  \citenamefont {{Nomoto}}, \citenamefont {{Pearce}}, \citenamefont {{Peer}},
  \citenamefont {{Pian}}, \citenamefont {{Prantzos}}, \citenamefont
  {{Raffelt}}, \citenamefont {{Reimer}}, \citenamefont {{Rhode}}, \citenamefont
  {{Ryde}}, \citenamefont {{Schmidt}}, \citenamefont {{Silk}}, \citenamefont
  {{Shustov}}, \citenamefont {{Strong}}, \citenamefont {{Tanvir}},
  \citenamefont {{Thielemann}}, \citenamefont {{Tibolla}}, \citenamefont
  {{Tierney}}, \citenamefont {{Tr{\"u}mper}}, \citenamefont {{Varshalovich}},
  \citenamefont {{Wilms}}, \citenamefont {{Wrochna}}, \citenamefont
  {{Zdziarski}},\ and\ \citenamefont {{Zoglauer}}}]{2012ExA....34..551G}%
  \BibitemOpen
  \bibfield  {author} {\bibinfo {author} {\bibfnamefont {J.}~\bibnamefont
  {{Greiner}}}, \bibinfo {author} {\bibfnamefont {K.}~\bibnamefont
  {{Mannheim}}}, \bibinfo {author} {\bibfnamefont {F.}~\bibnamefont
  {{Aharonian}}}, \bibinfo {author} {\bibfnamefont {M.}~\bibnamefont
  {{Ajello}}}, \bibinfo {author} {\bibfnamefont {L.~G.}\ \bibnamefont
  {{Balasz}}}, \bibinfo {author} {\bibfnamefont {G.}~\bibnamefont
  {{Barbiellini}}}, \bibinfo {author} {\bibfnamefont {R.}~\bibnamefont
  {{Bellazzini}}}, \bibinfo {author} {\bibfnamefont {S.}~\bibnamefont
  {{Bishop}}}, \bibinfo {author} {\bibfnamefont {G.~S.}\ \bibnamefont
  {{Bisnovatij-Kogan}}},  \emph {et~al.},\ }\href {\doibase
  10.1007/s10686-011-9255-0} {\bibfield  {journal} {\bibinfo  {journal}
  {Experimental Astronomy}\ }\textbf {\bibinfo {volume} {34}},\ \bibinfo
  {pages} {551} (\bibinfo {year} {2012})},\ \Eprint
  {http://arxiv.org/abs/1105.1265} {arXiv:1105.1265 [astro-ph.HE]} \BibitemShut
  {NoStop}%
\bibitem [{\citenamefont {Gros}\ \emph {et~al.}(2003)\citenamefont {Gros},
  \citenamefont {Goldwurm}, \citenamefont {Cadolle-Bel}, \citenamefont
  {Goldoni}, \citenamefont {Rodriguez} \emph {et~al.}}]{Gros:2003wb}%
  \BibitemOpen
  \bibfield  {author} {\bibinfo {author} {\bibfnamefont {A.}~\bibnamefont
  {Gros}}, \bibinfo {author} {\bibfnamefont {A.}~\bibnamefont {Goldwurm}},
  \bibinfo {author} {\bibfnamefont {M.}~\bibnamefont {Cadolle-Bel}}, \bibinfo
  {author} {\bibfnamefont {P.}~\bibnamefont {Goldoni}}, \bibinfo {author}
  {\bibfnamefont {J.}~\bibnamefont {Rodriguez}},  \emph {et~al.},\ }\href
  {\doibase 10.1051/0004-6361:20031584} {\bibfield  {journal} {\bibinfo
  {journal} {Astron.Astrophys.}\ }\textbf {\bibinfo {volume} {411}},\ \bibinfo
  {pages} {L179} (\bibinfo {year} {2003})},\ \Eprint
  {http://arxiv.org/abs/astro-ph/0311176} {arXiv:astro-ph/0311176 [astro-ph]}
  \BibitemShut {NoStop}%
\bibitem [{\citenamefont {{Schoenfelder}}\ \emph {et~al.}(1993)\citenamefont
  {{Schoenfelder}}, \citenamefont {{Aarts}}, \citenamefont {{Bennett}},
  \citenamefont {{de Boer}}, \citenamefont {{Clear}}, \citenamefont
  {{Collmar}}, \citenamefont {{Connors}}, \citenamefont {{Deerenberg}},
  \citenamefont {{Diehl}}, \citenamefont {{von Dordrecht}}, \citenamefont {{den
  Herder}}, \citenamefont {{Hermsen}}, \citenamefont {{Kippen}}, \citenamefont
  {{Kuiper}}, \citenamefont {{Lichti}}, \citenamefont {{Lockwood}},
  \citenamefont {{Macri}}, \citenamefont {{McConnell}}, \citenamefont
  {{Morris}}, \citenamefont {{Much}}, \citenamefont {{Ryan}}, \citenamefont
  {{Simpson}}, \citenamefont {{Snelling}}, \citenamefont {{Stacy}},
  \citenamefont {{Steinle}}, \citenamefont {{Strong}}, \citenamefont
  {{Swanenburg}}, \citenamefont {{Taylor}}, \citenamefont {{de Vries}},\ and\
  \citenamefont {{Winkler}}}]{1993ApJS...86..657S}%
  \BibitemOpen
  \bibfield  {author} {\bibinfo {author} {\bibfnamefont {V.}~\bibnamefont
  {{Schoenfelder}}}, \bibinfo {author} {\bibfnamefont {H.}~\bibnamefont
  {{Aarts}}}, \bibinfo {author} {\bibfnamefont {K.}~\bibnamefont {{Bennett}}},
  \bibinfo {author} {\bibfnamefont {H.}~\bibnamefont {{de Boer}}}, \bibinfo
  {author} {\bibfnamefont {J.}~\bibnamefont {{Clear}}}, \bibinfo {author}
  {\bibfnamefont {W.}~\bibnamefont {{Collmar}}}, \bibinfo {author}
  {\bibfnamefont {A.}~\bibnamefont {{Connors}}}, \bibinfo {author}
  {\bibfnamefont {A.}~\bibnamefont {{Deerenberg}}}, \bibinfo {author}
  {\bibfnamefont {R.}~\bibnamefont {{Diehl}}},  \emph {et~al.},\ }\href
  {\doibase 10.1086/191794} {\bibfield  {journal} {\bibinfo  {journal}
  {Astrophys.J.Suppl.}\ }\textbf {\bibinfo {volume} {86}},\ \bibinfo {pages}
  {657} (\bibinfo {year} {1993})}\BibitemShut {NoStop}%
\bibitem [{\citenamefont {Thompson}\ \emph {et~al.}(1993)\citenamefont
  {Thompson}, \citenamefont {Bertsch}, \citenamefont {Fichtel}, \citenamefont
  {Hartman}, \citenamefont {Hofstadter} \emph {et~al.}}]{Thompson:1993zz}%
  \BibitemOpen
  \bibfield  {author} {\bibinfo {author} {\bibfnamefont {D.}~\bibnamefont
  {Thompson}}, \bibinfo {author} {\bibfnamefont {D.}~\bibnamefont {Bertsch}},
  \bibinfo {author} {\bibfnamefont {C.}~\bibnamefont {Fichtel}}, \bibinfo
  {author} {\bibfnamefont {R.}~\bibnamefont {Hartman}}, \bibinfo {author}
  {\bibfnamefont {R.}~\bibnamefont {Hofstadter}},  \emph {et~al.},\ }\href
  {\doibase 10.1086/191793} {\bibfield  {journal} {\bibinfo  {journal}
  {Astrophys.J.Suppl.}\ }\textbf {\bibinfo {volume} {86}},\ \bibinfo {pages}
  {629} (\bibinfo {year} {1993})}\BibitemShut {NoStop}%
\bibitem [{\citenamefont {Funk}\ and\ \citenamefont
  {Hinton}(2013)}]{Funk:2012ca}%
  \BibitemOpen
  \bibfield  {author} {\bibinfo {author} {\bibfnamefont {S.}~\bibnamefont
  {Funk}}\ and\ \bibinfo {author} {\bibfnamefont {J.~A.}\ \bibnamefont
  {Hinton}},\ }\href {\doibase 10.1016/j.astropartphys.2012.05.018} {\bibfield
  {journal} {\bibinfo  {journal} {Astropart.Phys.}\ }\textbf {\bibinfo {volume}
  {43}},\ \bibinfo {pages} {348} (\bibinfo {year} {2013})},\ \Eprint
  {http://arxiv.org/abs/1205.0832} {arXiv:1205.0832 [astro-ph.HE]} \BibitemShut
  {NoStop}%
\bibitem [{\citenamefont {Hofmann}(2006)}]{Hofmann:2006wf}%
  \BibitemOpen
  \bibfield  {author} {\bibinfo {author} {\bibfnamefont {W.}~\bibnamefont
  {Hofmann}},\ }\href@noop {} {\  (\bibinfo {year} {2006})},\ \Eprint
  {http://arxiv.org/abs/astro-ph/0603076} {arXiv:astro-ph/0603076 [astro-ph]}
  \BibitemShut {NoStop}%
\bibitem [{\citenamefont {{S{\'a}nchez-Conde}}\ \emph
  {et~al.}(2011)\citenamefont {{S{\'a}nchez-Conde}}, \citenamefont {{Cannoni}},
  \citenamefont {{Zandanel}}, \citenamefont {{G{\'o}mez}},\ and\ \citenamefont
  {{Prada}}}]{2011JCAP...12..011S}%
  \BibitemOpen
  \bibfield  {author} {\bibinfo {author} {\bibfnamefont {M.~A.}\ \bibnamefont
  {{S{\'a}nchez-Conde}}}, \bibinfo {author} {\bibfnamefont {M.}~\bibnamefont
  {{Cannoni}}}, \bibinfo {author} {\bibfnamefont {F.}~\bibnamefont
  {{Zandanel}}}, \bibinfo {author} {\bibfnamefont {M.~E.}\ \bibnamefont
  {{G{\'o}mez}}}, \ and\ \bibinfo {author} {\bibfnamefont {F.}~\bibnamefont
  {{Prada}}},\ }\href {\doibase 10.1088/1475-7516/2011/12/011} {\bibfield
  {journal} {\bibinfo  {journal} {JCAP}\ }\textbf {\bibinfo {volume} {12}},\
  \bibinfo {eid} {011} (\bibinfo {year} {2011})},\ \Eprint
  {http://arxiv.org/abs/1104.3530} {arXiv:1104.3530 [astro-ph.HE]} \BibitemShut
  {NoStop}%
\bibitem [{\citenamefont {Pinzke}\ and\ \citenamefont
  {Pfrommer}(2010)}]{Pinzke:2010st}%
  \BibitemOpen
  \bibfield  {author} {\bibinfo {author} {\bibfnamefont {A.}~\bibnamefont
  {Pinzke}}\ and\ \bibinfo {author} {\bibfnamefont {C.}~\bibnamefont
  {Pfrommer}},\ }\href {\doibase 10.1111/j.1365-2966.2010.17328.x} {\bibfield
  {journal} {\bibinfo  {journal} {Mon.Not.Roy.Astron.Soc.}\ }\textbf {\bibinfo
  {volume} {409}},\ \bibinfo {pages} {449} (\bibinfo {year} {2010})},\ \Eprint
  {http://arxiv.org/abs/1001.5023} {arXiv:1001.5023 [astro-ph.CO]} \BibitemShut
  {NoStop}%
\bibitem [{\citenamefont {Harvey}\ \emph {et~al.}(2015)\citenamefont {Harvey},
  \citenamefont {Massey}, \citenamefont {Kitching}, \citenamefont {Taylor},\
  and\ \citenamefont {Tittley}}]{Harvey:2015hha}%
  \BibitemOpen
  \bibfield  {author} {\bibinfo {author} {\bibfnamefont {D.}~\bibnamefont
  {Harvey}}, \bibinfo {author} {\bibfnamefont {R.}~\bibnamefont {Massey}},
  \bibinfo {author} {\bibfnamefont {T.}~\bibnamefont {Kitching}}, \bibinfo
  {author} {\bibfnamefont {A.}~\bibnamefont {Taylor}}, \ and\ \bibinfo {author}
  {\bibfnamefont {E.}~\bibnamefont {Tittley}},\ }\href {\doibase
  10.1126/science.1261381} {\bibfield  {journal} {\bibinfo  {journal}
  {Science}\ }\textbf {\bibinfo {volume} {347}},\ \bibinfo {pages} {1462}
  (\bibinfo {year} {2015})},\ \Eprint {http://arxiv.org/abs/1503.07675}
  {arXiv:1503.07675 [astro-ph.CO]} \BibitemShut {NoStop}%
\end{thebibliography}%

\end{document}